\documentclass[ aip
 amsmath,amssymb,
 reprint]{revtex4-1}
\usepackage{amsmath}
\usepackage{multirow}
\usepackage{placeins}
\usepackage{graphicx}
\usepackage{dcolumn}
\usepackage{bm}
\usepackage[T1]{fontenc}
\usepackage[font=small,labelfont=bf,tableposition=top]{caption}
\usepackage{subcaption}
\usepackage{graphicx}
\usepackage{dcolumn}
\usepackage{bm}
\usepackage[utf8]{inputenc}
\usepackage{mathptmx}
\usepackage{etoolbox}
\usepackage{hyperref}

\newcommand{\figref}[1]{Fig.\ \ref{#1}}

\newcommand{\tabref}[1]{Tab.\ \ref{#1}}
\newcommand{\secref}[1]{Section \ref{#1}}
\newcommand{\appref}[1]{Appendix \ref{#1}}

\DeclareCaptionLabelFormat{andtable}{#1~#2  \&  \tablename~\thetable}
\makeatletter
\def\@email#1#2{%
 \endgroup
 \patchcmd{\titleblock@produce}
  {\frontmatter@RRAPformat}
  {\frontmatter@RRAPformat{\produce@RRAP{*#1\href{mailto:#2}{#2}}}\frontmatter@RRAPformat}
  {}{}
}%
\makeatother

\begin{document}
\preprint{AIP/123-QED}

\title[Information dynamics of microstructural variables around the 2017 Bitcoin markets crash]{
Information dynamics of price and liquidity around the 2017 Bitcoin markets crash}
\author{Vaiva Vasiliauskaite}
\affiliation{Computational Social Science, ETH Z\"urich, Switzerland. Email: vvasiliau@ethz.ch }

\author{Fabrizio Lillo}
\affiliation{Dipartimento di Matematica, Universit\`a di Bologna and Scuola Normale Superiore, Pisa, Italy }

\author{Nino Antulov-Fantulin}
\affiliation{Computational Social Science, ETH Z\"urich, Switzerland}

\date{\today}
\begin{abstract}
We study information dynamics between the largest Bitcoin exchange markets during the bubble in 2017-2018. 
By analysing high-frequency market-microstructure observables with different information theoretic measures for dynamical systems, we find temporal changes in information sharing across markets. In particular, we study time-varying components of predictability, memory, and (a)synchronous coupling, measured by transfer entropy, active information storage, and multi-information. 
By comparing these empirical findings with several models we argue that some results could relate to intra-market and inter-market regime shifts, and changes in direction of information flow between different market observables.
\end{abstract}
\maketitle


Bitcoin is a cryptocurrency that was originally designed as a medium of exchange~\cite{nakamoto2019bitcoin}, however there is still no strong consensus as to whether it is a currency, a commodity or an asset~\cite{BAUR2018177,glaser2014bitcoin,gronwald2019bitcoin}.  
Cryptocurrencies are usually transmitted and created via distributed peer-to-peer networks with well-defined cryptographic protocols that record the system's state via public ledger (blockchain). Thus blockchain (or another type of a ledger that maintains knowledge of distributed consensus) is at the heart of digital currencies. The blockchain technology promises benefits such as proving the existence of an asset as well as keeping track of its current and all past ownerships, both in a distributed manner. It also plays a role in crypto-asset price discovery~\cite{DB20}.

Electronic coins, such as Bitcoin, are ``chains'' of digital signatures: an owner transfers the coin by digitally signing a hash of the previous transaction and the public key of the next owner, adding them to the end of the coin. The exchange of cryptocurrencies to fiat money (USD, EUR, GBP, etc.)\ occurs in cryptocurrency exchange markets which are based on electronic double auctions, operating using limit order books. 
Here, real-time market data, such as transaction volume, amount of bid and ask orders, and exogenous information, such as news and social media mentions, are sources of information that feed into trading decisions. Within the crypto-market ecosystem, two distinguishable types of information sources exist: intra-market sources and inter-market sources. The first source is related to internal dynamics of a market's limit order book, e.g., its liquidity and volatility, that may influence trading decisions~\cite{DB15} as well as one another~\cite{glosten1985bid}. The second source is related to communication between different markets. Therefore, any two crypto-markets are ``connected'': explicitly, when there is a mutually traded currency (or an arbitrage opportunity), or implicitly, when prices of currencies amongst several markets are correlated. 

\begin{figure}[ht]
    \centering
    \includegraphics[width=1\linewidth]{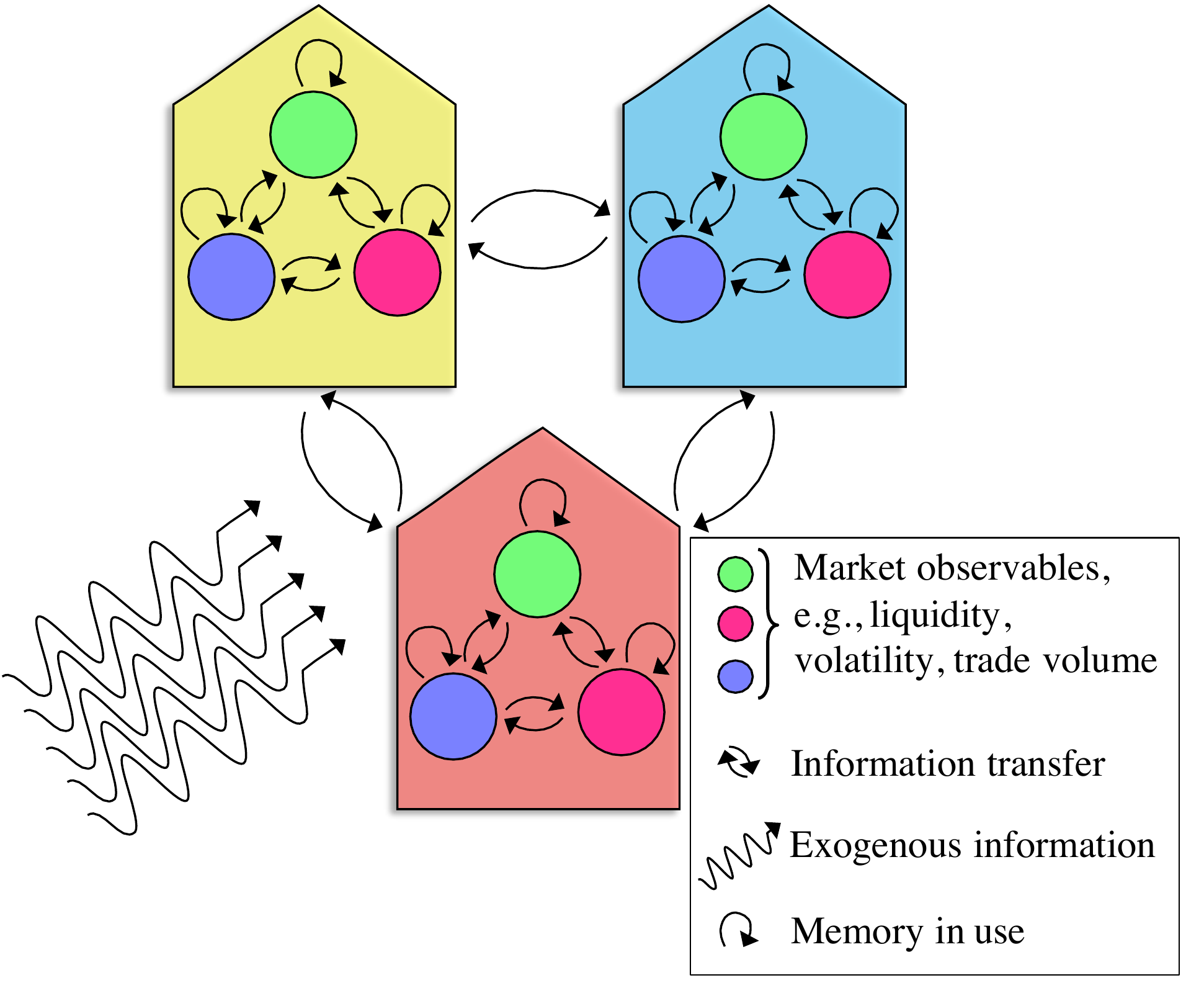}
    \caption{Information is shared between markets (house icons) and within markets. Different nodes indicate separate market observables, discussed in \secref{sec:data}. Past information about one market observable (e.g., volatility) can be used by the same market observable (a self-loop), as well as by other market observables (edges between different colour nodes in the same market). Information is also shared across markets (edges amongst market icons), and the system as a whole is also exposed to unknown exogenous information (curly arrows).}
    \label{fig_information_sources}
\end{figure}

The blockchain ensures that, in the long run, price of a cryptocurrency develops synchronously across exchanges, i.e., the law of one price holds~\cite{DB20}. However, even without the presence of blockchain, one would expect price synchronisation across different venues due to arbitrage~\cite{MS20}. Hence, Bitcoin price is processed collectively, in a distributed manner. Its dynamics is affected both by internal feedback and by exogenous information (e.g., public news), as illustrated in \figref{fig_information_sources}. It is not clear, however, what type of information is the most relevant for the price formation process. Often, these multiple influences lead to very large fluctuations (high volatility in technical terms) of exchange rates between cryptocurrencies and fiat money, price bubbles, and sudden price crashes. These properties make cryptocurrencies exchange rates a unique laboratory to empirically study the collective dynamics leading to market instabilities, which are well known to be ubiquitous in all financial markets. 

Although at low frequencies, the prices of Bitcoin in different markets (and, oftentimes, prices of other cryptocurrencies in relation to Bitcoin) develop in apparent synchrony, the law of one price does not hold at very high frequencies~\cite{MS20,PD19}. Furthermore, at sufficiently high frequency, one observes lagged relationships between the prices of the asset(s), i.e., past information about the change in price of one cryptocurrency is informative for predicting the future price change of another cryptocurrency. Such mechanisms are detectable with information transfer measures: linear relationships can be detected via Granger causality~\cite{G69}, while more general nonlinear dependencies can be detected via transfer entropy~\cite{S00,PKHS01}, that is one of  constituents of a process's entropy. Entropy itself relates to computation: at each point in time, \emph{computation} of the next state of the process. When a system is composed of multiple interacting units, information transfer could be thought of as communication or signalling, and the system as a whole as processing information to determine its collective behaviour at each time step via distributed computation. A study of \emph{information dynamics} aims to decompose this computation into unique elements, namely, transferred, stored, and modified information, and their changes in space and time~\cite{BHL16,L12}.

Spatio-temporal patterns of information dynamics (in particular, information transfer and multi-information) within the system of interacting markets was observed to increase in the financial crisis periods, meaning that the system appears to be more synchronised~\cite{afsharizand2020market}. Such synchronisation of a system could be a precursor of a ``phase transition''~\cite{BBSHdD18,peron2011collective}---a dramatic dynamical shift that occurs due to exogenous or endogenous events that perturb the system. Inefficiencies and delays of transfer of information amongst system's constituents open possibilities for arbitrage, risk, distress spillover across markets. They may also be precursors of price bubbles. Although the information theoretic measures have been shown to signify dramatic dynamical changes in the system, both where the phase transition can be pinpointed exactly~\cite{LPZ08,LPP11}, as well as when they are discussed in a more qualitative manner~\cite{harre2009phase}, the observed dynamical patterns are difficult to interpret. This is particularly challenging if the data is incomplete, unrepresentative, noisy, e.g., sampled at a low~\cite{abeliuk2020predictability}, inconsistent frequency, with not all relevant system's constituents taken into account, using erroneous measurement tools. 

Therefore, in this paper, our aim is twofold. First, we study several econometric models that couple market microstructure variables. Our aim is to find out whether a particular coupling and its (sudden or slow) change are detectable with information dynamics tools. We then analyse the persistence of information dynamical patterns that signify a particular regime shift. We then analyse information dynamics patterns across markets during one turbulent event in the crypto-market ecosystem, namely, the Bitcoin bubble in 2017-18. We remind that the bubble occurred in December 2017 when Bitcoin price went up from around $6,463\$$ on the $1^{\textrm{st}}$ of November, $2017$ to, at that time, an all-time-high of $19,716\$$ on the $17^{\textrm{th}}$ of December, see the left panel of \figref{fig:price}. After this crash date the price dropped to $11,414\$$ before the end of the year, descending for weeks and months afterwards. Furthermore, ripple effects of the Bitcoin bubble also penetrated to other cryptocurrencies, and much research has been put in studying the drivers of the crypto crash, e.g.\ see~\cite{DP19}. This price bubble is also unique in that in 2017 Bitcoin was still a dominant cryptocurrency with the largest market capitalisation. At the time of writing, alternative coins have become more prominent. Assuming that high frequency market data related to Bitcoin traded against dollar and Tether captures majority of the dynamics that took place at that time, we concentrate our attention on information dynamics within and across markets using this particular set of data.

The paper is organised as follows. First, in \secref{sec:data}, we describe the data used, as well as market microstructure observables, related to price as well as liquidity in markets, that we use for inference of information dynamics measures in \secref{sec:micro_observables}. A methods section \secref{sec:TE} includes description of information theoretical tools, experimental setup for data analysis, and analysis of models. We report the data analysis results in \secref{sec:results} and conclude the paper with a discussion in \secref{sec:discussion}.

\section{Data}\label{sec:data}

The data for this paper are tick-level trading and order book data obtained from~\footnote{Kaiko---a cryptocurrency market data provider for institutional investors and enterprises. See \href{https://www.kaiko.com}{kaiko.com}.}. Data of trades is provided at a millisecond frequency, and for observables based on this data, we aggregate the information to a minute level frequency. The limit order book data consists of one snapshot in each minute. Since the snapshots are not taken at exactly the same second of a minute in different markets, we align the data from the limit order books and from trades to ensure the lack of non-causal information flows.
We describe this procedure in \appref{app_data}. Note that the alignment procedure ensures that causal order of events is respected when we treat snapshots of a limit order book as a discrete-time process that is studied at a frequency that is at most one minute.

\begin{figure*}[ht]
     \centering
     \includegraphics[width =0.32 \linewidth]{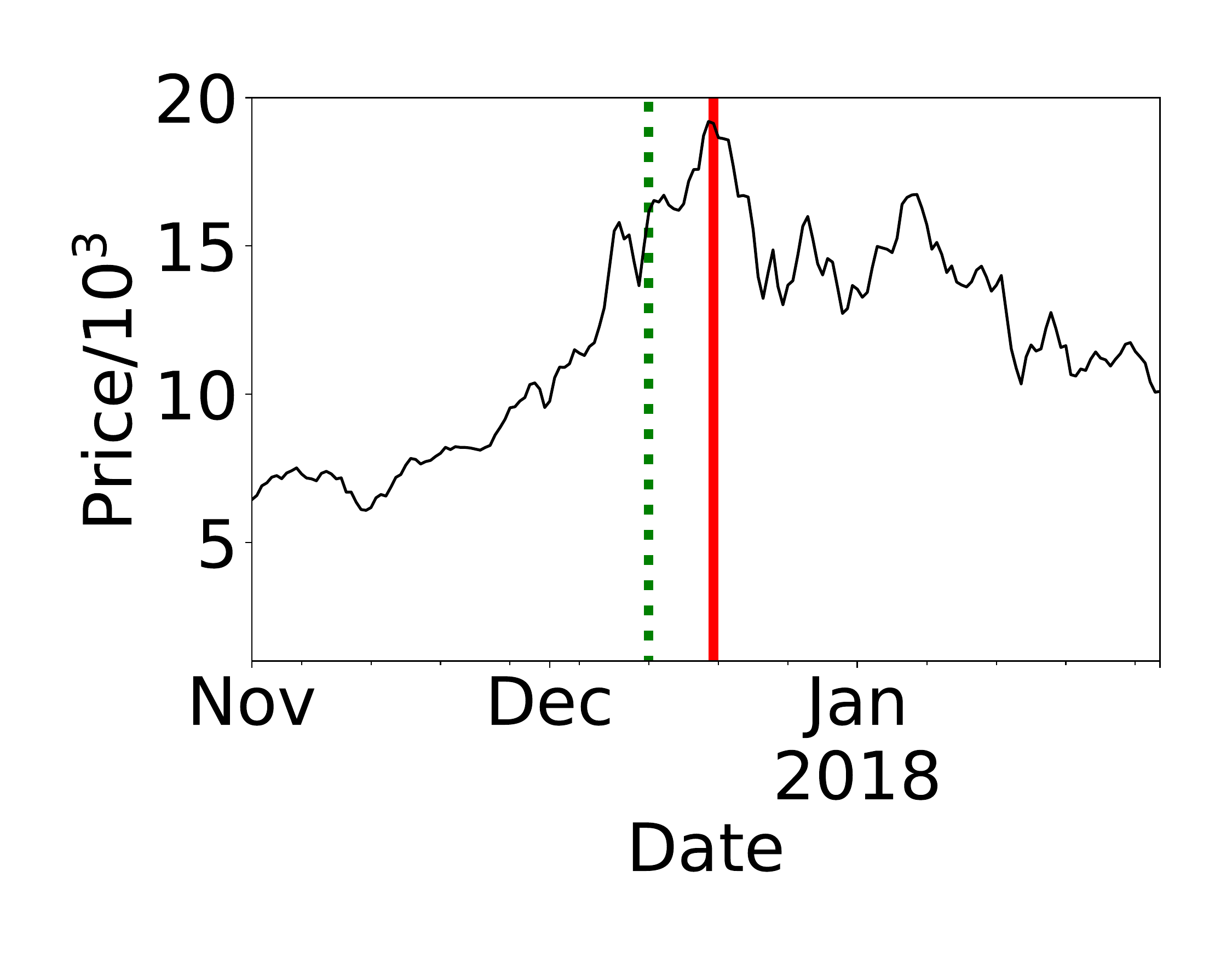}
    \includegraphics[width = 0.32\linewidth]{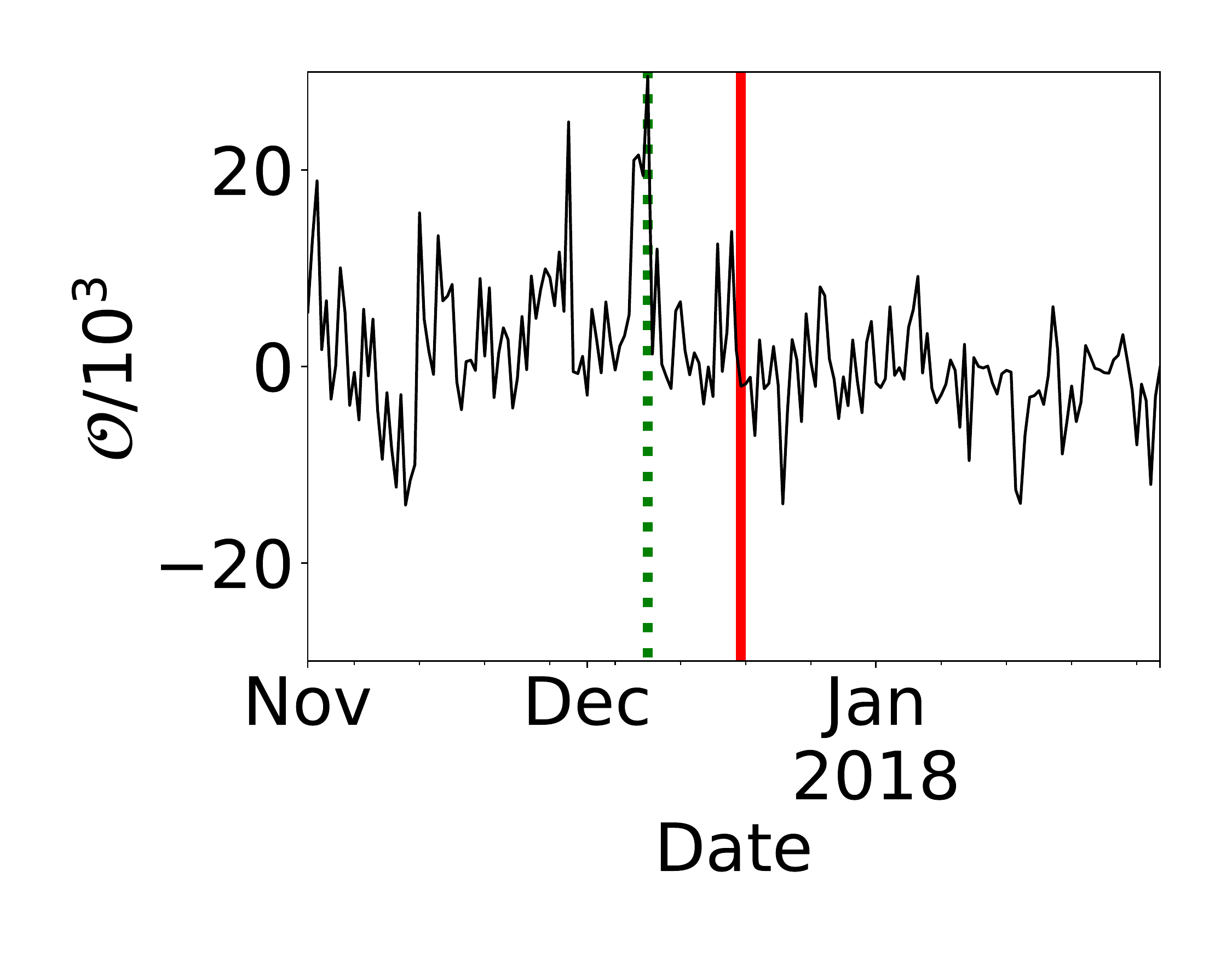}
    \includegraphics[width = 0.32\linewidth]{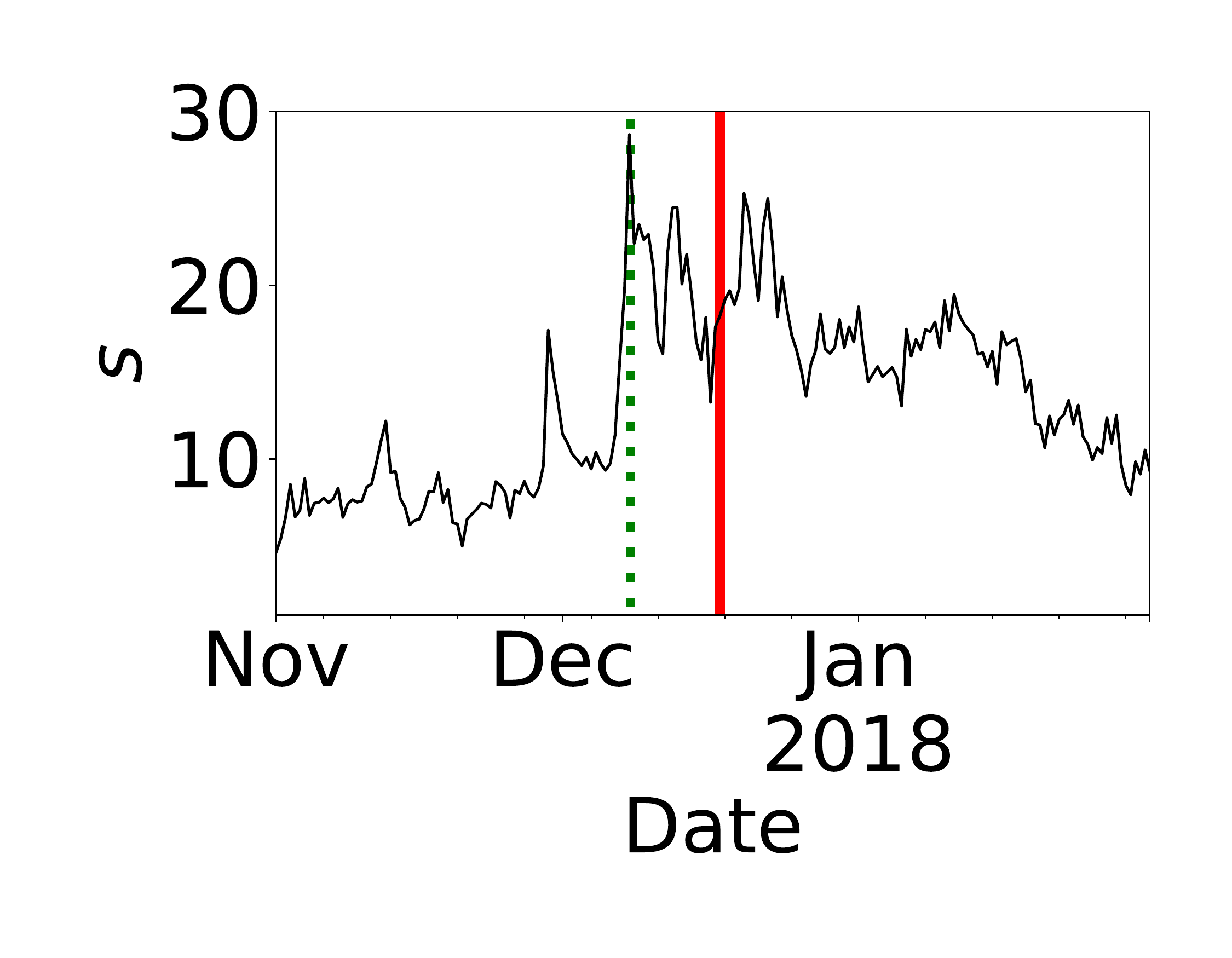}
     \caption{Left: average mid-price of Bitcoin (black).
     Middle: realised order imbalance expressed in quote currency, aggregated across 12-hour windows and all considered trading venues (black). Right: average 12-hour spread (black). Red lines indicate the peak of the bubble, and the green dotted lines indicate the dates at which each observable reached the maximum value (for price, we show the date of the maximum value of the price returns). From left to right, each dotted green line indicates the following dates: 2017-12-10, 2017-12-07, 2017-12-08. The red line is positioned at 2017-12-17.}
     \label{fig:price}
 \end{figure*}

For the proceeding analysis, we restrict our attention to the most liquid and the largest cryptocurrency --- Bitcoin (BTC), traded against either a fiat currency of US dollar (USD) or Tether (USDT), a \emph{stablecoin}, designed to be worth $\$1.00$ at all times. More specifically, we consider Bitcoin traded against USD in the following venues: Gemini, BTC-e, Bitstamp, Coinbase, Kraken, HitBTC, Bitfinex, and against USDT in Binance, Bittrex, and Poloniex. The time period under our study ranges from $1^{\textrm{st}}$ of November, $2017$ to $1^{\textrm{st}}$ of February, $2018$. This period involves a price bubble observed in Bitcoin as well as other cryptocurrencies. The left panel of \figref{fig:price} shows the price dynamics.

\subsection*{Microstructural variables}\label{sec:micro_observables}
In order to study the price and liquidity dynamics in the different venues, we introduce three market microstructure variables. 

\paragraph{Price} quantifies the value of an asset. Here we define price as the mid-price at each snapshot of the order book, namely $p_t=\frac{p^a_t+p^b_t}{2}$, where $p^a,p^b$ are, respectively, the best ask and bid price at time $t$. We define the price increments (hereafter termed returns for simplicity) as $r_t=p_{t}-p_{t-1}$. Price dynamics is strongly asymmetric around market crashes and shows a slow (power law) relaxation of volatility~\cite{omori}.
    \paragraph{Order imbalance} quantifies the demand of liquidity takers. Specifically, suppose we have a set of executed trades, indexed over arbitrary integer index $i: i\in \mathbb{Z}$, and each trade also has an associated time stamp $t_i\in \mathbb{R}^+$, a sign $\epsilon_i\in \{-1,1\}$ and a volume $v_i\in\mathbb{R}^+$. If the trade is initiated by a buyer, then its sign is $\epsilon_i=+1$, while $\epsilon_{i}= -1$ for seller-initiated trades. We define order imbalance for the time interval $(t-\delta,t]$  as a sum of the signed volumes of all trades within the interval: 
\begin{equation}\label{e_order_imbalance}
    \mathcal{O}_t=\sum_{i| t-\delta<t_i \leq t  }\epsilon_{i}v_i.
\end{equation} 
Similarly, $\mathcal{O}$ can be expressed in quote currency: $ \mathcal{O}_{\$,t}=\sum_{i|  t-\delta<t_i \leq t  }\epsilon_{i}v_i p_i$ if $p_i$ is the transaction price. For the relation between order flow and price returns, see for example \cite{bfl,lillo2021}.
\paragraph{Spread} is a symmetric measure of market's liquidity (as small spread indicates that trades are easily executable), and is defined as $s_t=p_{t}^a-p^b_t$. Spread has been shown to display nonsymmetric dynamics around market crashes~\cite{ponzi}.

In \figref{fig:price}, we show the dynamics of price, order imbalance and spread for time period from $1^{\textrm{st}}$ of November, $2017$ to $1^{\textrm{st}}$ of February, $2018$. 

\section{Methods}\label{sec:TE}

\subsection{Information dynamics for stochastic processes}
Consider a system, composed of $N$ stochastic processes $\{X^{\alpha}\}_{\alpha\in\Omega}$, together forming a multivariate process $\mathbf{X}^{\Omega}$. Each stochastic process $X^{\alpha}$ is a collection of random variables $\{X_t^{\alpha}\}_{t\in \mathbb{N}^+}$ with $Q$ the total number of observations. For each random variable $X_t$, its realised value is defined as $x_t$. Note that $\Omega$ denotes the finite set of markets in this paper.
 In the definitions of information dynamics measures, we will use subscripts $X$, $Y$ to refer to information dynamics amongst some two stochastic processes (e.g.\ $X=X^{\alpha}$, $Y=X^{\beta}$). These information dynamics measures will be defined using length-$l$ and length-$k$ collections of random variables, for each time $t$: $\textbf{Y}^{(l)}_{t-\delta}= \{Y_{t-\delta-l},Y_{t-\delta-l+1},...,Y_{t-\delta}\}$, $\textbf{X}^{(k)}_{t-1} = \{X_{t-1-k},X_{t-1-k+1},...,X_{t-1}\}$, where $\delta$ is a time delay.



\paragraph{Information-theoretic measures for static variables}
The fundamental quantity in this work is Shannon entropy of a random variable $X$, defined as $H(X)= -\sum_x p(x) \log p(x)$. Here $x$ is an instance of a random variable, and the sum is over of all possible values that $x$ can take. Conditional entropy of $X$ given $Y$ is the average uncertainty that remains about $X$ after learning the values of $Y$: $H(X|Y)= -\sum_{x,y} p(x,y) \log p(x|y)$. Mutual information between $X$ and $Y$ measures the average amount of information that is communicated in one random variable about another. $Y$: $I(X,Y)= H(X) - H(X|Y)$. The conditional mutual information between $X$ and $Y$ when $Z$ is known is defined as $I(X,Y|Z)= H(X|Z) - H(X|Y,Z)$. 

These information-theoretic measures for static variables can also be adapted to analyse stochastic processes. When temporal information is incorporated, one 
can quantify: how much information is shared between system's units at each point in time; how much information about the current state of $X$ is conveyed in the past states of $Y$; how much information about the current state of $X$ is conveyed in the past states of $X$. These questions can be addressed with information dynamics measures~\cite{BHL16,L12} that are summarised below.

\paragraph{Multi-information} Multi-information (MI) is defined as a measure of the deviation from independence of the components in the system~\cite{L12,TSE94}:
\begin{equation}\label{eq_mi}
    I_{\textbf{X}^{{\Omega}}}
    =\left(\sum_{\alpha=1}^N H(X_t^{\alpha})\right) - H(\textbf{X}^{{\Omega}}_t).
\end{equation}
Large value of MI is a signature of high synchronous inter-connectivity of a system. 

\paragraph{Transfer entropy} Transfer entropy (TE) encapsulates the ``distributed nature'' of computation. Schreiber~\cite{S00}, and, independently, Palu\v{s} et al.~\cite{PKHS01}, defined TE as the amount of information that a source process $Y$, and, in particular, its past state $\textbf{Y}^{(l)}_{t-\delta}$ provides about a target's state $X_{t}$ in the context of the target's immediate past state $\textbf{X}^{(k)}_{t-1}$~\cite{BHL16}: 
\begin{eqnarray}\label{eq_te}
T_{Y\rightarrow X}^{(k,l,\delta)}
&=& I(\textbf{Y}^{(l)}_{t-\delta},X_{t}|\textbf{X}^{(k)}_{t-1})   \\
&=&  H(X_{t}|\textbf{X}^{(k)}_{t-1})-H(X_t|\textbf{X}^{(k)}_{t-1},\textbf{Y}^{(l)}_{t-\delta}).\nonumber
\end{eqnarray}
This formulation of information transfer quantifies pairwise relationships between variables. When a system is composed of more than two stochastic processes, \eqref{eq_te} is known as \emph{apparent transfer entropy}, as it does not account unobserved sources, potentially leading to over-estimation of the total amount of entropy transferred within the system.

\paragraph{Conditional transfer entropy} 
To discount redundant joint influences of two sources $Y,Z$ on a target $X$, \eqref{eq_te} is generalised 
to a measure of transfer entropy from $Y$ to $X$ given that $Z$ can also provide information about $X$:
\begin{eqnarray}
T_{Y\rightarrow X | Z}^{(k,l,m,\delta)}
&=& I(\textbf{Y}^{(l)}_{t-\delta},X_{t}|\textbf{X}^{(k)}_{t-1},\textbf{Z}^{(m)}_{t-1})   \\ \nonumber
&=&  H(X_{t}|\textbf{X}^{(k)}_{t-1}, \textbf{Z}^{(m)}_{t-1})-H(X_t|\textbf{X}^{(k)}_{t-1},\textbf{Y}^{(l)}_{t-\delta},\textbf{Z}^{(m)}_{t-1}). \\ \nonumber
\end{eqnarray}
Finally, the conditional TE can be generalised to account for a multivariate set of processes with the next definition.

\paragraph{Collective transfer entropy} To find the total amount of information that was collectively transferred to a target $X$
from all potential sources in the system, we use \emph{collective TE}, $T_X$, which accounts for redundancies and synergies among information that sources provide to a target~\cite{L12}. To compute $T_X$, we consider a set $\textbf{X}^{{\Omega}}$. The true effect of one source variable $Y$ to the target variable $X$ in the universe of $\textbf{X}^{{\Omega}}$ is computed through conditional transfer entropy: $T_{Y\rightarrow X| \textbf{X}^{\Omega\backslash \{Y\}}_{X}}$, where $\textbf{X}^{\Omega\backslash \{Y\}}_{X}=\{Z\in \textbf{X}^{\Omega}\backslash \{X, Y\}\}$. However $\sum_Y T_{Y\rightarrow X| \textbf{X}^{\Omega\backslash \{Y\}}_{X}}$ would not be equal to the total amount of information transferred to $X$, as only unique information~\cite{williams2010nonnegative} would be accounted for. To calculate the collective TE (see~\cite{L12} for more details), we sum incrementally conditional TE terms. Let us consider an ordered set $\textbf{X}^{\Omega }_{X}=\{Z\in \textbf{X}^{\Omega}\backslash \{X\}\}:Z^1,...,Z^\beta,...,Z^{N-1}$ and its subset $\textbf{X}^{\Omega,\beta}_{X}=\{Z^{\alpha}\in \textbf{X}^{\Omega}_{X} | \alpha\leq \beta\}$. Collective transfer entropy is then defined as 
\begin{equation}\label{eq_coll_te}
    T_X^{(k)}
    =\sum_{\beta=1}^{N-1} T_{Y^{\beta}\rightarrow X|\textbf{X}^{\Omega,\beta-1}_{X}}. 
\end{equation}

\paragraph{Active information storage} Overall, the predictability of the next state of a process $X$ is characterised by its entropy, whose non-overlapping constituents are the collective transfer entropy $T_X^{(k)}$, the \emph{active information storage} (AIS),
$A_X^{(k)}$, defined as \emph{memory of the process that is actively in use}~\cite{L12}:
\begin{equation}
    A_X^{(k)}
    = I(\bold{X}^{(k)}_{t-1},X_t)
    = H(X_{t}) - H(X_{t}|\bold{X}_{t-1}^{(k)} ).
\end{equation}
AIS is information in the past that contributes to the computation in the next state of $X$. 

\paragraph{Local information dynamics} The measures defined above can also be considered in a point-wise fashion, as they are expectation values of local measures at each observation at time $\tau$ as follows~\cite{LPZ08,L12}:
\begin{eqnarray}\label{eq_local}
    I_{ \textbf{X}^{\Omega}}&=&\langle i_{\textbf{X}^{\Omega}}(\tau)\rangle_\tau=\frac{1}{Q}\sum_{\tau=1}^Q
       \log\frac{p(x_{\tau}^{1},x_{\tau}^{2},...,x_{\tau}^{N})}{\prod_{\alpha=1}^N p(x_{\tau}^{\alpha})},  \\ 
    T_{Y\rightarrow X}^{(k,l,\delta)}&=&\langle t_{Y\rightarrow X}^{(k,l,\delta)}(\tau)\rangle_\tau =\frac{1}{Q}\sum_{\tau=1}^{Q}\log \frac{p(x_{\tau+1}|\textbf{x}_{\tau}^{(k)},\textbf{y}_{\tau}^{(l)})}{p(x_{\tau+1}|\textbf{x}_{\tau}^{(k)})}, \nonumber \\
    A_{ X}^{(k)}&=&\langle a_{X}^{(k)}(\tau)\rangle_\tau= \frac{1}{Q}\sum_{\tau=1}^Q\log \frac{p(x_{\tau+1},\textbf{x}_{\tau}^{(k)})}{p(x_{\tau+1})p(\textbf{x}_{\tau}^{(k)})}.\nonumber 
\end{eqnarray}


\subsection{Experimental setup}\label{sec:setup}

\paragraph{Information dynamics} We use Java Information Dynamics Toolkit -- JIDT~\cite{L14} to compute all information dynamics quantities. For the TE estimator, we choose Kraskov, St{\"o}gbauer, Grassberger K-nearest neighbour estimator (KSG)~\cite{KSG04}, which is minimally parameterised and has been shown to be robust for a wide range of data. To ensure our estimates are reliable, we studied sample-size bias of transfer entropy using the approach of~\cite{HN19} as described in \appref{ksg}. We found that $K=4$ for nearest neighbours estimator produces reliable results for our dataset. All values of information dynamics are reported in units of nats.

Throughout the paper, we considered the case where $l=1$. To ensure that we do not over-estimate information transfer for \emph{active information storage}---the memory of a process in use---of $X$, we chose $k = \underset{\kappa\in [1,60]}{\mathrm{argmax }} (A_X^{(\kappa)})$ for results in \secref{sec_res_one_observable} and $k = \underset{\kappa\in [1,10]}{\mathrm{argmax }} (A_X^{(\kappa)}) $ for results in \secref{sec_info_flow_obsr}. We then used this $k$ when computing $T_{Y\rightarrow X}$ as well as $T_{X}$. 
Lastly, as was demonstrated in~\cite{WPPSSLLV13}, when two processes are coupled via non-zero delay $u$, $T_{Y\rightarrow X}^{(\delta)}$ is maximised for $\delta=u$. Therefore, we consider $T_{Y\rightarrow X}^{(\delta)}$ for $\delta\in [1,10]$, and select $\delta$ for which $T_{Y\rightarrow X}^{(\delta)}$ is maximal. Such procedure ensures that non-instantaneous coupling across markets is also captured. For the results in \secref{sec_info_flow_obsr} we do not do this and consider a fixed $\delta=1$.

\paragraph{Rolling windows} For results discussed in \secref{sec_res_one_observable} we considered weekly subsets of the time series, moving each sliding window by three days with respect to the previous one. In such a way we achieve a twofold advantage. First, we reveal temporal patterns in information dynamics. Secondly, we ensure that the data is locally stationary (we expect the time series associated with markets around the crash to be globally non-stationary, however, at sufficiently small time windows intervals we observe stationary time series). In \secref{sec_info_flow_obsr} we considered two non-overlapping time series windows: the data preceding the price crash date (December $17^{\textrm{th}}, 2017$) and the remainder.

\paragraph{TE Significance test} All reported results are significant, where significance is tested against the null hypothesis that no information transfer exists. For all results, we used $100$ surrogate time series to evaluate significance, and, unless otherwise stated, we chose the significance level of $0.05$. For the results of \secref{sec_res_one_observable} in each time window a Benjamini–Yekutieli procedure~\cite{BY01} was also performed.

\paragraph{Stationarity test} For each observable's time series in all rolling windows, we considered the Augmented Dickey-Fuller (ADF) test~\cite{said1984testing} whose null hypothesis is a presence of a unit root, and we use significance level of $0.05$. We found that we can reject the null hypothesis that there is a unit root with a 5\% of significance level for the test statistics for all of the time windows of $s, \mathcal{O}$ time series. Contrary, we found the presence of unit-roots in price time series, so we used price returns (i.e.\ equal to applying difference operation $r_t=p_t-p_{t-1}$), after which all time windows passed the ADF test. 

\subsection{Statistics of information dynamics}\label{sec_stats}

\begin{figure*}
    \centering
    \includegraphics[width=\linewidth]{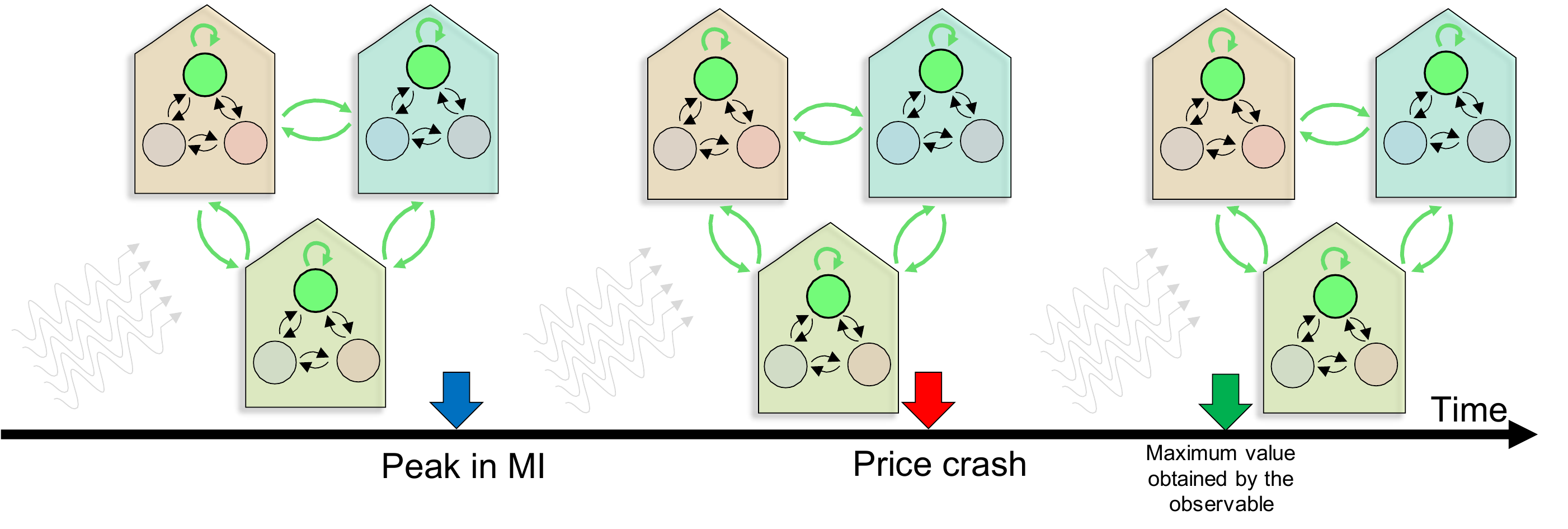}
    \caption{Illustration of analysis described in \secref{sec_res_one_observable}. We choose one observable (illustrated with green node) for each market. We then divide the observable's time series into overlapping sliding windows and compute information dynamical quantities (MI, TE) between markets using this one observable. The connectivity between markets via green observable's ``channel'' is illustrated with green edges linking different markets. We also compute the AIS for each market; this is illustrated with a self-loop for green node. Given a set of individual market's, or pairs of markets information dynamics quantities, we then aggregate these quantities into global statistics for each time window. The timeline illustrates three critical points: the price crash date, the date when we observe the maximum value obtained by the (green) observable, and the peak in MI for a given same observable. In \figref{fig_MI_TE} these points are shown in vertical lines of corresponding colours.}
    \label{fig_illustration_one_observable}
\end{figure*}

\begin{figure}
    \centering
    \includegraphics[width=0.8\linewidth]{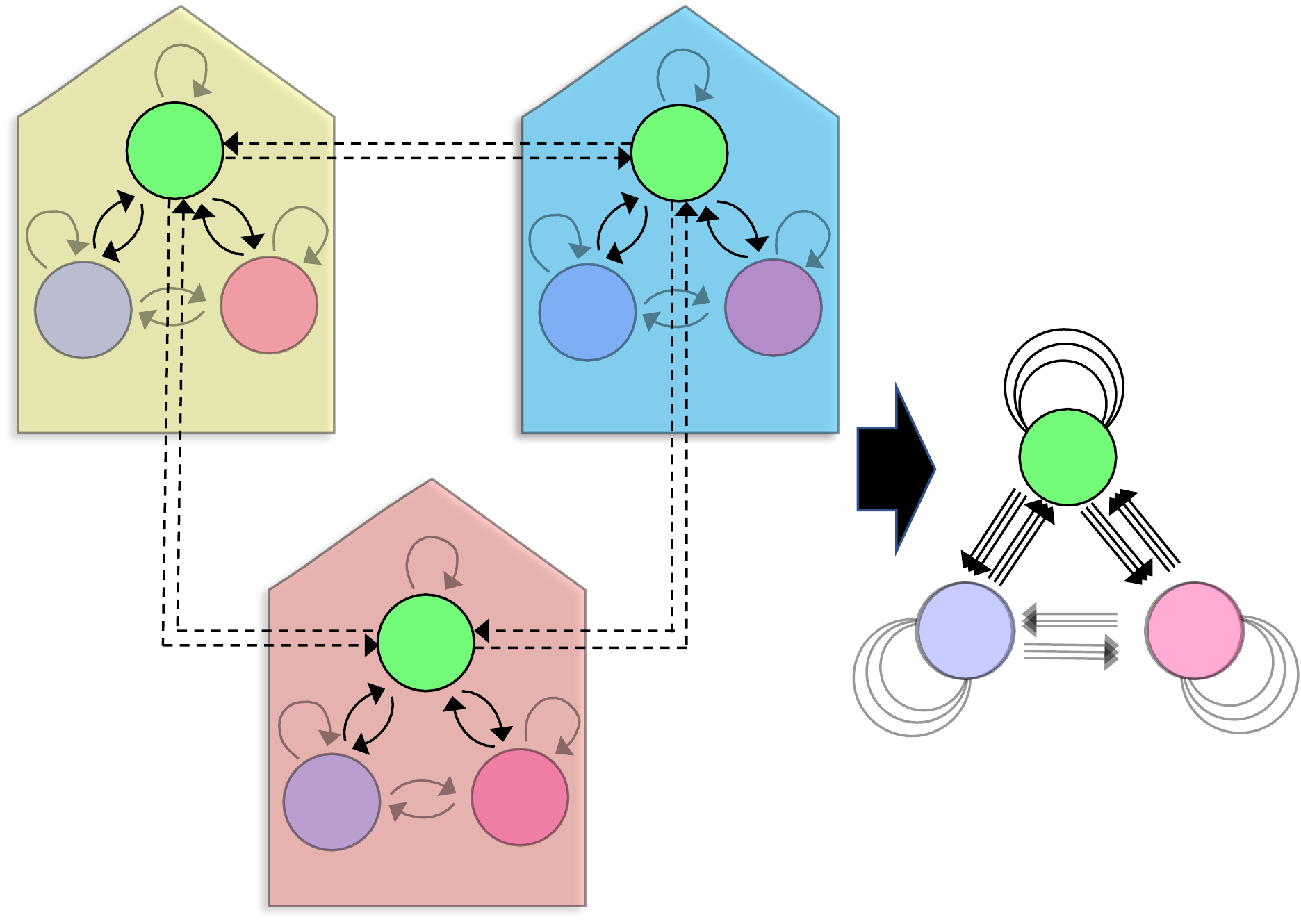}
    \caption{Illustration of analysis described in \secref{sec_info_flow_obsr}. Each market is characterised by a set of market microstructure observables (nodes of the same colour illustrates the same observable). In the section we discuss the information transfer between observables. TE between the same observable is illustrated as an edge between nodes of the same colour and results into self-loops in a resulting multigraph where nodes represent microstructure observables. We allow for multiple loops where each edge represents a link based on a single pair of markets. TE between different observables is illustrates as the flow from one colour nodes to another colour nodes. In a resulting directed multigraph these connections are edges between nodes. We again allow for multiple edges where each indicates information transfer from one observable in a market to another observable in the same market, calculated for all markets. The illustration emphasises how connectivity in the triadic multidigraph is obtained for a ``green observable''. }
    \label{fig_illustration_multi_observables}
\end{figure}

In the data analysis part, we will consider information dynamics, described in \secref{sec:TE}, first using one of the observables for each market. In doing so, we assume that the strongest coupling occurs through the same observable (spread in $\alpha$ has a stronger effect on spread in $\beta$ than on price returns in $\beta$), see \figref{fig_illustration_one_observable} for an illustration. Secondly, we also consider the extent to which market variables are internally coupled via different observables (e.g., quantifying how much spread in $\alpha$ has an effect on price returns in $\alpha$). See \figref{fig_illustration_multi_observables} for an illustration. To interpret the results, we aggregate information dynamics measures into several spatio-temporal global and local metrics.

For each information dynamics metric, namely, MI, AIS, and TE, the argument $w$ denotes the time series window, based on which the reported values are computed. We will also denote one microstructure observable (e.g., $s$), with the same capital letter that defines a random variable for a market, e.g., $T_{X^{\alpha}\rightarrow X^{\beta}}$ denotes transfer entropy from a market $\alpha$ to a market $\beta$ considering the same market micrstructure observable $X$, while $T_{X^{\alpha}\rightarrow Y^{\beta}}$ would indicate that different microstructure observables are considered.

\paragraph{Global metrics} To quantify the global, system-describing information dynamics, we aggregate results obtained for all markets within a time series window. In particular, in \secref{sec_res_one_observable} we study the extent to which a microstructural variable in one market affects the same microstructural variable in another market. Therefore the extent of this effect is captured with a sum of transfer entropies \eqref{eq_te} between all pairs of markets, namely, the \emph{total apparent transfer entropy}:
\begin{equation}
    T^{\textrm{app,sys}}_X=\sum_{\alpha,\beta|\alpha\neq\beta } T_{X^{\alpha}\rightarrow X^{\beta}}
\end{equation}
and the \emph{total collective transfer entropy}
\begin{equation}
    T^{\textrm{coll,sys}}_X=\sum_{\alpha}T_{X^{\alpha}}.
\end{equation}
These two metrics reflect on the total amount of information transfer within the system. 

Similarly, we also study \emph{markets' synchronisation within one variable} via the multi-information: in the definition \eqref{eq_mi} we consider $\textbf{X}^{\Omega}$ as a union of random variables that describe a single market observable in $N$ markets, denoting it as $I_{\textbf{X}^{\Omega}}^{\textrm{sys}}$ in the further discussion.

Lastly, we report the \emph{average active information storage}, $ A_X ^{\textrm{sys}}$, defined as 
\begin{equation}
  A_X^{\textrm{sys}}  = \frac{1}{N}\sum_{\alpha=1}^N A_{X^{\alpha}}=\langle A_{X^{\alpha}} \rangle_{\alpha},
\end{equation}
i.e., it is the average AIS per market, at a each time series window.

\paragraph{Market-specific statistics} To analyse the market-specific information dynamics measures, we will consider averages of a measure obtained from different time windows  $w\in[1,W]$. The amount of information transfer received by each individual market $\alpha$, termed an \emph{average collective transfer entropy} is defined as
\begin{equation}
    T^{\textrm{coll}}_{X^\alpha}=\frac{1}{W}\sum_{w=1}^W T_{X^{\alpha}}(w)=\langle T_{X^{\alpha}} \rangle_{w},
\end{equation}
and an individual \emph{market's average AIS} is defined as
\begin{equation}
    A_{X^{\alpha}}=\frac{1}{W}\sum_{w=1}^W A_{X^{\alpha}}(w)= \langle A_{X^{\alpha}} \rangle_{w}.
\end{equation}

\paragraph{Comparison between inter- and intra- market connectivity}
In \secref{sec_info_flow_obsr} we consider the amount of information transfer between microstructure observables within one market. Therefore, we compute the apparent information transfer of the form $T_{X^{\alpha}\rightarrow Y^{\alpha}}$, and contrast the amount of information transfer within the market to the amount of information sharing across markets. To make a fair comparison, we normalise the average values of TE per possible link in a transfer entropy network where nodes are market observables. 

For the comparative analysis of inter- and intra- market interactions, consider a directed graph that consists of three nodes that represent our observables. We map information transfer results onto this triadic graph, in which multiple edges are allowed between node pairs. Furthermore, each node is allowed to have multiple self-loops. The directed multi-graph representation depicts information transfer amongst different observables in a market (a somewhat endogenous information flow). In opposition to this type of internal information sharing, we also allow for inter-market connectivity, as discussed in \secref{sec_res_one_observable} (this can be thought of as exogenous information flow). Such links are represented as loops in the directed multi-graph, as information is transferred from an observable in one market to the same observable in another.
By appropriate edge averaging, we will turn this directed multi-graph into a directed graph that depicts averaged information transfer.

Each loop represents a significant transfer entropy link from one market to another, therefore, at most, there can be $N(N-1)$ self-loops for each node-observable, since each market can have $2(N-1)$ TE links ($N-1$ incoming and $N-1$ outgoing). 
We calculate the averaged TE for an observable $X$ between all pairs of markets $\alpha$, $\beta$ as:
\begin{eqnarray}
    \omega^{\textrm{self}}_X &=& \langle T_{X^\alpha\rightarrow X^{\beta}}\rangle_{\alpha,\beta}=\sum_{\alpha={1}}^N\sum_{\beta=1|\beta\neq\alpha}^NT_{X^{\alpha}\rightarrow X^{\beta}} .
\end{eqnarray}
We also compute the average strength of incoming links, $\omega^{\textrm{in}}$, and outgoing links $\omega^{\textrm{out}}$:
\begin{eqnarray}
 \omega^{\textrm{in}}_X &=& \langle T_{Y^{\alpha}\rightarrow X^{\alpha}} \rangle_{\alpha,Y}  \nonumber \\
 &=& \frac{1}{N(N-1)} \sum_{Y\neq X}\sum_{\alpha={1}}^N T_{Y^{\alpha}\rightarrow X^{\alpha}}
 \\  \nonumber
 \omega^{\textrm{out}}_X &=& \langle T_{X^{\alpha}\rightarrow Y^{\alpha}} \rangle_{\alpha,Y}  \nonumber \\ 
 &=& \frac{1}{N(N-1)}\sum_{Y\neq X}\sum_{\alpha={1}}^N T_{X^{\alpha}\rightarrow Y^{\alpha}}
\end{eqnarray}
Here $X$ represents one type of observable, e.g., $s$, and $Y$ represents a different observable, e.g., $r$.

\subsection{Models of non-linear information transfer in complex systems}\label{sec:models}

Here we define two case models of stochastic processes coupled via lagged non-linear interaction. By analysing the information-theoretic, spatio-temporal signatures of these models, we will be able to explain in the information dynamics trends observed using real market microstructure data.

\paragraph{Vector auto-regressive model of regime shifts}
Let us first consider two coupled auto-regressive processes:
\begin{eqnarray}\label{eq_ar_def}
X_{t} &=& \alpha_1 X_{t-1} +\beta_1 \varepsilon_{1,t} +K(t)\varepsilon_{t} \\
Y_{t} &=& \alpha_2 Y_{t-1} +\beta_2 \varepsilon_{2,t} +K(t)\varepsilon_{t} +C(t)\left|X_{t-1}\right|^d,\nonumber 
\end{eqnarray}
i.e., we have a time-delayed directional coupling $X\rightarrow Y$, whose time-varying strength is $C(t)$. In addition to this, let us assume the presence of a common, time-varying hidden driver whose strength is $K(t)$. $\varepsilon$ terms denote independent Gaussian noise. Here $d$ is a constant, that we use to vary the linearity of the causal link from $X$ to $Y$. Note that the hidden driver term $K(t)\epsilon_t$ is present in both variables without any delay. This model is useful to analyse causal interaction amongst system's sub-units when they are ``of the same type'', i.e., representing the same market observable in some pair of markets.

 We consider several potential systemic changes, and local information dynamics signatures associated with them. A regime shift itself is defined as a significant change in the strength of a certain type of coupling between two variables, observed at a certain point in time $t_C$. In particular, here we consider a time-varying coupling strength $f$ modelled via logistic function 
\begin{equation}\label{eq_sigmoid}
    f(t) = \frac{s}{1+e^{-b(t-t_C)}}+C.
\end{equation}
Here $s$ relates to the maximum value of the function, and $b$ to the sharpness of the transition. For example, a \emph{regime shift of a causal driver} may be modelled as a change of a coupling strength term $C(t)$. 

\begin{figure*}[ht]
    \centering
    a)
    \includegraphics[width = 0.3\linewidth]{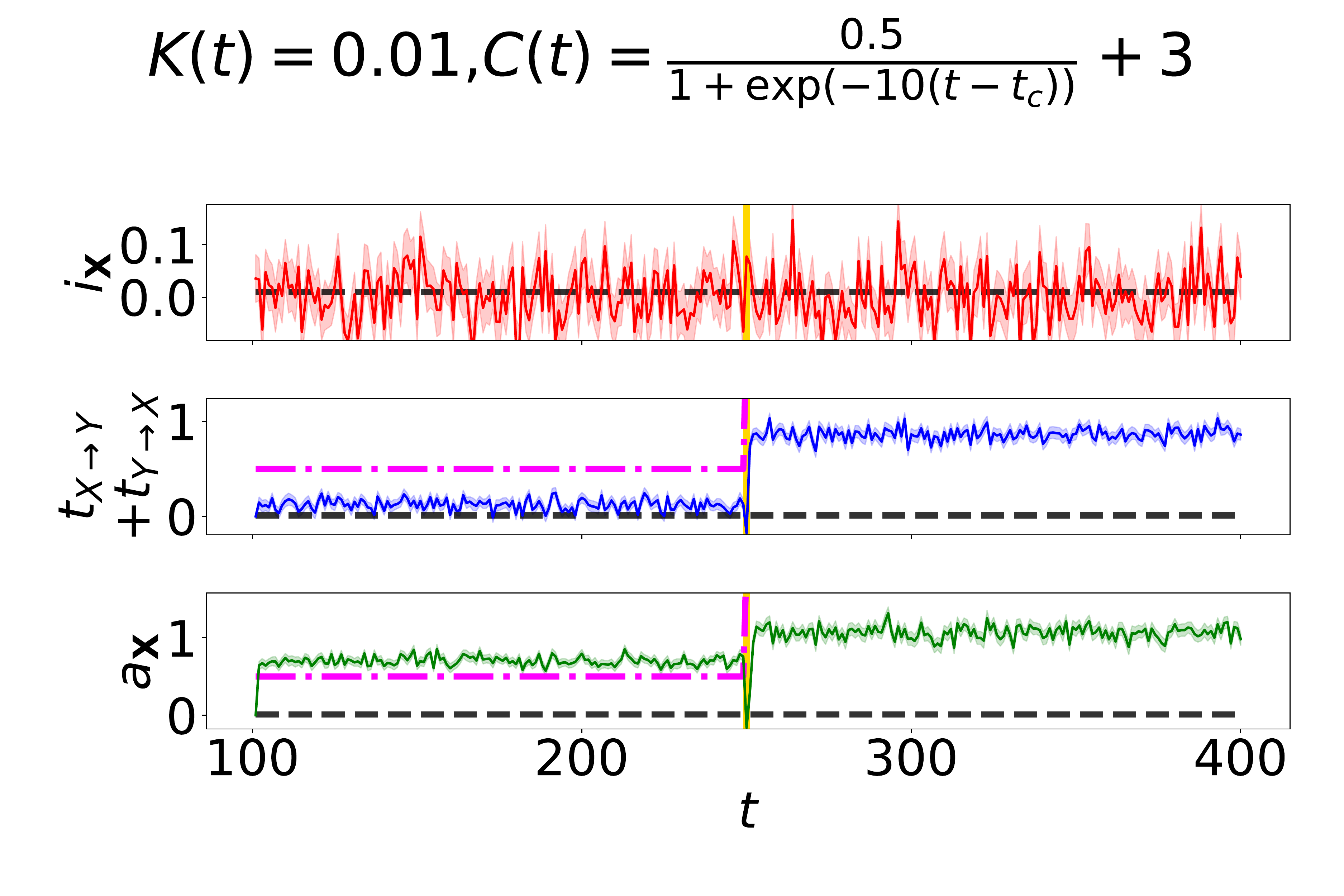}
    b)
    \includegraphics[width = 0.3\linewidth]{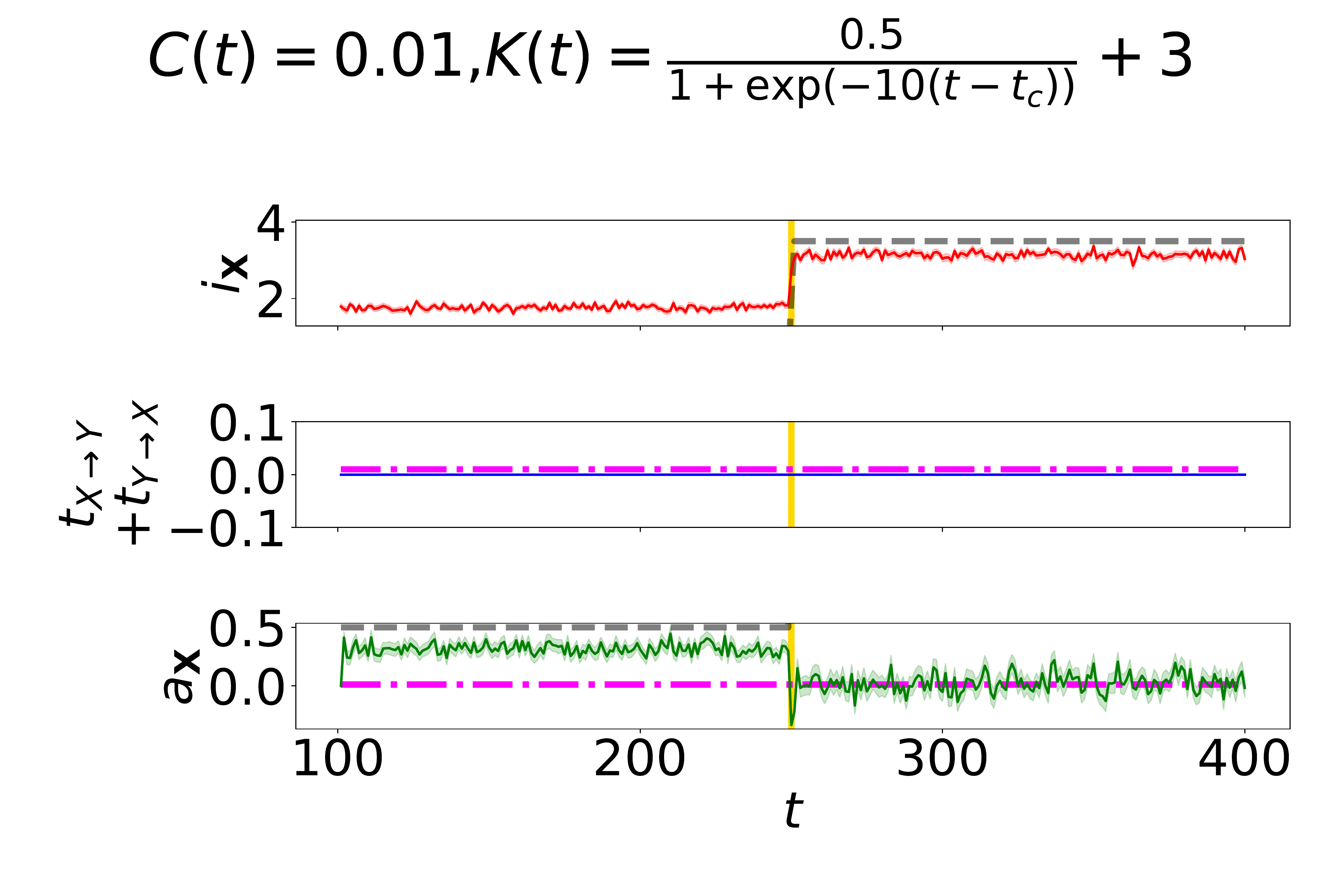}
    c)
    \includegraphics[width = 0.3\linewidth]{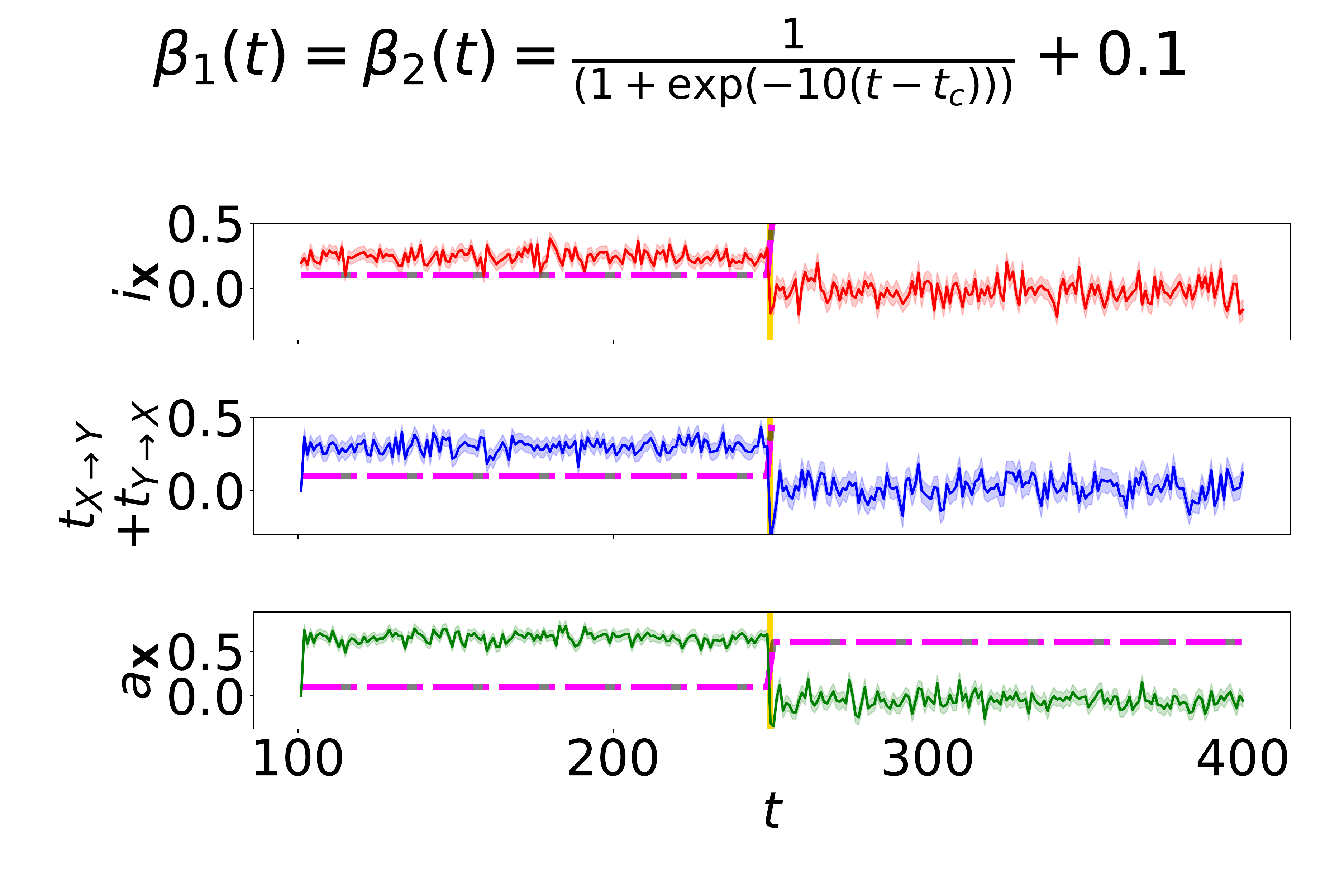}
    \caption{Causally coupled auto-regressive processes and local information theoretic signatures of the response of a system to three types of regime shifts: the change in a) strength of a causal driver, b) strength of a hidden driver c) intrinsic uncertainty. Figures report mean at each time step and the standard error in the mean, obtained from 100 independent simulations. Yellow vertical line indicates $t_C$, purple line shows causal coupling strength $C(t)$, and grey dashed line shows the strength $K(t)$ of a hidden driver. The parameters $\alpha_1=\alpha_2=0.2, d=0.5$ in all figures. In a) and b), $\beta_1=\beta_2 = 1$, while in c) $\beta_1=\beta_2$ are functions of time, defined using \eqref{eq_sigmoid} and depicted in the figure in grey dashed and magenta dash-dot lines, while $K(t)$ and $C(t)$ are constants, equal to $0.01$ and $0.5$ respectively.}
    \label{fig_case_1_main}
\end{figure*}

As \figref{fig_case_1_main}(a) shows, \emph{in case of a regime shift in a causal driver, we observe a significant change in the absolute values of local information dynamical measures} \footnote{Note that the results reported in \figref{fig_case_1_main} are local information dynamics values, and each local value is estimated for each individual run. For non-stationary time series, an appropriate treatment would be to obtain each local value from an ensemble of time series available [43]. However, in financial time series, usually only one trajectory is available. What we aimed to show with our controlled vector auto-regressive experiments of regime shifts in \figref{fig_case_1_main} was that certain regime shifts are detectable even from a single time series realisation, and the information theoretic signature is visible for a wide range of such non-stationary time series (see Appendix).}. This situation is characterised by large TE and large AIS in high coupling regime and no significant change in MI. In the Appendix, we also show that a change in MI is also possible, when $K(t)$'s absolute value is large: its value is high when AIS and TE values are low and vice versa.

Similarly the case of a regime shift of a hidden (common, simultaneous) driver can be modelled via variation in the coupling strength $K(t)$. The result, shown in \figref{fig_case_1_main}(b) suggests that \emph{the change in a hidden driver's strength is signified by high MI when coupling is strong, and vice versa when coupling is weak}. Note that changes in AIS and TE are possible: the former is possible when $K(t)$ is sufficiently high-valued, while the latter is possible when $C(t)$ is high-valued. 

Lastly, we studied a case when $\beta_1=\beta_2=\beta(t)$, defined using \eqref{eq_sigmoid} i.e., the system's overall uncertainty is a time functional, while $C(t)=K(t)=0.3=$const. In this case, we find that all three information dynamic measures can be mutually low-valued the side of transition that models high-uncertainty regime, and mutually low-valued on the other, where the system's intrinsic uncertainty is low, see \figref{fig_case_1_main}(c).

In the \appref{app_signatures} we also show that the results are robust for a wide range of parameter values $s,b,C$. Although the analysis ensures that the results shown in \figref{fig_case_1_main} are not coincidental, further statistical proof is needed to prove that signatures are persistent in a general spectrum of nonlinearly coupled autoregressive systems. We also note that it is more than possible that a combination of these effects comes into play in real data, and such convolution and its signature is not studied any further in the current work.

\paragraph{GARCH model for price returns and spread}
Next, we analyse a model of coupling between different market microstructure variables, using a variant of generalised autoregressive conditional heteroskedasticity~\cite{GARCH,glosten1985bid}.
Our model considers possibly bidirectional coupling between price returns $r$ and spread $s$ when spread is coupled to volatility (variance) of price. Therefore we have coupled time series of the form:
\begin{eqnarray}\label{eq_garchspread}
    r_{t} &=& \sigma_t \varepsilon_{1,t}, \quad \varepsilon_{1,t}\sim \mathcal{N}(0,1) \\  \nonumber
    \sigma_{t}^2 &=& w+\alpha r_{t-1}^2 +\beta \sigma_{t-1}^2+\gamma s_{t-1}^2, \\ \nonumber
    s_t^2 &=& as^2_{t-1}+b\sigma_{t-1}^2 + c\epsilon_{2,t}^{2} \quad \epsilon_{2,t}\sim \mathcal{N}(0,1) .
\end{eqnarray}
 Compared to the standard GARCH(1,1)~\cite{GARCH} (where ``(1,1)'' indicates that there is one variance dependence term, and the second indicates that there is one ARCH term $\epsilon$, $w$ denotes constant base level of volatility, $\alpha, \beta$ determine the influence of past squared returns and past volatility), we have several additional parameters, $a,b,c,\gamma$. The new parameters have the following economic interpretation: $a$ is the (bare, i.e., not volatility driven) persistence of the spread, which is known to be strongly autocorrelated (especially for small tick stocks)~\cite{ponzi}; $c$ measures the dispersion of spread and $\epsilon_t$ is the associated innovation noise; $b$ links past volatility to future spread. It might be connected with asymmetric information, since market makers adjust spread according to volatility (see, for example~\cite{glosten1985bid}); $\gamma$ links past spread with future volatility. If the spread is large, it is likely that the limit order book is sparse (as empirically shown in~\cite{lillo2005key}). But a sparse, i.e., illiquid, book leads to a more volatile price, since any order can create a large price change~\cite{Mike}.

Note that in this formulation, a link from price to spread is indirect, and in fact price returns affect spread at the two-step delay. Therefore in the following figures we will scan through $\delta\in[1,2]$ and use the one for which the observed TE is larger. Depending on parameter values $b,\gamma$ we can have bi- or uni-directional couplings $T_{s\rightarrow r}$ or $T_{r\rightarrow s}$. Each datapoint in the proceeding results is considered at a significance level of $0.01$, and results are obtained from $100$ independent simulations.
 
\begin{figure}[ht]
    \centering Expected link $s\rightarrow r$\\ 
    \includegraphics[width = 0.45\linewidth]{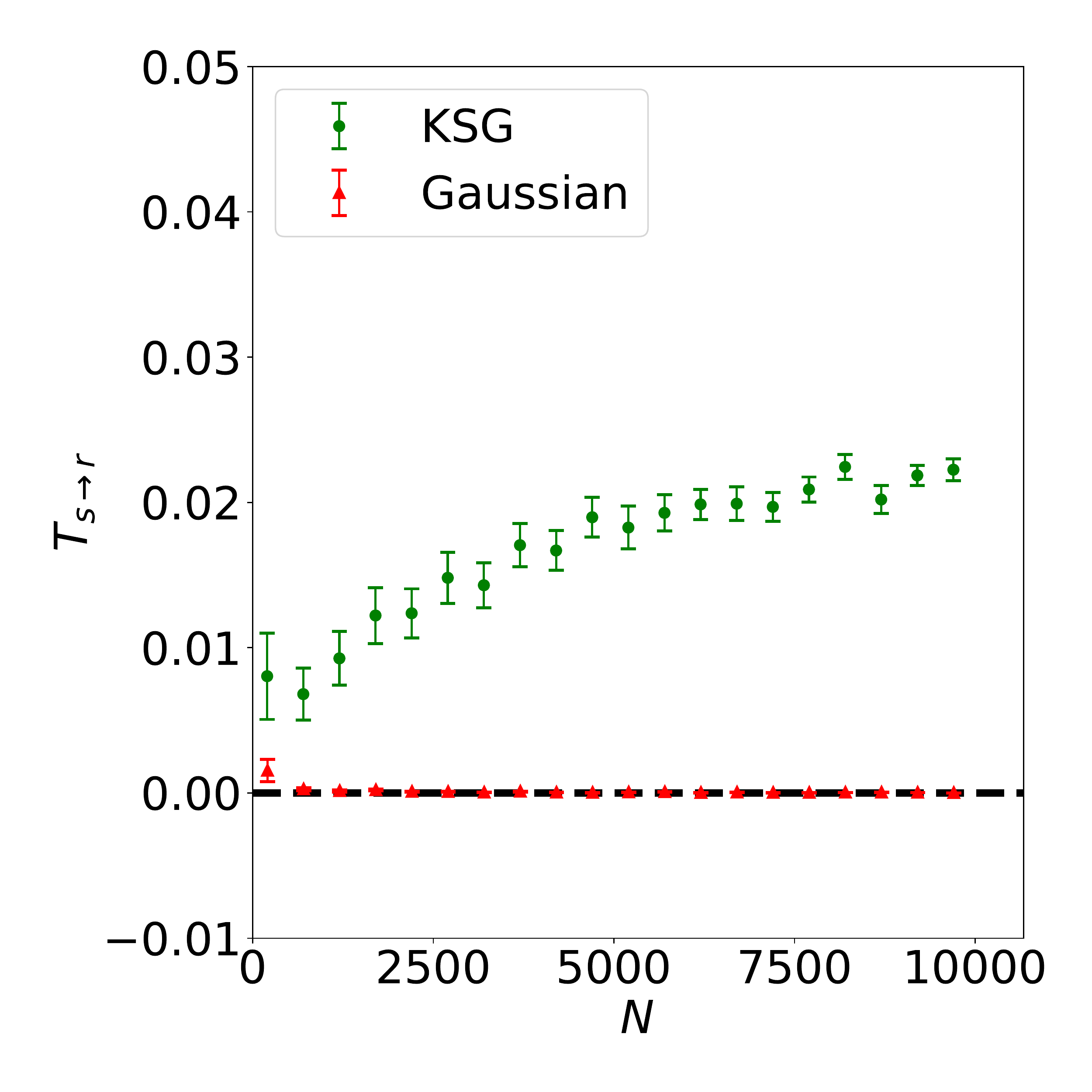}\includegraphics[width = 0.45\linewidth]{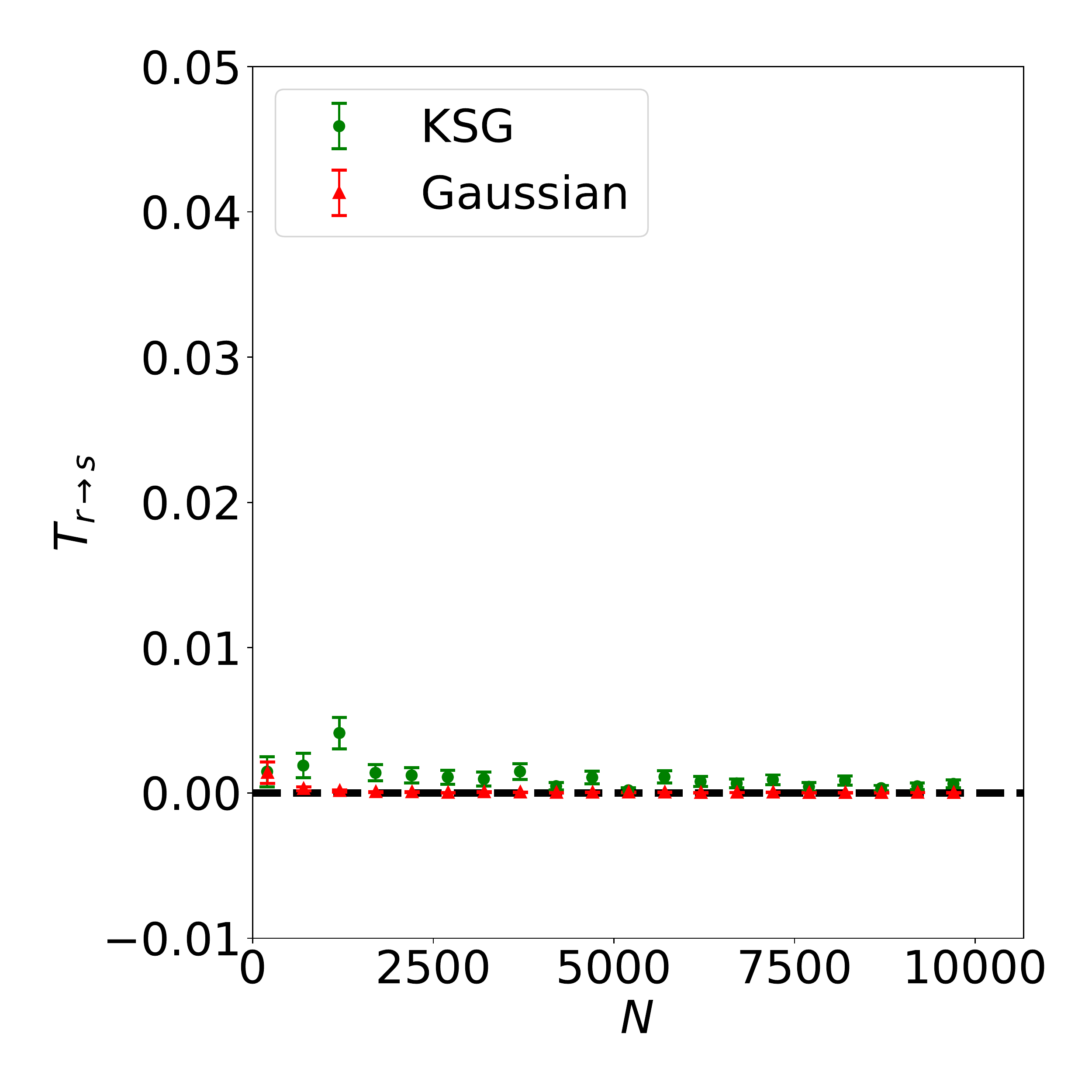}\\
    \centering Expected link $r\rightarrow s$\\
   \includegraphics[width = 0.45\linewidth]{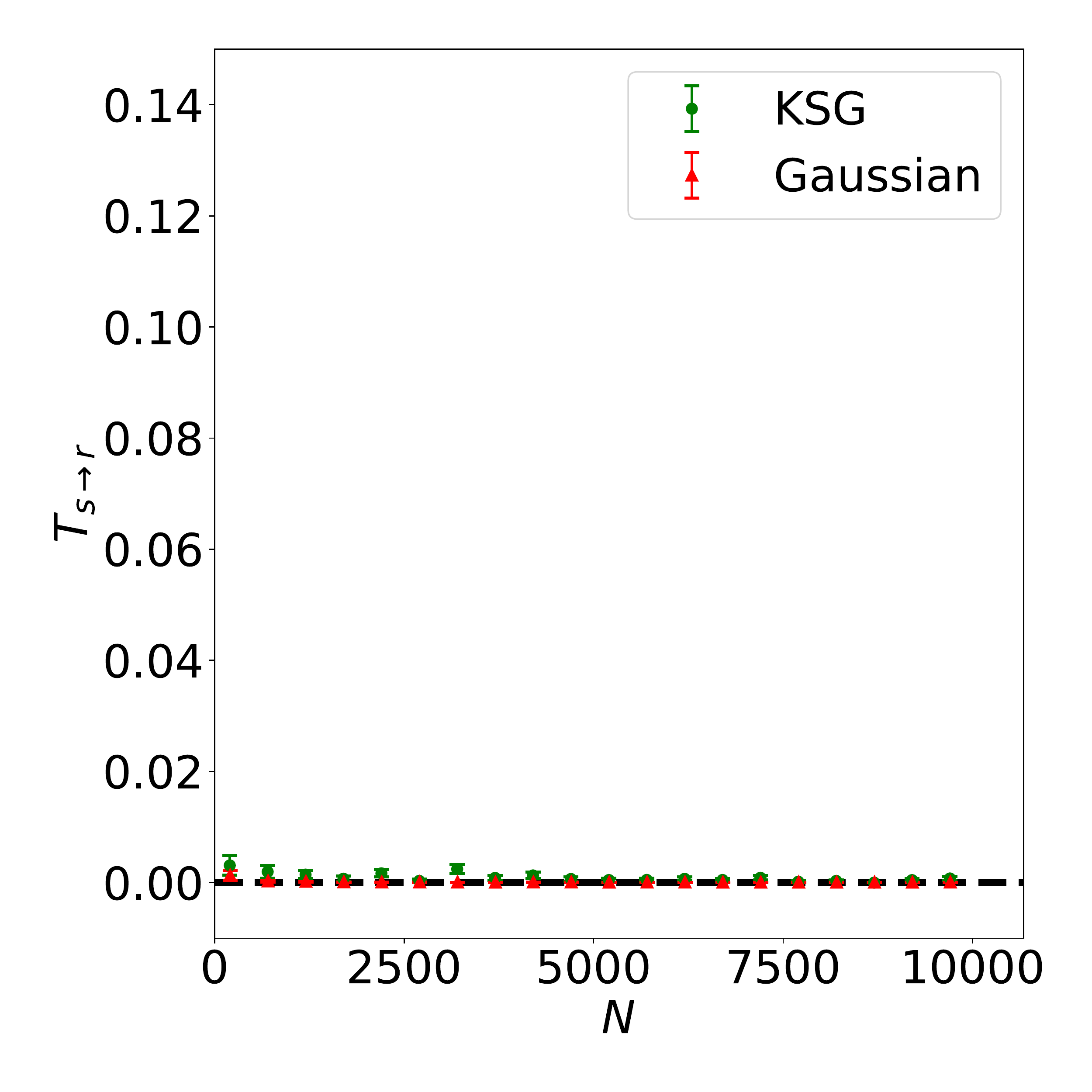}\includegraphics[width = 0.45\linewidth]{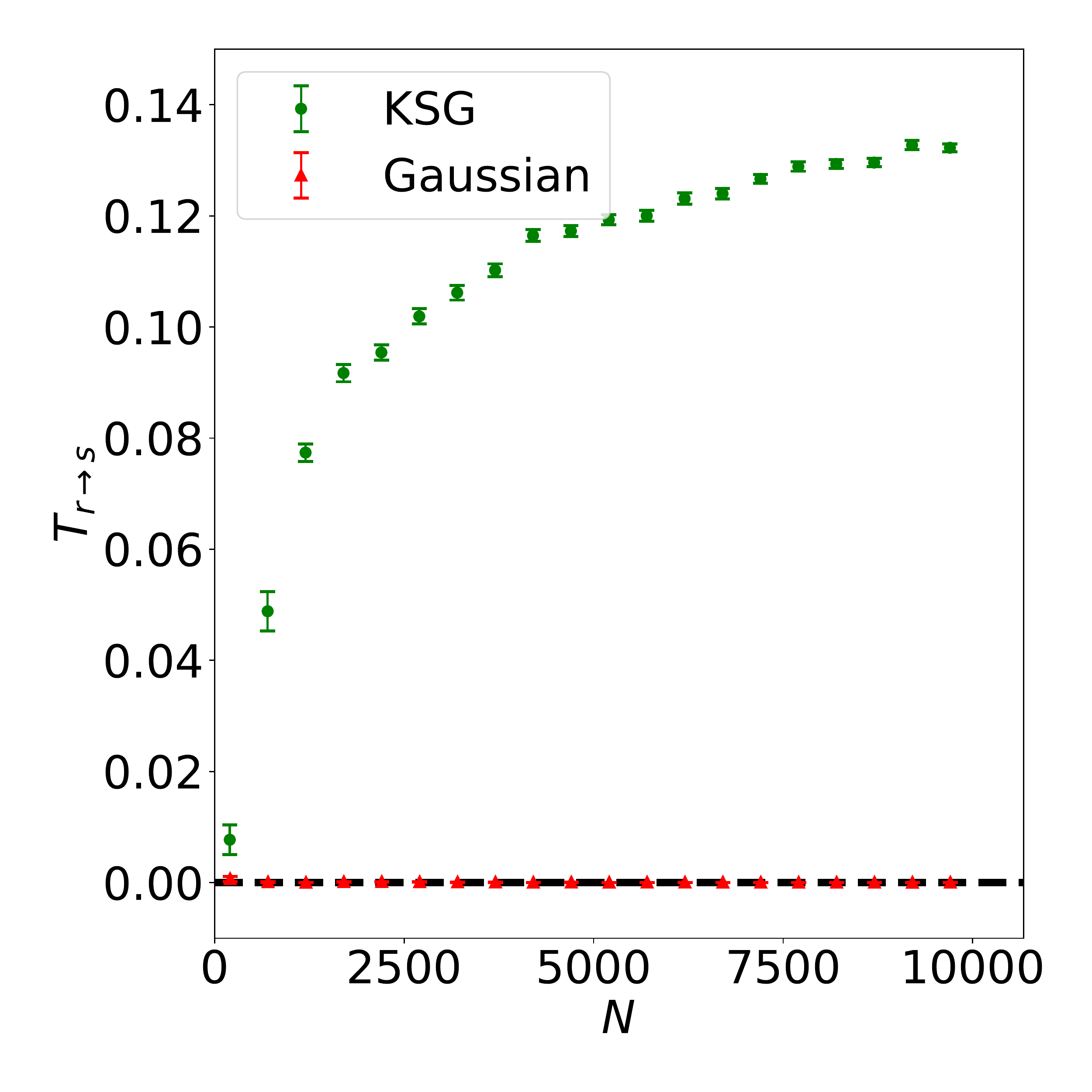}
    \caption{Top: TE from spread to returns (left) and reverse (right), $\alpha=0.1$, $\beta=0.4$, $a=0.8$, $b=0.0$, $c=0.1$, $\gamma=0.9$. Bottom: TE from spread to returns (left) and reverse (right), $\alpha=0.1$, $\beta=0.1$, $a=0.1$, $b=0.9$, $c=0.1$, $\gamma=0.0$.} 
    \label{fig_spread_to_returns_main}
\end{figure}

Parameters $\alpha,\beta,a,b,c ,w,\gamma$ affect stationarity and damping in the system therefore we need to choose such parameter values that system would be stationary throughout. To find a stationary combination of the parameters, we perform a moment analysis of the two processes (see \appref{app_toy_example} for more details). In \figref{fig_spread_to_returns_main}, we show that an expected links from spread to returns (top) as well as from returns to spread (bottom) are detectable using KSG estimator for TE, whereas Gaussian estimator cannot detect the non-linearly coupled GARCH variables. We also note that at small time series lengths, spurious bidirectional coupling may be observed, and the observed coupling strength converges to a true one when the time series are sufficiently long.

Since for stationary time series transfer entropy is an expectation value of the logarithm of ratios of conditional probability density functions, defined as per \eqref{eq_local}, in GARCH model \eqref{eq_garchspread}, these probability density functions can be estimated using numerical integration and samples of a joint process $\lbrace r_t,s_t \rbrace$. In case of causal link from spread to returns, TE is defined as an expectation value of $\log\frac{p(r_t|r_{t-1},s_{t-1})}{p(r_t|r_{t-1})}$. Here we have an exact probabilistic model, namely
\begin{eqnarray*}
 p(r_t=x| r_{t-1}, s_{t-1}, \sigma_{t-1}) &=&\\
  \mathcal{N}(x;0, w+\alpha r_{t-1}^2 +\beta \sigma_{t-1}^2+\gamma s_{t-1}^2), 
\end{eqnarray*}
where $\mathcal{N}(x;\mu=0,\sigma^2=w+\alpha r_{t-1}^2 +\beta \sigma_{t-1}^2+\gamma s_{t-1}^2))$ is a Normal probability density function parameterised with $\mu,\sigma$. Similarly, one can obtain the analytical expression for $p(r_t| r_{t-1})$ via marginalisation. Similarly, transfer entropy from returns to spread is the expectation value of $ \log \frac{f(s_t| s_{t-1}, s_{t-2}, r_{t-1}, r_{t-2})}{f(s_t| s_{t-1}, s_{t-2})}$. In \appref{app_toy_example} we show that the denominator, denoted by $\phi(s_t=y)$ can be written as follows
\begin{eqnarray*}
\phi(y) =  \frac{2y}{\Gamma(1/2)\sqrt{2c(y^2 - c^{*}_t)}}\exp{\left( -\frac{y^2 - c^{*}_t}{2c} \right)} \bold{1}_{ [{\sqrt{c^{*}_t},\infty }\big) } (y).
\end{eqnarray*}
Here $c^{*}_t= a s_{t-1}^2 + b(w+\alpha r_{t-2}^2 +\beta \sigma_{t-2}^2+\gamma s_{t-2}^2)$ and $\bold{1}_{\mathcal{A}}$ stands for an indicator function defined over a subset of real numbers, $\mathcal{A}$. Again, $f(s_t| s_{t-1}, s_{t-2})$ can be obtained by marginalisation.

In the \appref{app_toy_example} we show that the values of information transfer that we obtained empirically and report in \figref{fig_spread_to_returns_main} are in
agreement with the transfer entropy obtained using numerical integration.

\section{Empirical results}\label{sec:results}

In this section, we report results of information sharing across markets when a single observable is considered, and when information communication between different observables is taken into account. 

\subsection{Inter-market information dynamics for a single observable}\label{sec_res_one_observable}
\begin{figure*}[ht]
    \centering    
    \includegraphics[width=0.3\linewidth]{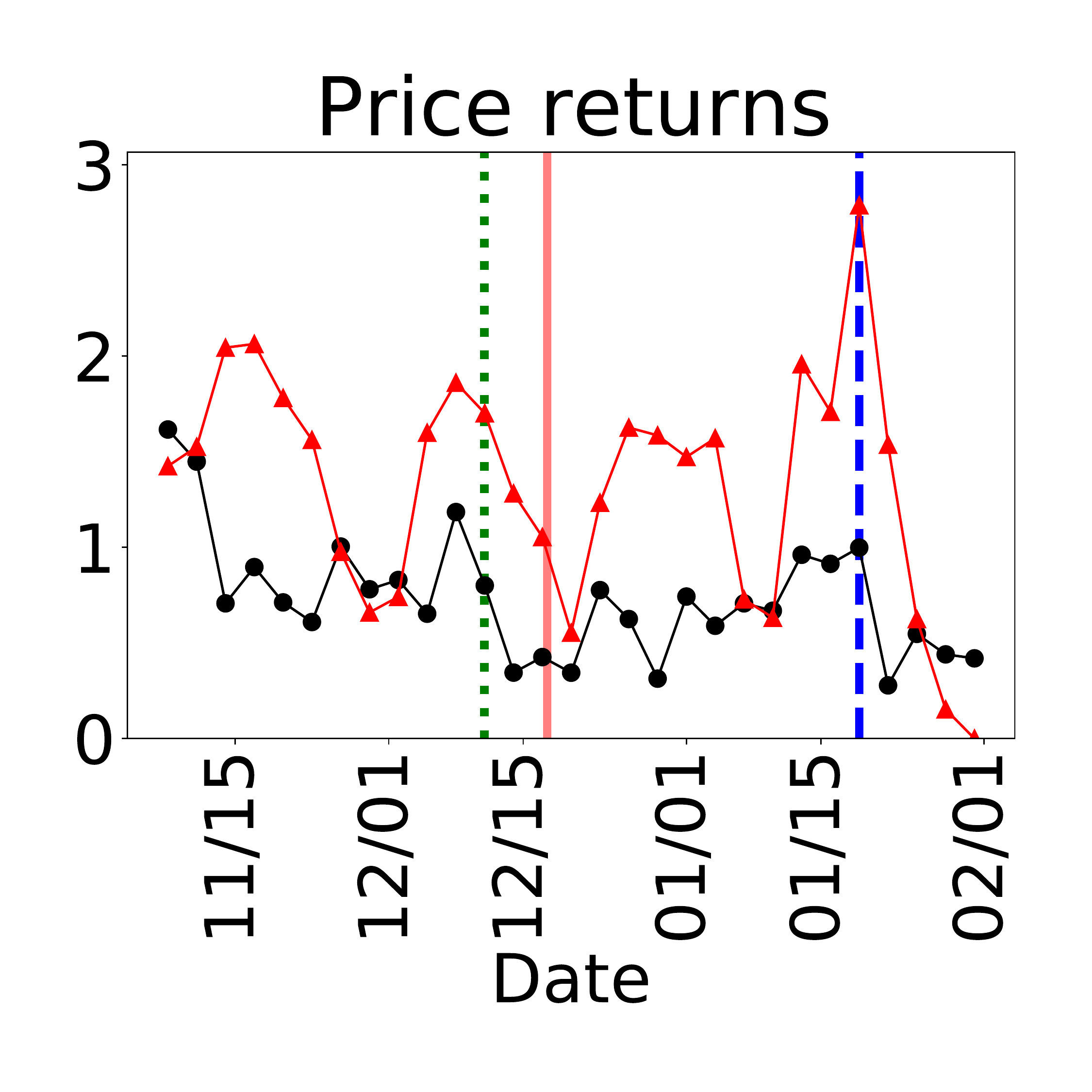}
    \includegraphics[width=0.3\linewidth]{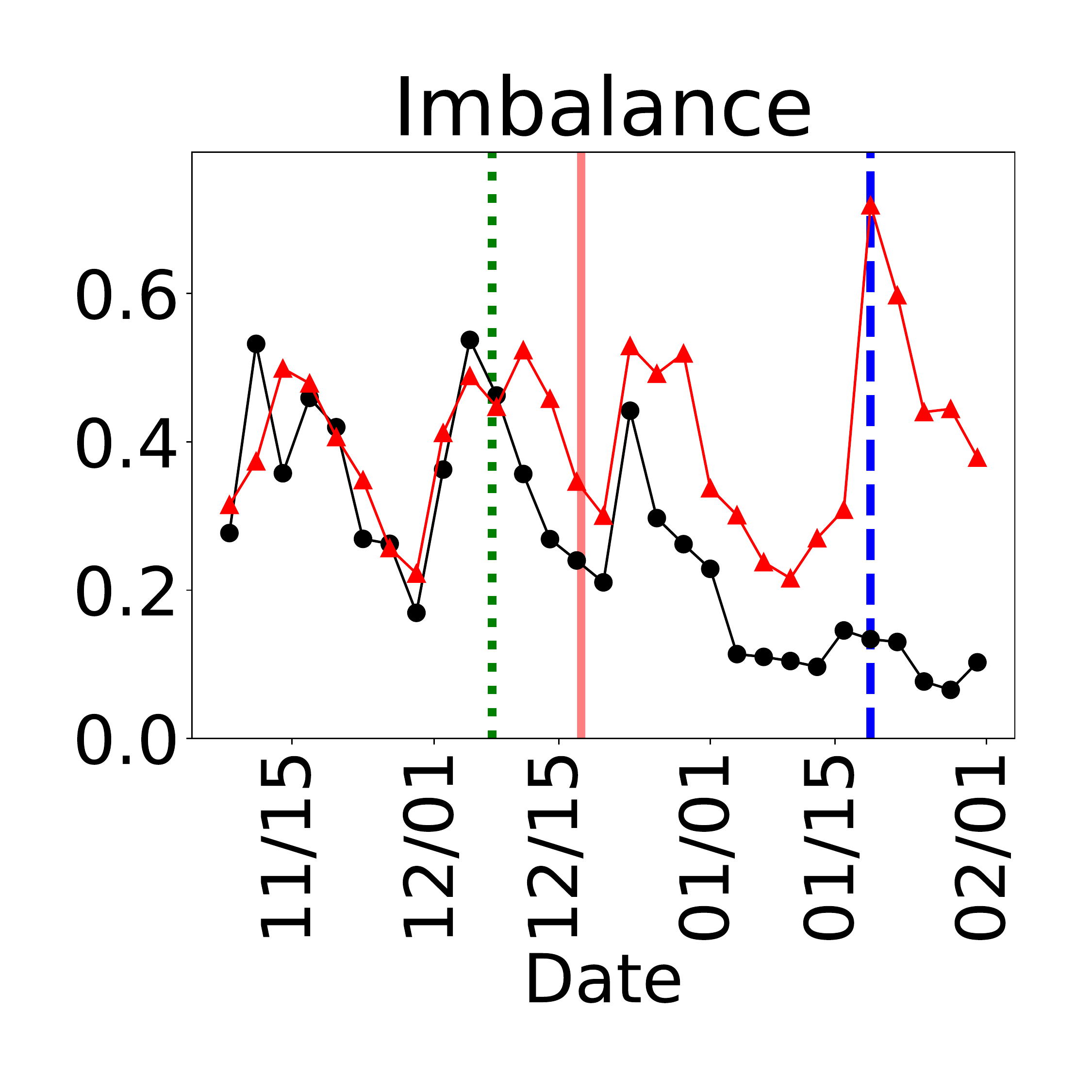}
    \includegraphics[width=0.3\linewidth]{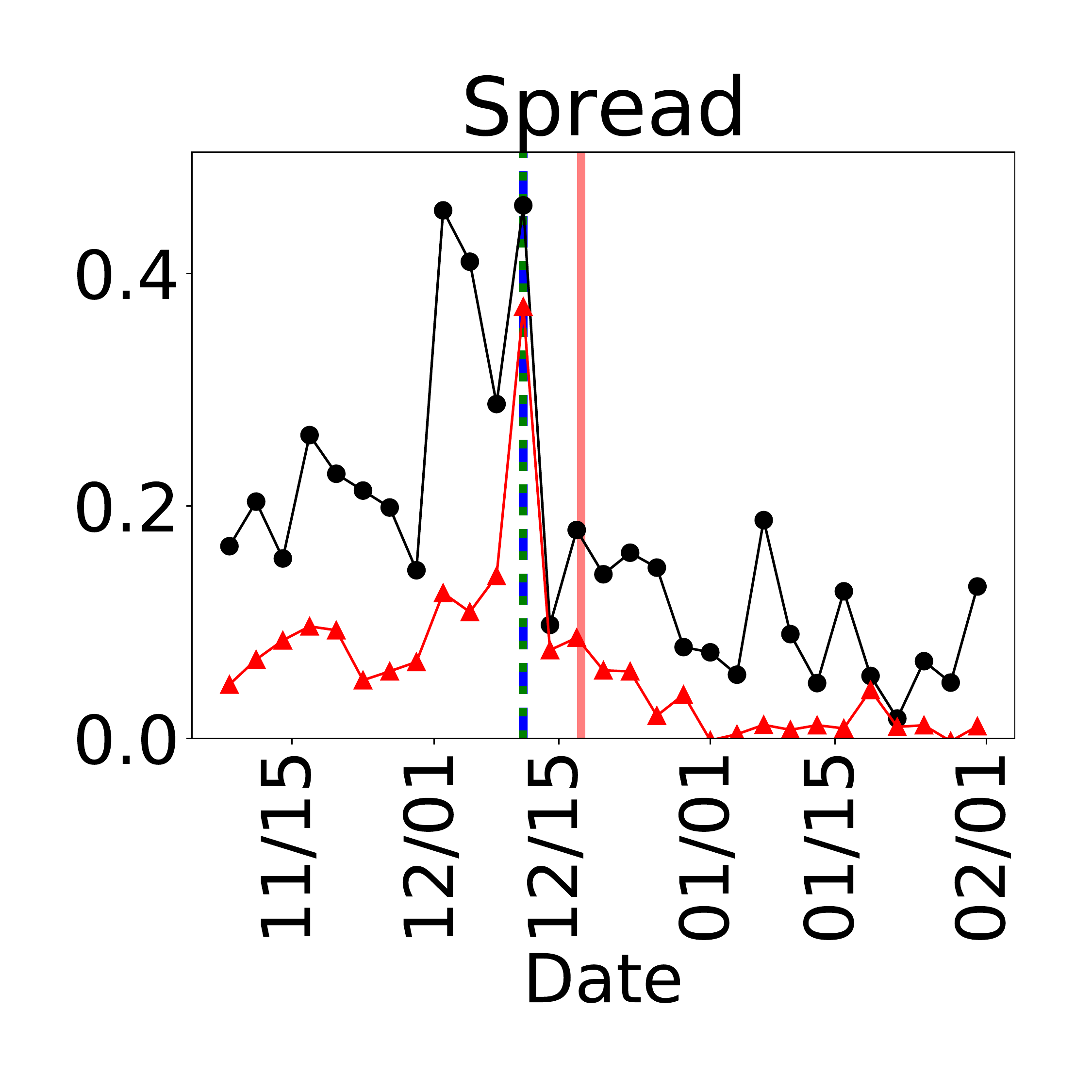}\\
    \includegraphics[width=0.3\linewidth]{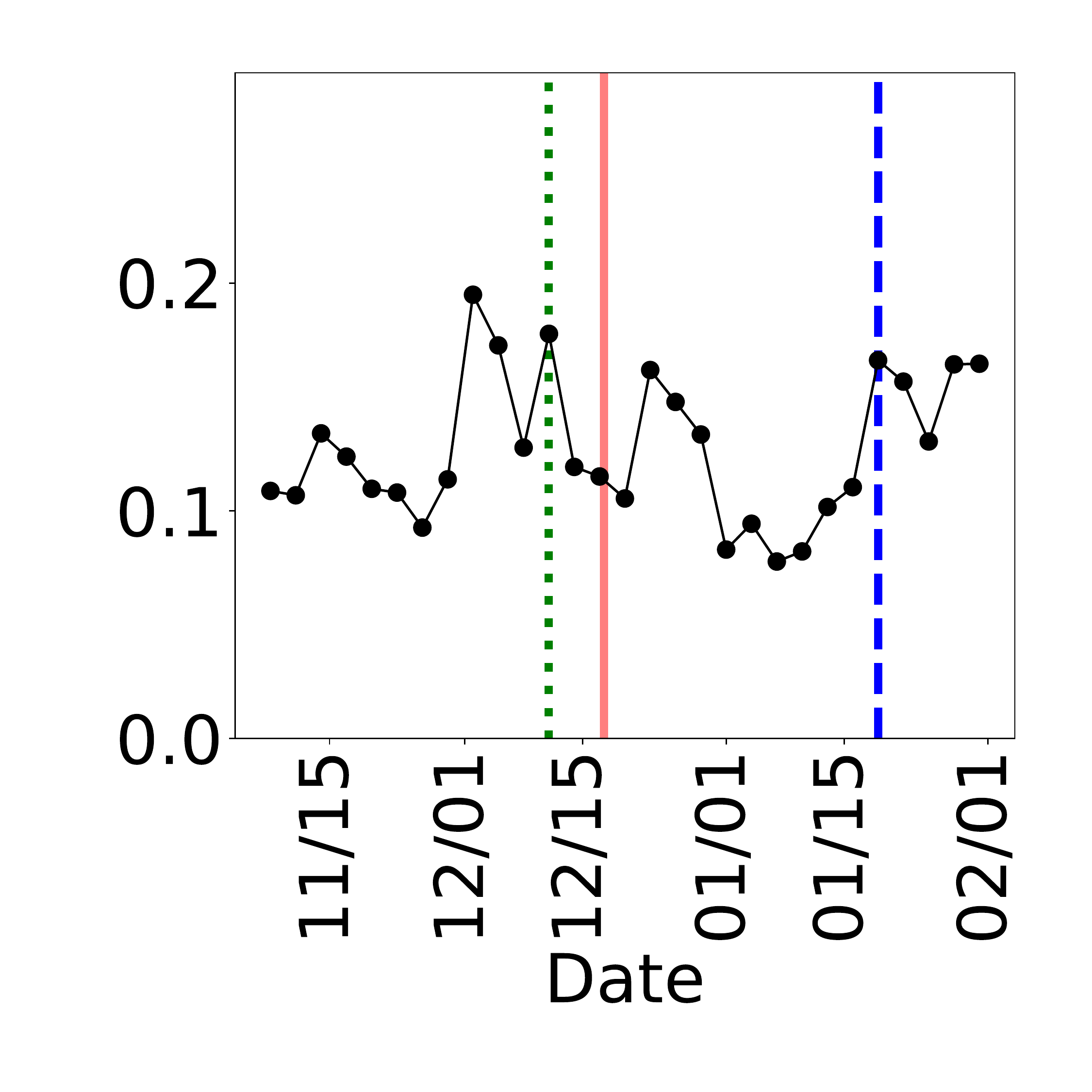}
    \includegraphics[width=0.3\linewidth]{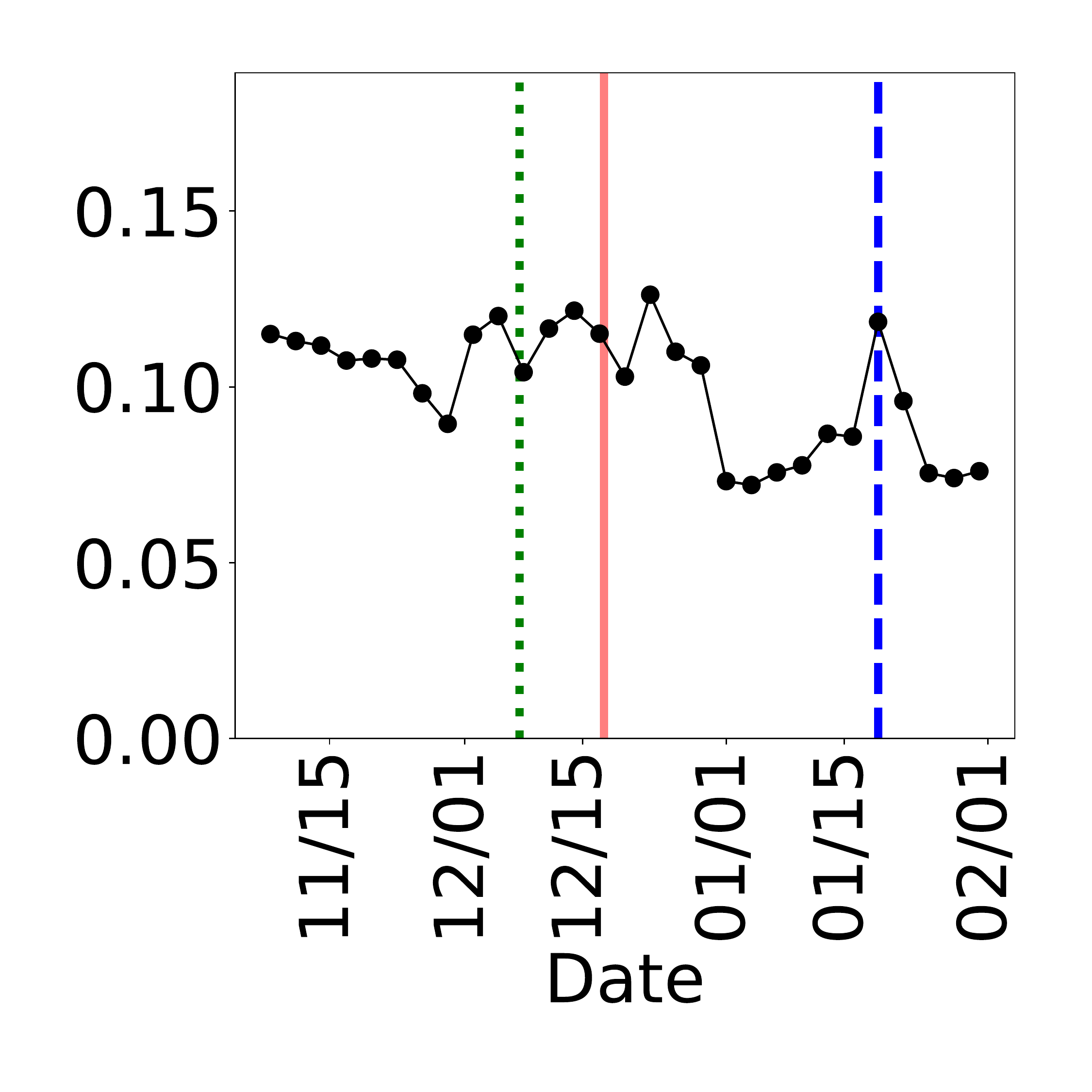}
    \includegraphics[width=0.3\linewidth]{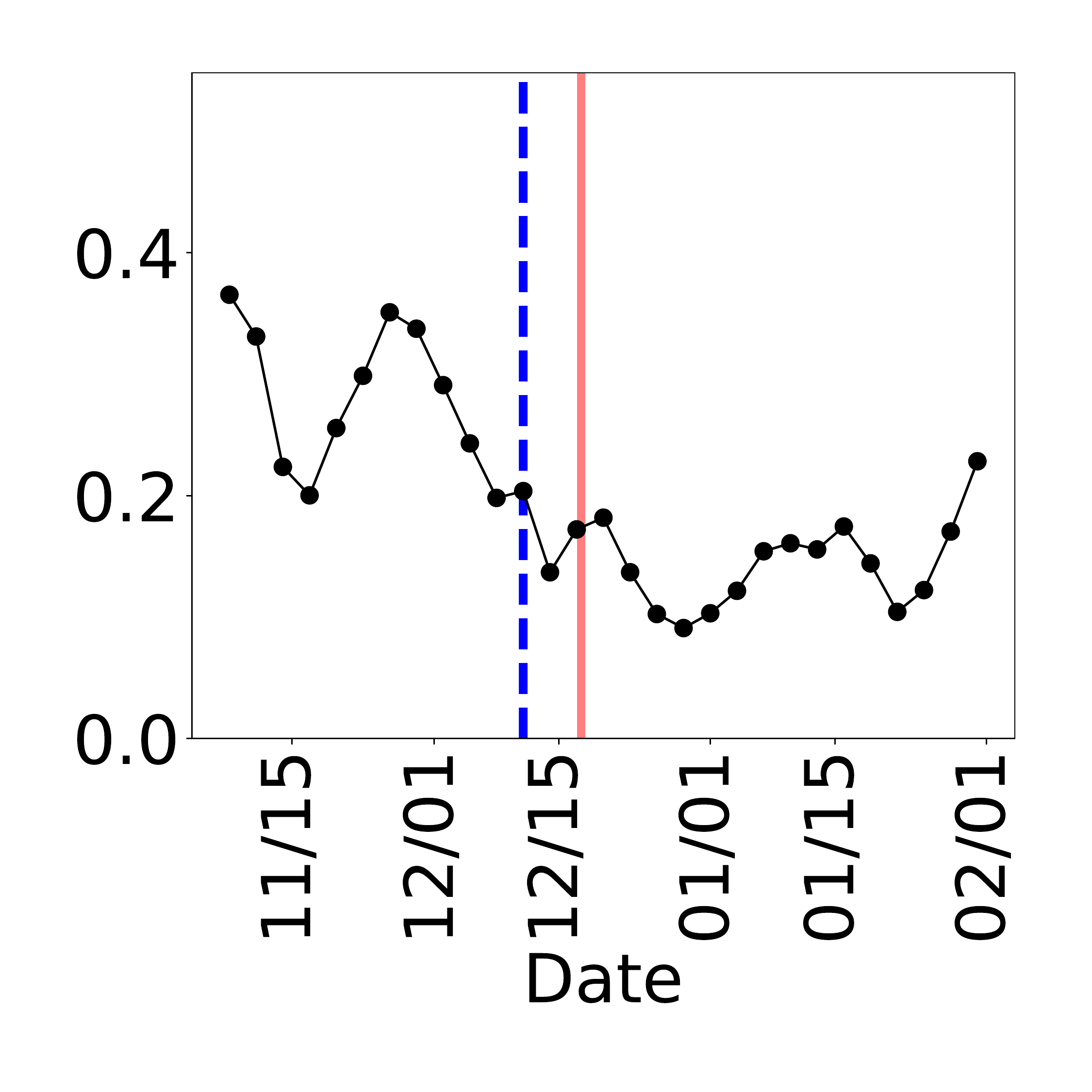}
    \caption{Top: Multi-information $I^{\textrm{sys}}$ (denoted with red triangles) and the total apparent transfer entropy $T^{\textrm{app,sys}}$ (denoted with black circles) across all pairs of markets using the following time series (from left to right): price returns, order imbalance, and spread. The blue dashed vertical lines indicate ``critical points'': the date of the peak value obtained by each variable is indicated by green dotted line; the red lines indicate the peak of the bubble, and the blue dashed lines highlight the dates at which multi-information was maximal. Bottom: average active information storage $A^{\textrm{sys}}$.}
    \label{fig_MI_TE}
\end{figure*}

Firstly, we consider information dynamics individually for each of the observables: $r$, $s$, $\mathcal{O}$. Here we look at the following: the total apparent transfer entropy, $T_r^{\textrm{app,sys}},T_s^{\textrm{app,sys}},T_{\mathcal{O}}^{\textrm{app,sys}}$, multi-information, $I_r^{\textrm{sys}},I_s^{\textrm{sys}},I_{\mathcal{O}}^{\textrm{sys}}$, and average active information storage, $A_r^{\textrm{sys}},A_s^{\textrm{sys}},A_{\mathcal{O}}^{\textrm{sys}}$. When constructing each of these measures, only significant information dynamics values are considered (the significance is discussed in \secref{sec:setup}). The results obtained for each sliding window are reported in \figref{fig_MI_TE}: top figures show the total apparent transfer entropy as well as multi-information; bottom figures report average active information storage. Overall, we observe that week by week, there were large fluctuations in the amount of information sharing and synchronisation across markets. The patterns of information dynamics for each microstructural variable differ, therefore we discuss them individually.

\paragraph{Spread}

The temporal evolution of information processing in spread suggests that a ``critical point'' exists nearing price peak and the peak in spread itself. Such point in time is thought to be characterised by a peak in MI amongst agents that constitute the system~\cite{matsuda1996mutual,barnett2013information,johansen1999predicting}. Authors of~\cite{harre2009phase} claim that stock market crashes exhibit a peak in MI at the point in time when one would expect a significant regime change to take place. A peak in MI observed in \figref{fig_MI_TE} for spread suggests that within that week, a mutual hidden driver could have increased in strength for this market microstructure observable (a similar case was simulated in \figref{fig_case_1_main}(b)).

\begin{figure*}[!ht]
 \centering \includegraphics[width = 0.75\linewidth]{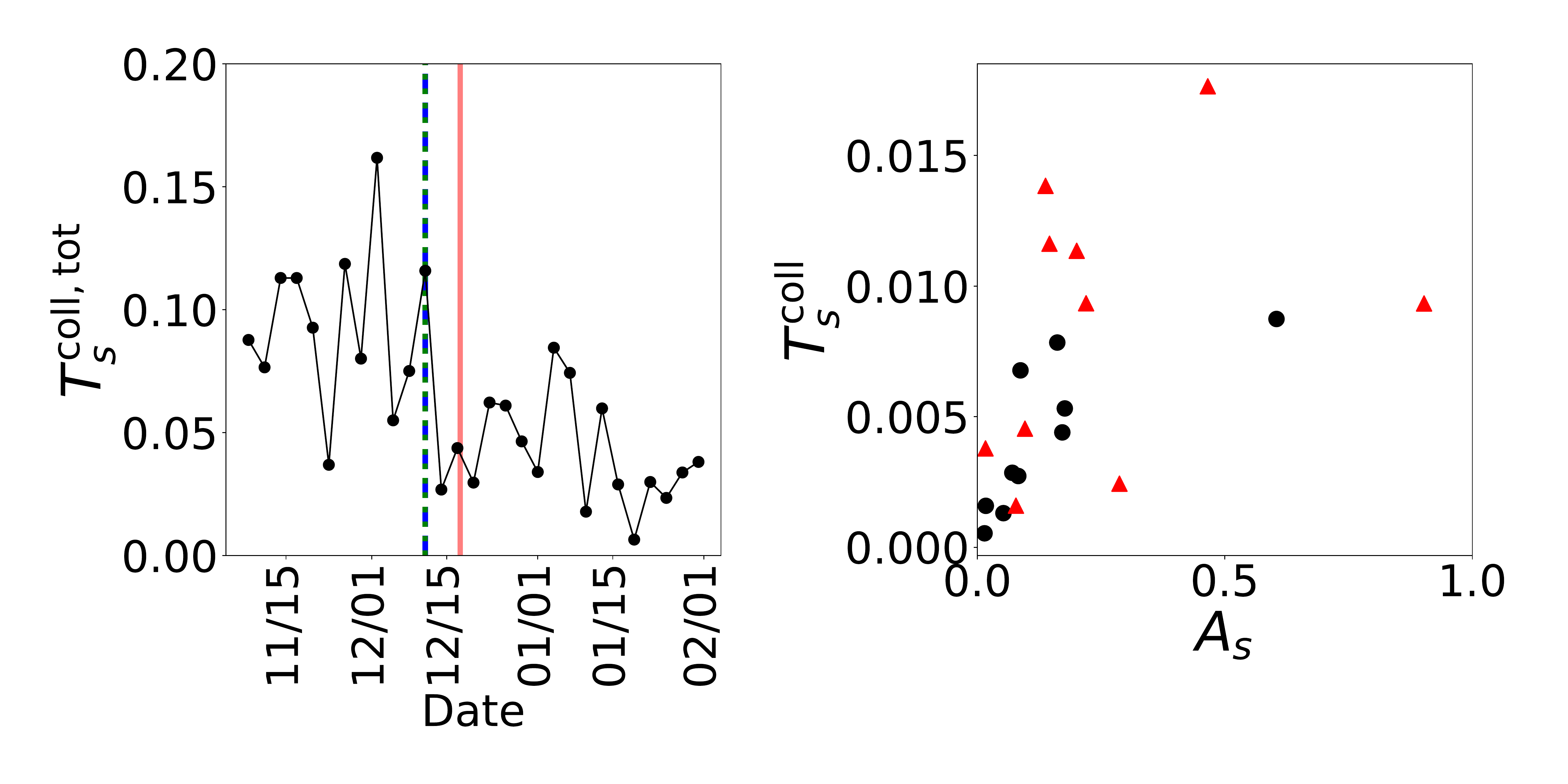}
    \begin{tabular}{c|p{2.5cm}|p{2.5cm}||p{2.5cm}|p{2.5cm}}
  $\alpha$ &  $ A_{s^{\alpha}}\cdot 10^{-2}$\newline before Dec. 17th &  $ A_{s^{\alpha}}\cdot 10^{-2}$\newline after Dec. 17th & $ T_{s^{\alpha}}^{\textrm{coll}}\cdot 10^{-3}$\newline before Dec.\ 17th &  $ T_{s^{\alpha}}^{\textrm{coll}}\cdot 10^{-3}$\newline after Dec.\ 17th \\
\hline\hline
BTC-e & 30.1 & 23.1 & 9.34 & 0.87 \\
Binance & 15.18 & 3.32 & 17.65 & 0.13 \\
Bitfinex & 4.13 & 2.35 & 1.59 & 0.16 \\
Bitstamp & 7.12 & 5.58& 11.62 & 0.68 \\
Bittrex & 8.15 & 5.87& 9.34 & 0.53 \\
Coinbase & 9.17 & 4.88 & 3.79 & 0.05 \\
Gemini & 11.4 & 9.54 & 4.54 & 0.29 \\
HitBTC & 23.0 & 9.58 & 2.44 & 0.44 \\
Kraken & 10.32 & 5.6 & 11.35 & 0.78 \\
Poloniex & 7.42 & 5.03 & 13.84 & 0.27 \\

\end{tabular}
    \captionlistentry[table]{A table beside a figure}
    \captionsetup{labelformat=andtable}
    \caption{Top left: total collective TE. Vertical lines have the same meaning as in \figref{fig_MI_TE}. Top right: relationship between average collective TE and average market's AIS. Black circles show this relationship averaging over time series windows before the price crash date; red triangles show the relationship averaging over the time series windows after the price crash date. Bottom: average market spread's AIS and average market spread's collective TE, obtained averaging over windows before the crash date (first and third columns), and averaging over windows after the crash date (second and fourth columns).      }
    \label{tab_ais_collte}
  \end{figure*}

In information dynamics amongst spread variables we see a large amount of total apparent transfer entropy, as well as large average AIS, persisting for weeks preceding the price crash. This is in line with a situation simulated in \figref{fig_case_1_main}(a), therefore we conclude that \emph{spread system variables transit from a strong-coupling regime before the crash to a weak-coupling regime after the crash}. Our intuition is further supported by similar results observed by considering  $T_{s}^{\textrm{coll,sys}}$, shown in top left of \figref{tab_ais_collte}. In the \tabref{tab_ais_collte} we report the average collective transfer entropy for the time windows before and after the price crash. As expected, the absolute values of the total collective transfer entropy $T_{s}^{\textrm{coll,sys}}$ are much smaller than the values of apparent transfer entropy. However, we 
still observe a similar shift from a strong-coupling regime to a weak-coupling regime, centered around the crash date.

In \tabref{tab_ais_collte} we analyse AIS and total collective TE for individual markets in the two market regimes (leading up to and following the price crash). The averages are calculated from either the observed AIS values in the windows that precede the crash or follow it. Markets with small $ A_s$  are, on average, ``information absorbers'' (e.g., Gemini, Coinbase): their order book spreads are responsive to either exogenous information sources or inter-market dynamics, or are generally intrinsically unpredictable. Other markets, such as BTC-e use a lot of information from their memory to compute the next market spread. Lastly, \figref{tab_ais_collte} shows that those markets that were shown to have large $T_s$, also are those with the largest $A_s$. On the other hand, spread in markets such as Gemini, Coinbase, Bitfinex, are driven by exogenous information as both $T_s^{\textrm{coll}}$ and $A_s$ is small. Therefore, given information available from market microstructure observables, they are the least predictable.
\tabref{tab_ais_collte} shows that Bitfinex, Gemini, and Coinbase are exchanges that experienced the smallest influx of information from other exchanges prior to crash. We also observed that these markets have small average AIS, suggesting that they are ``independent'' from other exchanges and could primarily be driven by exogenous information sources. On the contrary, Binance, Poloniex absorb the largest amount of information from other markets. They were also found to have large average AIS. Such markets could be labelled as ``information absorbers''. 

\paragraph{Price returns and order imbalance}

\figref{fig_MI_TE} shows that information dynamics patterns for $\mathcal{O},r$ are different from those of $s$. Firstly, for both observables, temporal evolution of MI is similar: somewhat symmetric around the crash date, minimal at the crash date, and fluctuating from high MI to low MI with a period of around 2-3 weeks. These observations suggest that similar processes were driving the system at either side of the price bubble. The peak of the bubble is characterised by a clear, although not dramatic, reduction in both TE and MI, and to a lesser extent, reduction in AIS. We observe that for $\mathcal{O}$ (and to a smaller extent, for $r$), all three measures---AIS, TE, MI---follow the same pattern, that is, when TE is large, AIS, MI are also large. We were able to simulate this scenario in \figref{fig_case_1_main}(c) by changing the autoregressive parameters responsible for the amount of uncertainty in the signals. When the amount of noise, unique for each system's constituents, is varied, the three information dynamics measures either increase or decrease in magnitude simultaneously as a response, being high-valued when the uncertainty is low. \emph{The minimal values of information dynamical quantities at the peak of the bubble suggest that the system was unpredictable, chaotic and not synchronised.}

\subsection{Information flow between market microstructure observables}\label{sec_info_flow_obsr}

Here we consider information sharing between different market microstructure observables and compare it to inter-market information sharing. We study two non-overlapping time series windows: the time series preceding the price crash date (2017-11-01 to 2017-12-17), and the remainder (2017-12-17 to 2018-01-31).

\begin{figure}[h!]
    \centering
    \includegraphics[width = 0.8\linewidth]{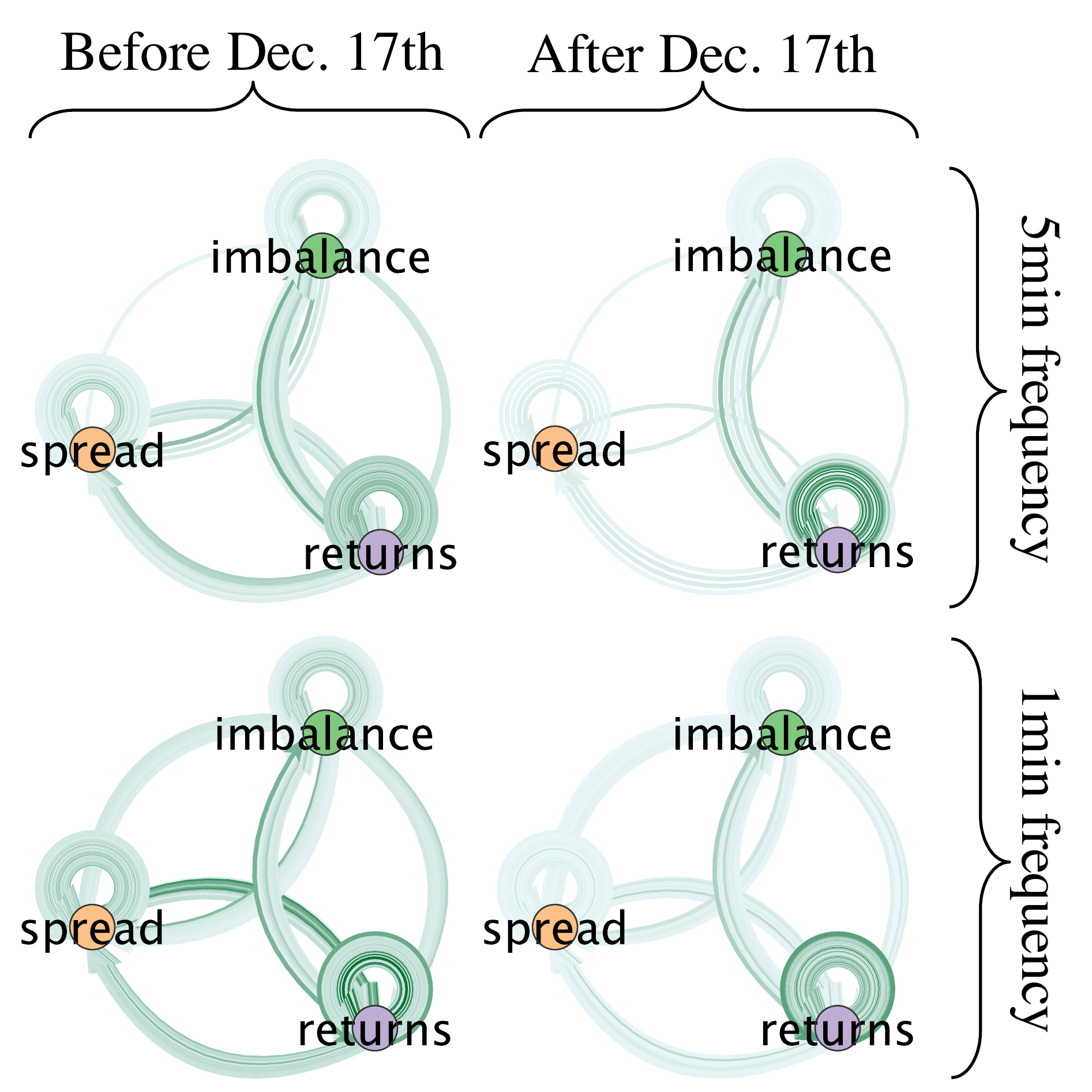}
        \hspace{1pt}
    \begin{tabular}{c|p{0.85cm}|p{1cm}|p{0.8cm}|p{0.8cm}|p{0.8cm}|p{0.8cm}}
    Freq. &Period &  Observ. & \parbox{0.1cm}{${\omega^{\textrm{in}}}$\\$\cdot10^{-3}$} & \parbox{0.1cm}{${\omega^{\textrm{out}}}$\\$\cdot10^{-3}$} & \parbox{0.1cm}{${\omega^{\textrm{self}}}$\\$\cdot10^{-3}$} & \parbox{0.1cm}{$ A^{\textrm{sys}}$\\$\cdot10^{-3}$} \\
    \hline\hline
\multirow{6}{*}{5min} &\multirow{3}{*}{\rotatebox{90}{\parbox{1.5cm}{Before Dec.\ $17^{\textrm{th}}$}}}
&$s$ & 6.5 & 3.07 & 3.55  &  272.93\\
&&$\mathcal{O}$ & 1.02 & 0.65 & 1.44  &  56.32\\
&&$r$ & 0.85 & 1.98 & 13.85  &  162.06\\
 \cline{2-7}
&\multirow{3}{*}{\rotatebox{90}{\parbox{1.3cm}{After Dec.\ $17^{\textrm{th}}$}}}&$s$ & 1.84 & 0.74 & 0.33  &  157.75\\
&&$\mathcal{O}$ & 1.19 & 0.21 & 1.01  &  41.73\\
&&$r$ & 0.22 & 1.45 & 11.23  &  80.64\\
\hline\hline
\multirow{6}{*}{1min} 
&\multirow{3}{*}{\rotatebox{90}{\parbox{1.5cm}{Before Dec.\ $17^{\textrm{th}}$}}}&$s$ & 10.56 & 15.92 & 9.83  &  570.48\\
&&$\mathcal{O}$ & 1.42 & 1.22 & 4.31  &  136.18\\
&&$r$ & 3.7 & 2.71 & 6.52  &  206.75\\
 \cline{2-7}
&\multirow{3}{*}{\rotatebox{90}{\parbox{1.3cm}{After Dec.\ $17^{\textrm{th}}$}}}&$s$ & 5.24 & 5.46 & 1.0  &  325.31\\
&&$\mathcal{O}$ & 1.49 & 0.79 & 3.55  &  104.89\\
&&$r$& 1.31 & 1.97 & 32.04  &  126.43\\
\end{tabular}
    \captionlistentry[table]{A table beside a figure}
    \captionsetup{labelformat=andtable}
    \caption{Top: Directed multi-graphs of information flow between the three market observables before the price crash (left) and after the price crash (right). Saturation of edges indicates the strength of information transfer. Self-loops represent transfer entropy from one market to another when considering the same observable. Bottom: Amount of apparent information transfer between market microstructure observables per possible link. The last column shows an average active information storage. The first part of the table reports results at $5$min-level frequency of re-sampling, whereas the second one is for a minute level frequency of re-sampling.  }
    \label{tab_triad}
\end{figure}

We report the results in \tabref{tab_triad}. Altogether, they suggest that \emph{the system's connectivity as a whole moved from a high-coupling regime to a low-coupling regime with one significant exception for price returns}. Namely, for price returns, where we see $\omega^{\textrm{self}}$ larger than $\omega^{\textrm{in}},\omega^{\textrm{out}}$ and $\omega^{\textrm{self}}$ prior to crash date is \emph{smaller} than after the crash. Lastly, we also observe a larger average AIS for returns prior to the crash. Furthermore, $\omega^{\textrm{in}},\omega^{\textrm{out}}$ is of the same order as $\omega^{\textrm{self}}$, suggesting that internal market dynamics played a significant role in price movement in the period.

For the other two observables, we observe that both inter- and intra- market connectivity becomes smaller after the crash. For spread, we observe that a relatively large inter-market connectivity drops by an order of magnitude after the crash, suggesting at that time spread is either responsive to changes in other market observables or it responds to unobserved variables. Seeing virtually no strong links in the multigraph that illustrates the post-crash market microstructure (\figref{tab_triad}, right), we consider the former case less likely. In comparison, order imbalance both received and absorbed a lot less information in contrast to spread. It may indicate that this variable responds to information that goes beyond our dataset. 

\section{Discussion}\label{sec:discussion}
In this paper we studied patterns of information dynamics in the Bitcoin system, describing it with market microstructure observables: price returns, order imbalance, and spread---three measures that quantify the state of market makers and market takers. Our analysis contrasted micro-level information processing within important Bitcoin trading venues during two distinguishable states: while price was, on average, ascending, and during the period in which price was, on average, dropping. We found persisting intra-market connectivity at the minute level frequency. Furthermore, we found that at high frequency,\emph{ markets are interconnected via order book spread, and the interconnectivity is enhanced prior to price crash, suggesting a potential regime shift}. On the contrary, the inter-market connectivity via price returns and order imbalance appeared symmetric around the price crash. We also observed the dip in all information dynamic measures at the time of the crash, suggesting that the system was asynchronous and unpredictable. We also observed information flow between different market observables. By contrasting the two market regimes, we found that the system shifted from a strongly interlinked state to a sparsely connected one.  

To supplement these findings, we analysed several simulated econometric models. With the model of auto-regressive non-linearly coupled variables that have time-varying drivers, we found several simplified mechanism that could explain the empirical observations. With the generalised auto-regressive conditional heteroskedasticity model (GARCH) coupled with spread dynamics, we studied a link between returns and spread, as well as potential of detecting it using means of information transfer. Our findings suggest that different types of regime changes can be distinguished when a collection of information dynamics measures is used together.

All in all, our findings suggest that prior to a price crash, Bitcoin system was in a strongly coupled state. In particular, this is clear when liquidity marker---spread---is considered. The drop in all forms of information at the point at which the price is maximal suggests a reduction in predictability of the system. It may also indicate that such a system is susceptible to various types of perturbations, however we leave such hypothesis for further research. We also make a note that the results of \figref{fig_MI_TE} (right column) for $s$ are reminiscent of a second order phase transition~\cite{matsuda1996mutual,barnett2013information} and could be a predictor of a financial crash~\cite{johansen1999predicting}. 
Of course other explanations are possible for this empirical observation and future studies could further clarify its microscopic origin.

Our findings, although interesting, have limited explanatory capacity due to the granularity of order-book data, presence of noise and effects due to unobserved (exogenous) system drivers. We assumed that the minute-level snapshots of the limit order book data are sufficiently high-frequency to reveal true interactions across and within markets. Of course, this data is an approximation of an, in reality, practically continuous-time dynamics of the markets. Finally, since we limited our analysis to Bitcoin traded against USD(T), analysis of larger market microstructure datasets, consisting of a bigger variety of currency pairs and forming larger information transfer networks would undoubtedly provide richer insights as to why we observe increased market co-integration at the time of a price bubble. Lastly, it would be important to repeat our analysis using other similar datasets, concerning market microstructure during price bubbles. Observing similar results would suggest that information dynamics framework may be a generally applicable tool for early warning signals of financial crashes.

\section*{Author contributions}
All authors designed the research. 
V.V.\ conducted experiments, analysed the data, and derived mathematical models. F.L.\ and N.A.F.\ derived mathematical models and supervised the work. All authors interpreted the results and wrote the manuscript.

\begin{acknowledgments}
This work is supported by the European Union – Horizon 2020 Program under the scheme “INFRAIA-01-2018-2019 – Integrating Activities for Advanced Communities”, Grant Agreement n.871042, “SoBigData++: European Integrated Infrastructure for Social Mining and Big Data Analytics” (http://www.sobigdata.eu).
\end{acknowledgments}

\section*{Data availability statement}
The data that support the analysis of the \secref{sec:models} and \appref{app_signatures}, \appref{app_toy_example} of this study have been synthetically generated by the authors and the results can be reproduced using the equations and parameters described in the article. 

\appendix

\section{Data alignment}\label{app_data}
As stated in the main text, the limit order books are snapshots sampled at approximately minute level frequency. However, the exact time of a snapshot is not the zeroth second of a minute, the intervals between consecutive snapshots in one market are not identical, due to different technological reasons. Furthermore, for a given minute, the times at which the snapshots are taken in different markets need not be exactly the same. The trading data on the other hand is stored in continuous time, as each trading event is being registered with its exact execution time. Derivatives obtained from each type of data therefore have to be carefully aligned for the consequent analysis to always respect the order of time.

To illustrate the data alignment problem, consider some two consecutive LOB snapshots in BTC-e market: the first limit order book snapshot is taken on 2017-11-01 at $t_1=$00:00:03.021 and the second one is taken approximately one minute later, at $t_2=$00:01:02.798. Further consider Binance market, where the first limit order book snapshots are taken at $t_1'=$00:00:01.267 and $t_2'$=00:01:01.274. One can already see that $t_1\neq t_1'$ and that $t_2-t_1\neq 60$s.

\begin{figure}
    \centering
    \includegraphics[width=0.8\linewidth]{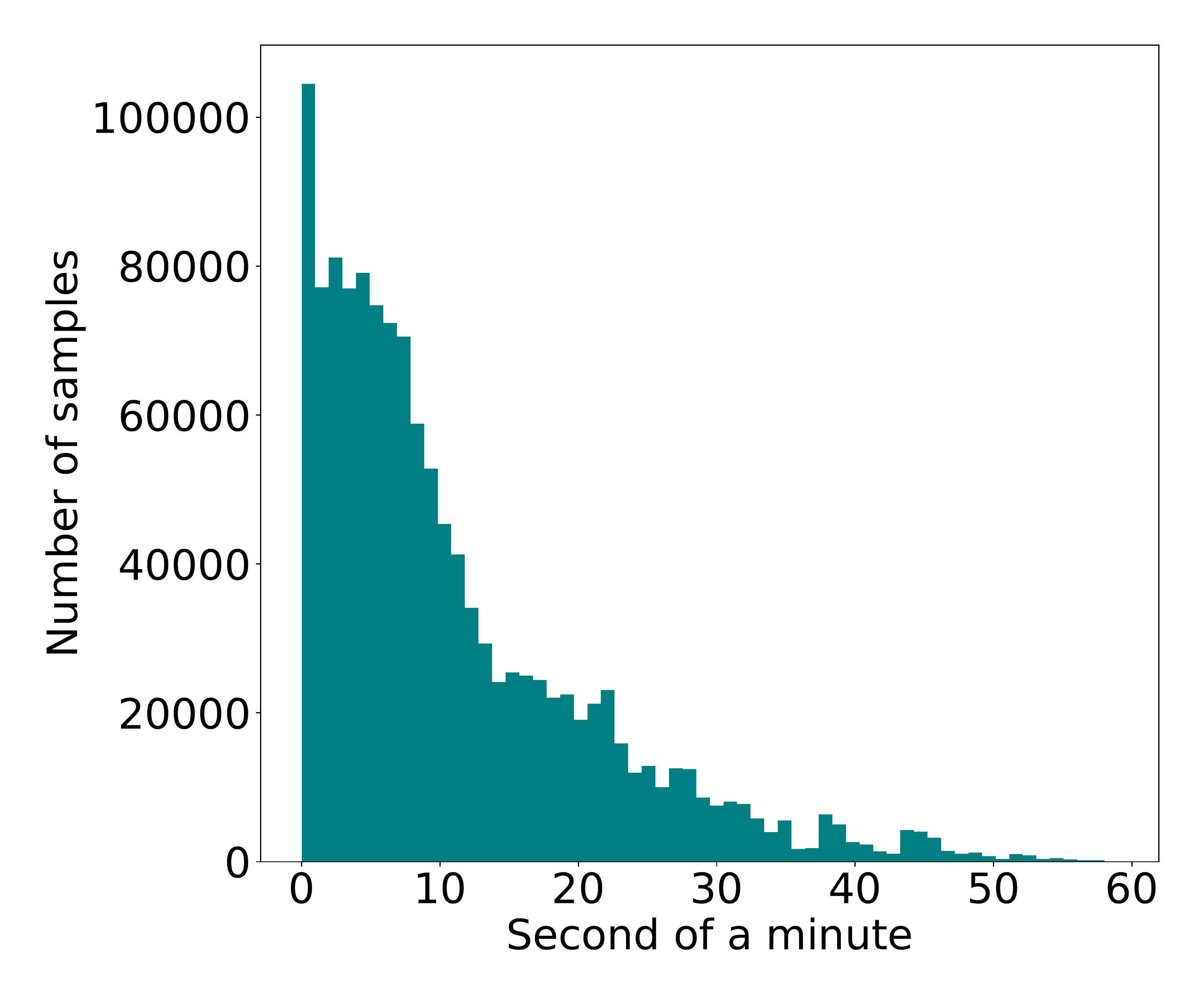}
    \caption{Histogram of the exact capture second of each minute snapshot of the limit order books used in this paper. }
    \label{fig_hist_lob}
\end{figure}

Let us first consider the data alignment for \textbf{analysis in \secref{sec_res_one_observable}}, where we consider information transfer between markets given one observable, e.g.\ spread. 

Firstly, for the observable based on trading data, we simply re-sample the trades at the minute level so that the trades that happened between 00:00:00.000 and 00:00:59.999 would represent the order imbalance at time $\tau$=00:01:00.000. 

For observables based on limit order books, we assume that the snapshot is \emph{representative of a limit order book that would be observed at the zeroth second of the minute}, namely, at precisely HH:mm:00.000 for every hour HH and minute mm. In \figref{fig_hist_lob} we show the histogram of the exact snapshot seconds within a minute of each limit order book. It is clear that the majority of the limit order books were captured at first halves of a minute. Since we chose to use information dynamics that assumes discrete time series, the limit order book snapshot times are first floored to the closest minutes, after which spread and price returns observables are extracted.

For the example snapshots we consider here $t_1$ and $t_2$ would be rounded to the closes smaller minutes: $\tau_1=\lfloor t_1 \rfloor = $00:00:00.000, $\tau_2=\lfloor t_2\rfloor =$00:01:00.000 from where we obtain $s_{\tau_1},s_{\tau_2}$ and $r_{\tau_2}$. 

\begin{figure}
    \centering
    \includegraphics[width=\linewidth]{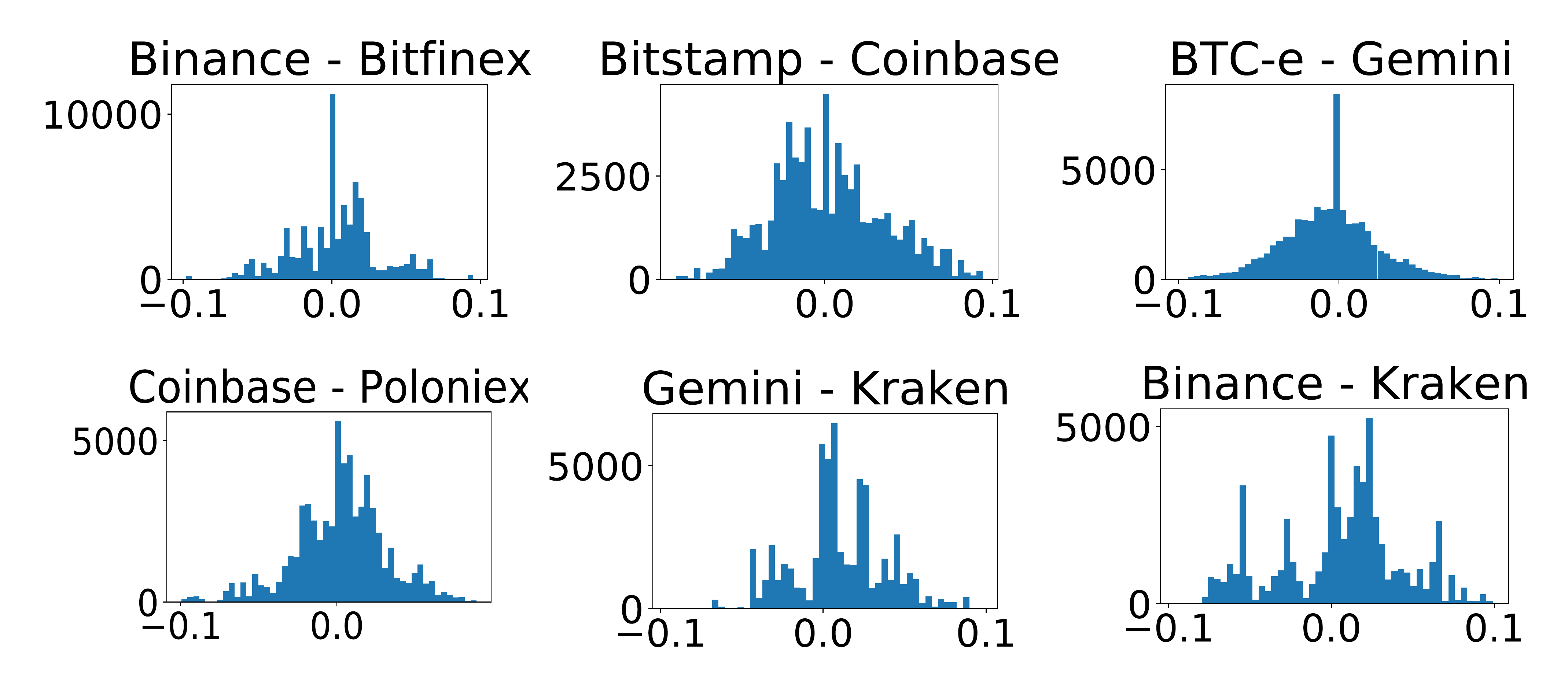}
    \caption{Differences between capture seconds at the same minute for a pair of markets. The title of each subfigure indicates $\alpha-\beta$ for the histogram that shows $t^{\alpha}-t^{\beta}$. The $x$-axis is in units of $1$min.}
    \label{fig_hists_lob}
\end{figure}

We note that the problems would occur only when there is a systematic delay between the exact snapshot times for a given minute in a pair markets. If for some pair of $\alpha,\beta$, $t^{\alpha}<t^{\beta}$ in the majority of minutes, one may observe a spurious link $\beta\rightarrow\alpha$. To test whether this could be the case, we computed the differences between the exact snapshot times at each minute between pairs of markets. We report several observed distributions and note that the distributions do not appear skewed and the mean of the distributions is close to 0, see \figref{fig_hists_lob}.

We could also observe spurious information transfer if mapping to the closest minute is not done systematically. For instance, information transfer from BTC-e (whose first snapshots are taken at $t_1,t_2$ and are later rounded to $\tau_1$, $\tau_2$) to Binance (whose first snapshots are taken at $t_1'<t_1,t_2'<t_2$ and are later rounded to $\tau_1$, $\tau_2$). One may think that TE in the direction from Binance to BTC-e would be spurious because the market representation of Binance at either $\tau$ relates to an earlier true date $t$. However, in estimating transfer entropy, we measure the amount of reduction of uncertainty about the state of some variable $\alpha$ at time $\tau_2$, given the state of another variable $\beta$ at an earlier time $\tau_1$, over and above the information about $\alpha$ at time $\tau_2$ that is already available from its own past at time $\tau_1$. Therefore, issues would arise only if we mapped, e.g., $t_1$ to $\tau_1$ and $t_1'$ to $\tau_2$ but not if we use flooring operator.
Note that this type of rounding, overall, ensures that not acausal, spurious links can occur when the system is probed at a frequency of at least $1$min. Therefore our results are limited to measuring information dynamics at the inter-minute granularity.

For the \textbf{second part of the analysis, presented in \secref{sec_info_flow_obsr}}, we consider information transfer between different observables. Therefore we need to ensure that we do not create acausal links. To compare the information flow from spread at $t_2$ to order imbalance within that minute, we must only consider the trades that occurred in the interval $(t_1,t_2]$, ((00:00:03.021,00:01:02.798] in the example) when calculating $\mathcal{O}$. Similarly, the price returns at $t_2$ must be the price difference between $t_2$ and $t_1$. Therefore we consider the time index of order books, where the frequency is approximately $1$min, and aggregate trading data based on that. For results, shown in \figref{tab_triad} where we consider lower frequency data, we take the first data point within each interval. 

\section{Daily order imbalance}\label{a_imbalance}

\tabref{tab:imbalance} reports the average of the daily sum of realised order imbalance, see Eq. \eqref{e_order_imbalance}, expressed in either quote or base currency. The table shows that before the crash date December $17^{\textrm{th}}$, 2017, the order imbalance was positive and significantly larger than after the crash date, when it was negative in all markets, except for Coinbase and Binance.
\begin{table}[h]
    \centering
    \begin{tabular}{c|c|c|c|c|p{20pt}|p{20pt}}
    & \multicolumn{2}{c}{before 17th Dec.}  &\multicolumn{2}{c}{after 17th Dec. }\\
    \hline\hline
      Exchange &  $\langle \mathcal{O}_{\$}\rangle / 10^5$ &$\langle \mathcal{O}\rangle / 10^3$&$\langle \mathcal{O}_{\$}\rangle / 10^5$ &$\langle \mathcal{O}\rangle / 10^3$ & sig. KS $\mathcal{O}_{\$}$ &sig. KS $\mathcal{O}$  \\\hline
Kraken & -263.08 & -0.06 & -3242.41 & -0.24 & ** &** \\
Bitstamp & 11106.43 & 0.9 & -2089.27 & -0.22 & ** &** \\
BTC-e & 1153.37 & 0.13 & -623.19 & -0.04 & ** &** \\
Bitfinex & 3922.05 & 0.13 & -21530.25 & -1.74 & ** &** \\
Gemini & 5227.95 & 0.42 & -252.24 & -0.04 & ** &** \\
Poloniex & 426.94 & -0.0 & -3735.7 & -0.29 & ** &** \\
Bittrex & 1371.6 & 0.14 & -1798.24 & -0.14 & ** &** \\
Coinbase & 30597.96 & 2.77 & 12309.85 & 0.85 & ** &** \\
Binance & 3221.5 & 0.22 & 2046.26 & 0.06 & ** &** \\
HitBTC & 89.72 & 0.0 & -599.99 & -0.04 & ** &** \\
\hline \hline 
    \end{tabular}
    \caption{Average order flow imbalance before and after Dec 17. The last two columns report $p$-value of two-sided Kolmogorov-Smirnov test, with the null hypothesis that 2 independent samples are drawn from the same continuous distribution. If p-value is high, then we cannot reject the hypothesis. For the test we considered $\langle \mathcal{O}\rangle $ before and after $17$th of December, 2017, and similarly for trade imbalance expressed in quote currency. $**$ indicates that $p$-value is below $0.01$ and the hypothesis is rejected.}
    \label{tab:imbalance}
  \end{table}

\section{Choice of $K$ in K-nearest neighbours}\label{ksg}

At the heart of transfer entropy is estimation of differential conditional mutual information, which in turn relies on the quality of probability distributions of our variables. Mutual information involves averages of logarithms of $P$, the underlying probability distribution. Since, for small $P$, $\log P\rightarrow \infty$, the ranges of values for our variables $X$,$Y$ where $P$ is small and hence cannot be sampled and estimated reliably from data contribute disproportionately to the value of information~\cite{HN19}. 

Information estimators that use continuity of real valued data to overcome issues related to undersampling, have proved to be the most successful for a wide range of datases and applications. Amongst those, one of the most successful such estimators is the Kraskov, St\"ogbauer, and Grassberger~\cite{KSG04}, which we will refer to as KSG. The KSG estimates information transfer based on the statistics of distances between neighbouring data points. To implement
this, KSG uses the $\max(\Delta x, \Delta y,\Delta z)$ metric to define the distance
between two points that are $(\Delta x, \Delta y,\Delta z)$ away from each other in the joint space
$\{x,y, z\}$. For each given test point, $n_z$ is the neighbour count strictly within $\varepsilon$ in the $z$ marginal space, and $n_{xz}$ and $n_{yz}$ are the neighbour counts strictly within $\varepsilon$
in the joint $\{x, z\}$ and $\{y, z\}$ spaces, respectively. Conditional mutual information is then defined in terms of these variables~\cite{BHL16}:
\begin{equation}
I(X , Y | Z) = \psi(K)-E\left[\psi(n_{xz})-\psi(n_{yz})+\psi(n_z)\right],
\end{equation}
where $\psi$ is a digamma function and averaging is over the samples. 

As pointed out in~\cite{HN19}, for any information estimator, it is essential to ensure that such estimators (there a KSG estimator for MI rather than conditional MI was considered) is minimally biased towards the sample size, and the choice of $K$ is optimal: that is, for independent sub-samples of the dataset the variance of obtained result is minimal. 

\begin{figure}[h]
    \centering
    \includegraphics[width=0.95\linewidth]{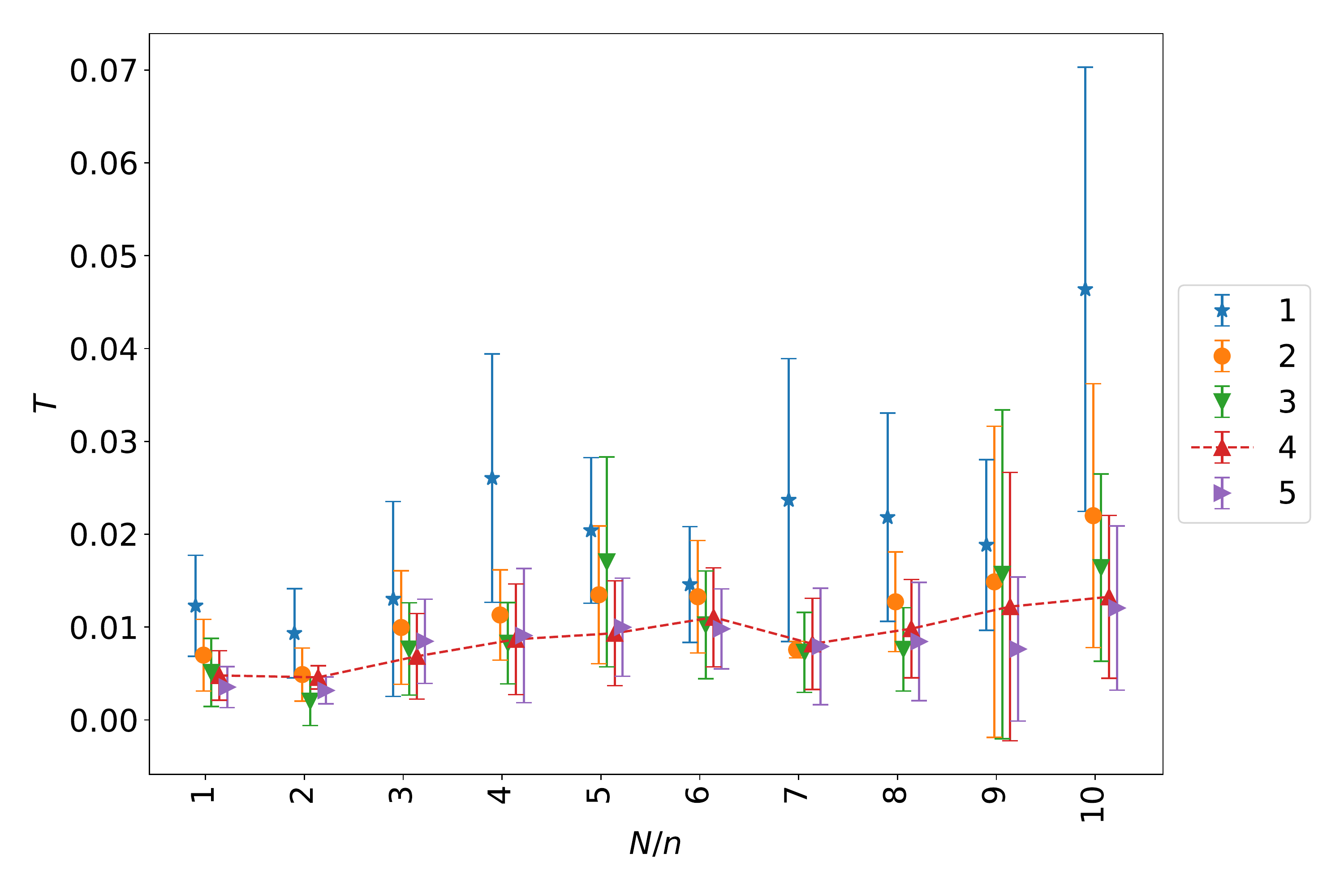}
    \caption{Bias and variance of transfer entropy for subsamples of size $n$ for spread time series from Bitstamp to HitBTC where we identified a significant information transfer of $0.0052$ nats using $K=4$ nearest neighbours.}
    \label{fig:knn}
\end{figure}

The details of this approach for the choice of $K$ are thoroughly described in~\cite{HN19}. Here we simply note that we tested several choices of nearest neighbours for a pair of sample time series for which we identified significant transfer entropy with estimator using $K=4$ nearest neighbours. As \figref{fig:knn} shows, this choice of $K$ has a small bias for the sample size: even if we slice the time series in 4-5 parts, the value of TE does not increase significantly (specify). Furthermore, the variance for values for individual subsamples is also minimal. Note for $K=1$ we have a strong sample-size dependent bias as well as large variance.

\section{Information transfer metrics around regime shifts in auto-regressive time series}\label{app_signatures}

Here we perform a parameter sensitivity analysis for the results, shown in \figref{fig_case_1_main}. Note that we do not consider the parameters $\alpha_1,\alpha_2,\beta_1,\beta_2$ as variables and set them to $\alpha_1=\alpha_2=0.2$ and $\beta_1=\beta_2=1$. In all figures, we also chose $d=0.5$. We note that for a definite conclusion regarding these information dynamical signatures, a thorough analysis of the relation to these parameters should also be conducted.

\begin{figure}[h!]
    \centering
    \includegraphics[width=\linewidth]{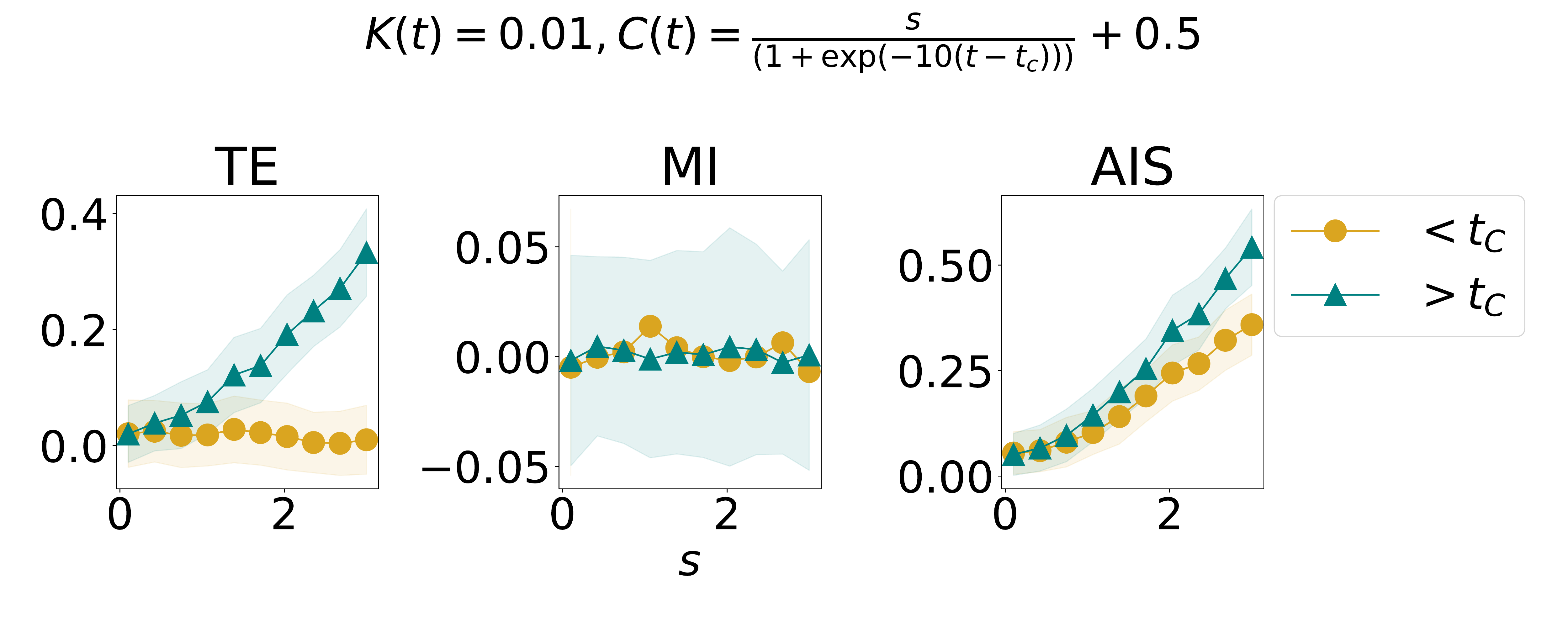}
    \includegraphics[width=\linewidth]{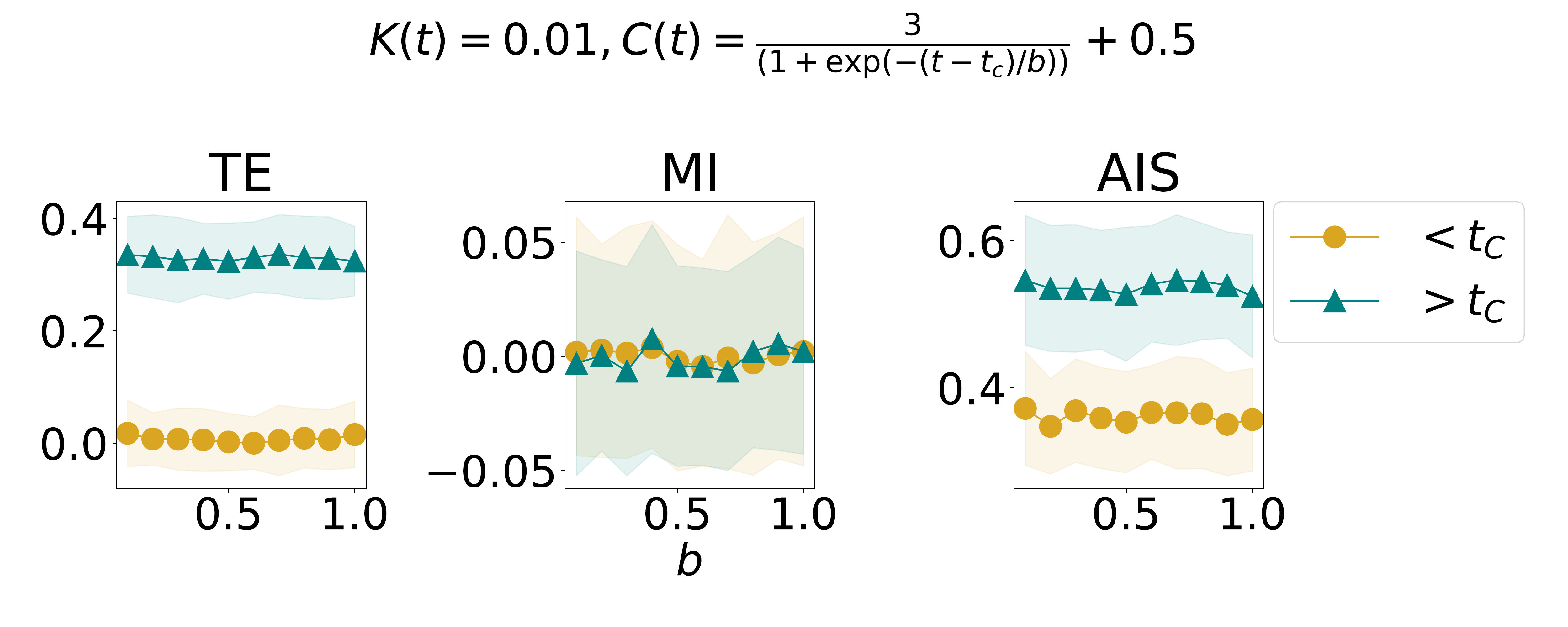}
    \includegraphics[width=\linewidth]{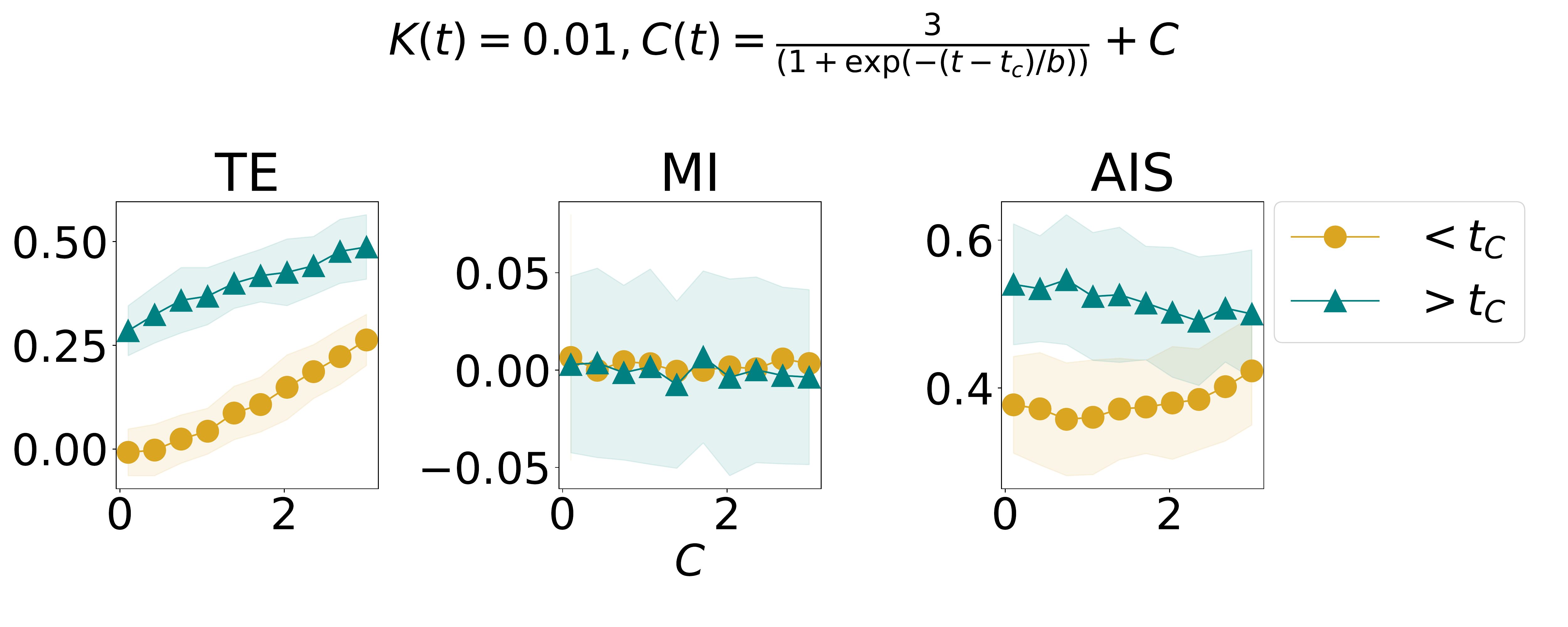}
    \includegraphics[width=\linewidth]{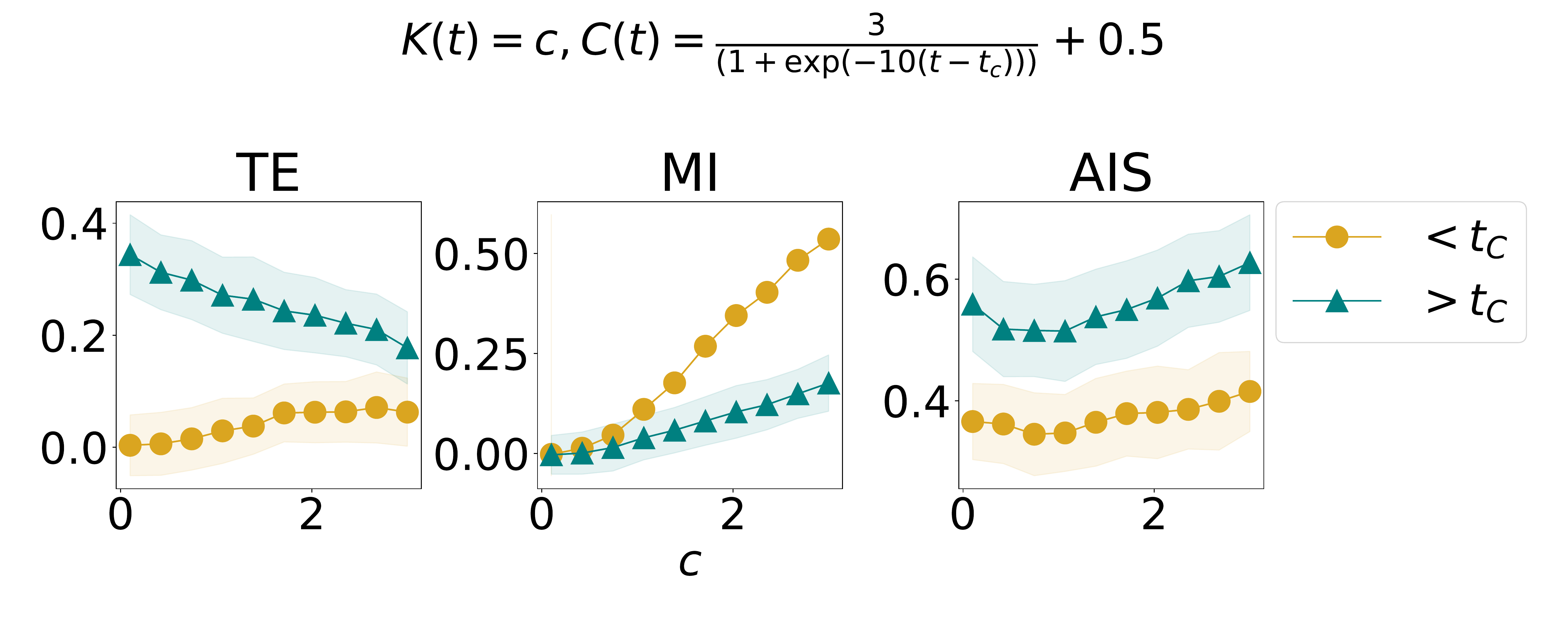}\caption{Expected values of information dynamical variables when $C(t)$ is a sigmoid functional of time, \eqref{eq_sigmoid}, parameterised by $b$, $s$, in the presence of a constant hidden driver of strength $c$. Information dynamic measures are significantly different in the two regimes, as defined by the values of the sigmoid function. }
    \label{causal_driver_varycb}
\end{figure}

In \figref{causal_driver_varycb} we show that a signature of high TE and AIS with no change in MI indicates strong causal coupling, and this statement is robust for a wide range of parameters. It is worth pointing out, however, that at high levels of $K$, a significant amount of MI is also detectable in a weak coupling regime.

The case of a shift in the hidden driver (illustrated in \figref{fig_case_1_main}(b)) is signed with a high MI value in a strong-coupling regime and vice versa in a weak coupling regime. We find that for a wide range of parameter values that characterise the strength and steepness of the regime transition function, as well as the strength of causal coupling, MI is significantly larger in the strong coupling regime. Although for some parameter values AIS is anti-correlated with MI, it is not a robust signifier based on the sensitivity analysis. We also note that, not surprisingly, the strength of the coupling is directly related to the magnitude of MI.

\begin{figure}[h!]
    \centering
    \includegraphics[width=\linewidth]{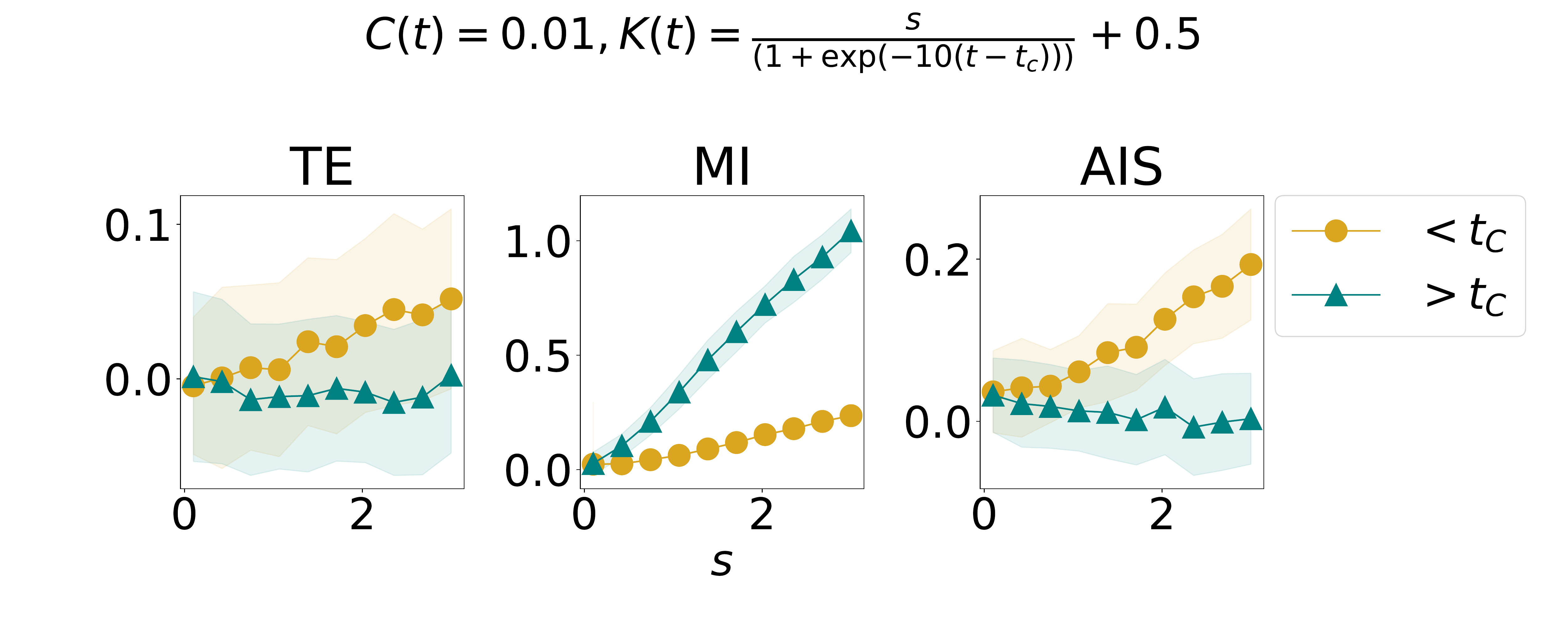}\\
    \includegraphics[width=\linewidth]{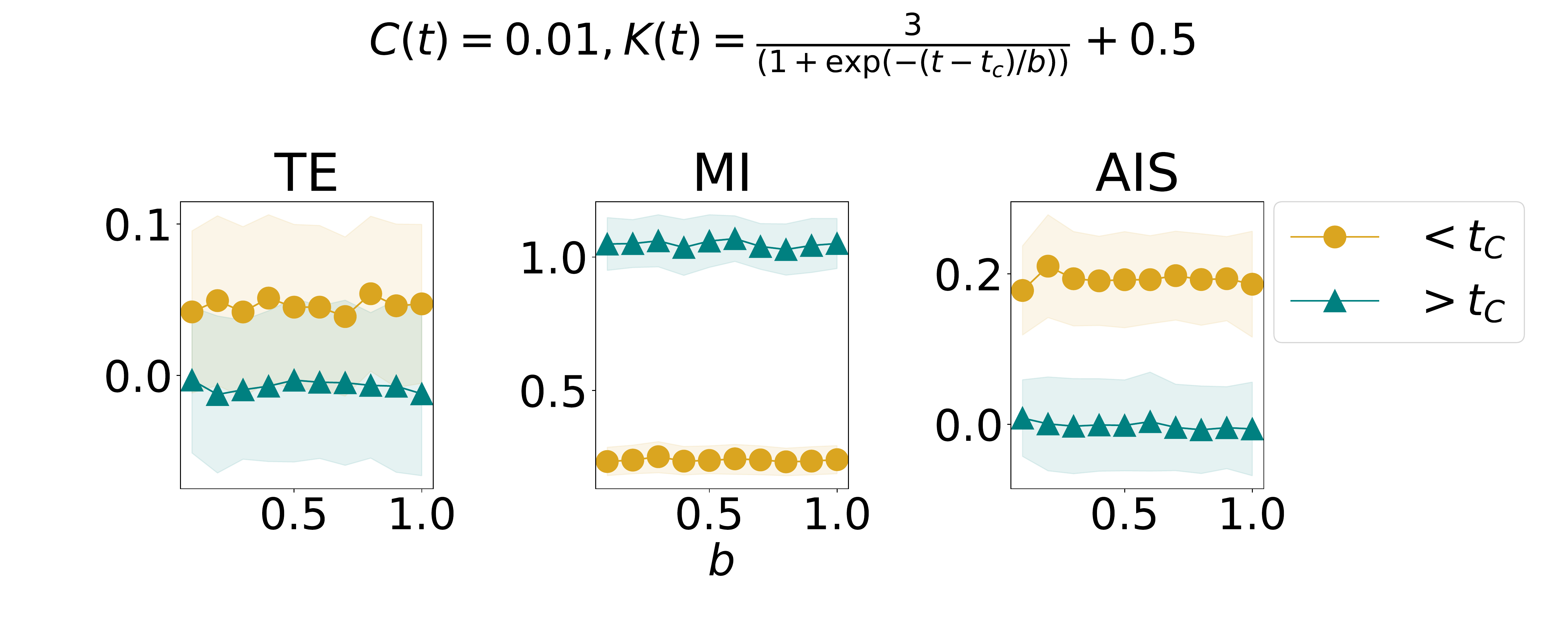}\\
    \includegraphics[width=\linewidth]{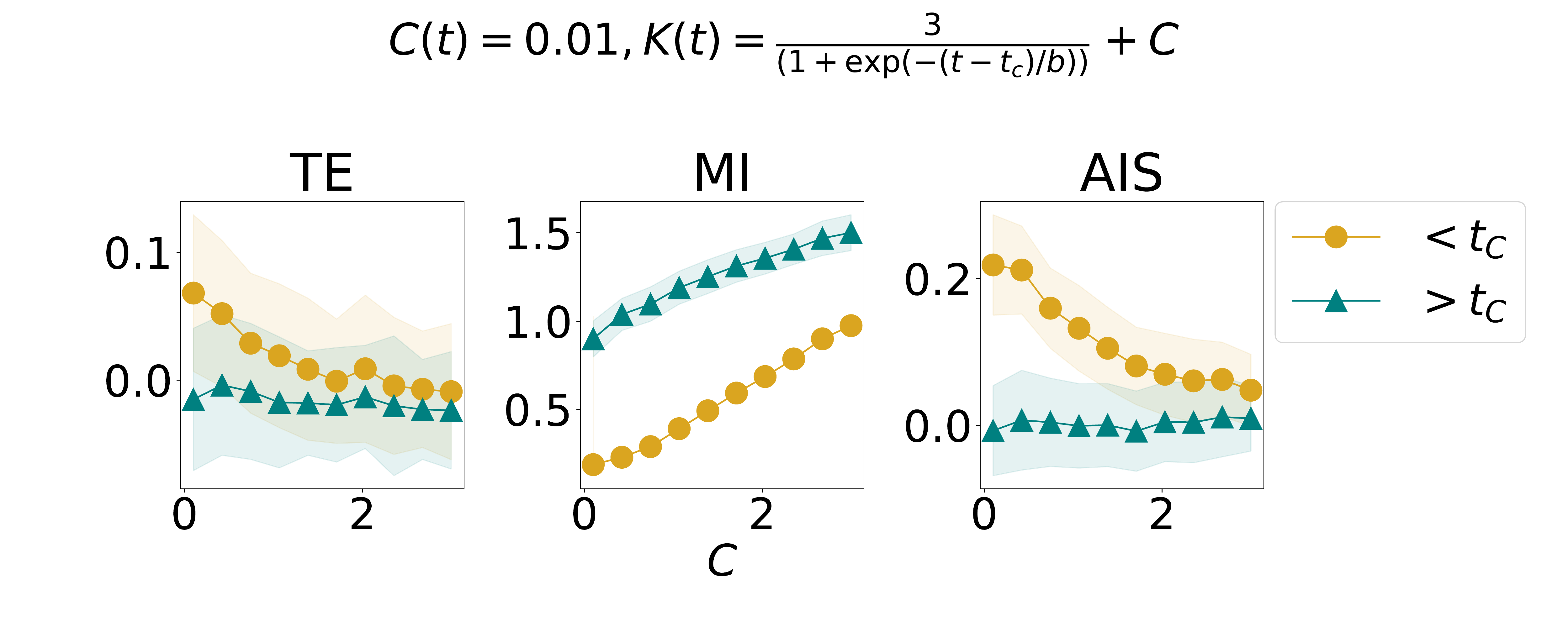}\\
    \includegraphics[width=\linewidth]{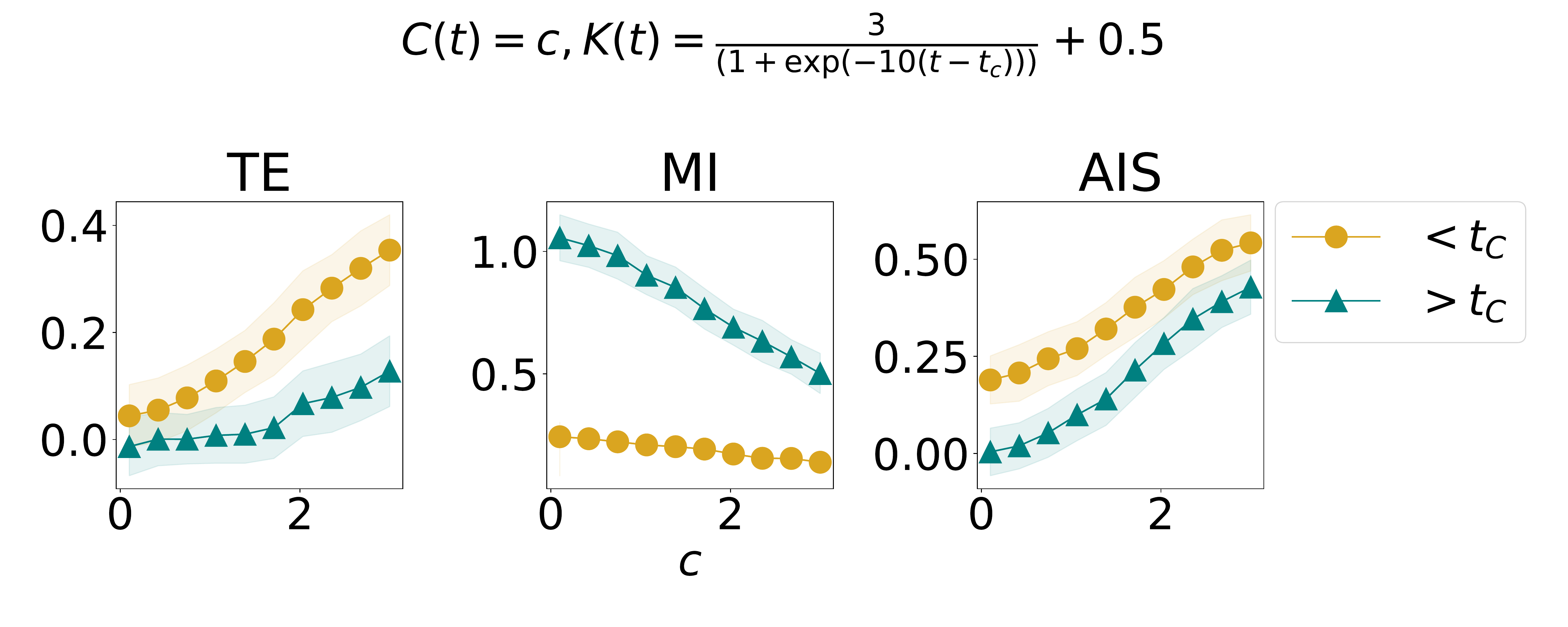}
    \caption{Expected values of information dynamical variables when $K$ is a sigmoid functional of time, parameterised by $s$ and $b$, $c$. Parameter $b$ quantifies the abruptness of the transition, whereas $s$ quantifies the magnitude of change in the two regimes centered at $t_C$. MI is high in the high-coupling regime, and the reverse is true in the low-coupling regime.}
    \label{hidden_driver_sensitivity}
\end{figure}

\begin{figure}[h!]
    \centering
    \includegraphics[width=\linewidth]{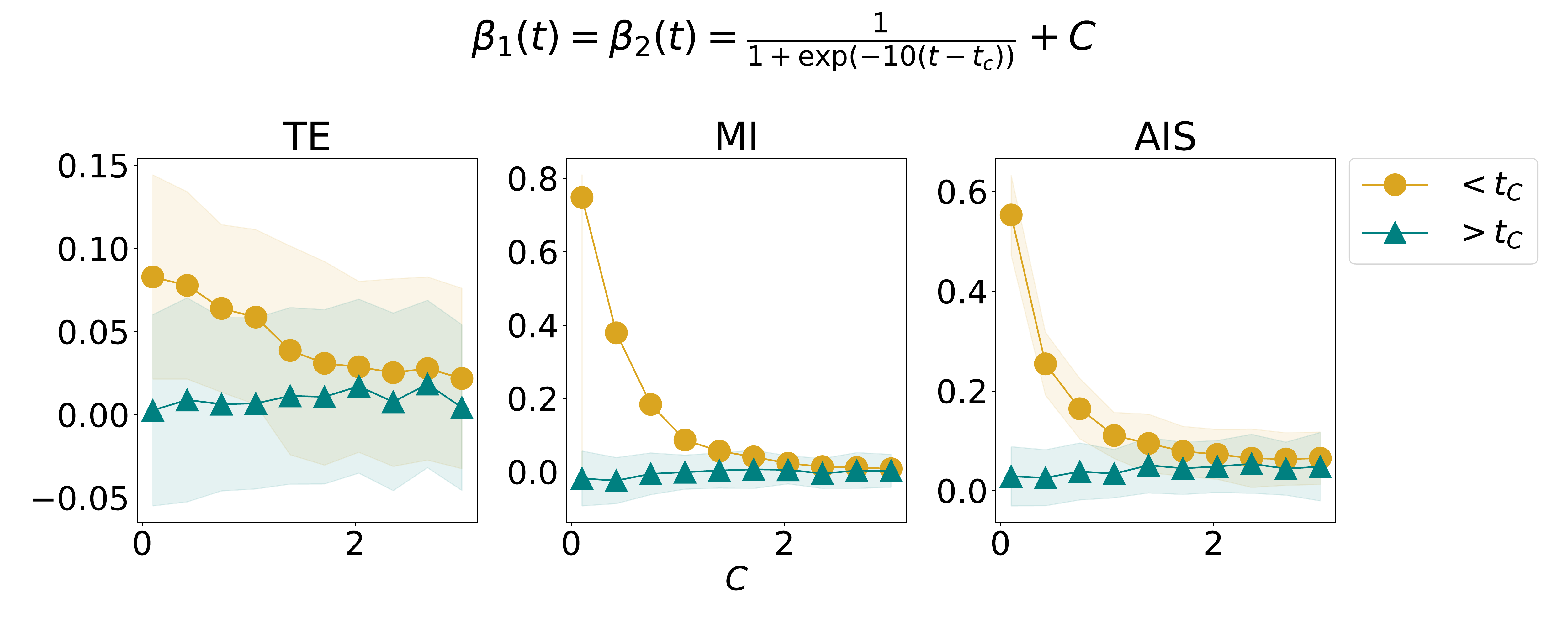}\\
    \includegraphics[width=\linewidth]{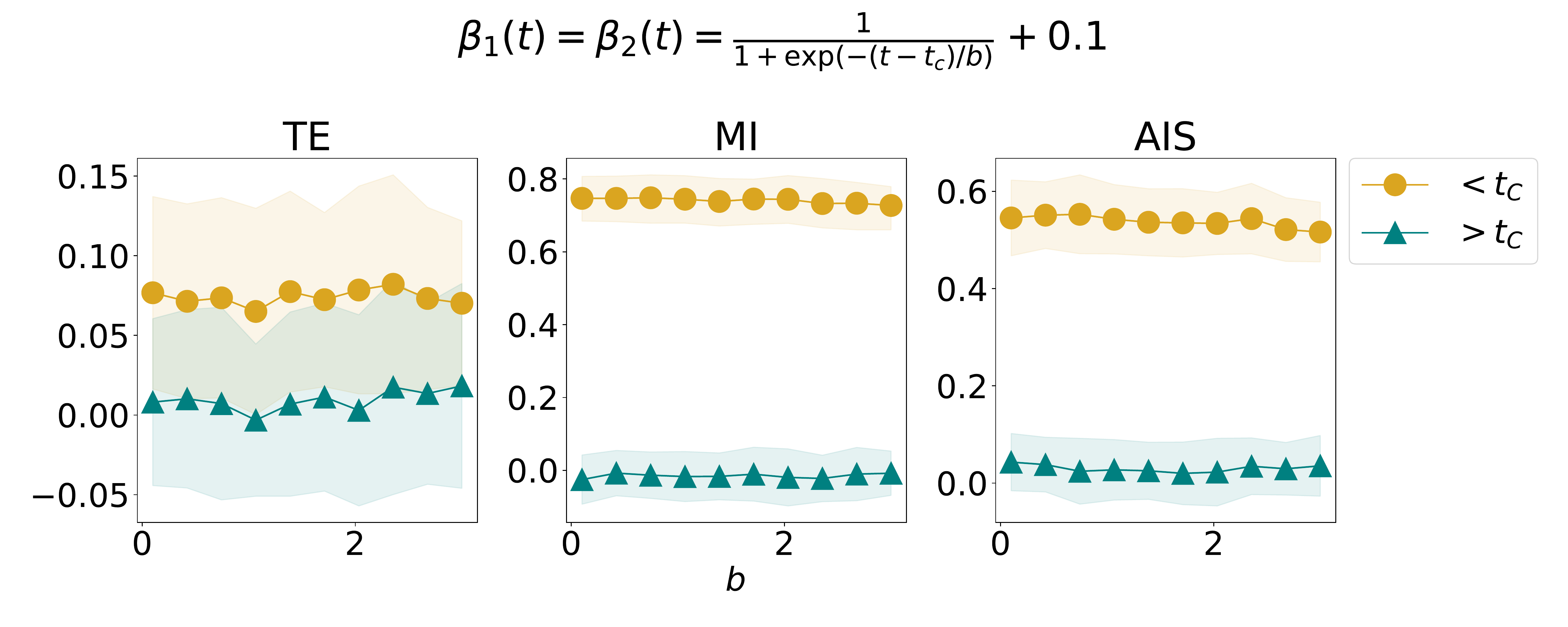}\\
    \includegraphics[width=\linewidth]{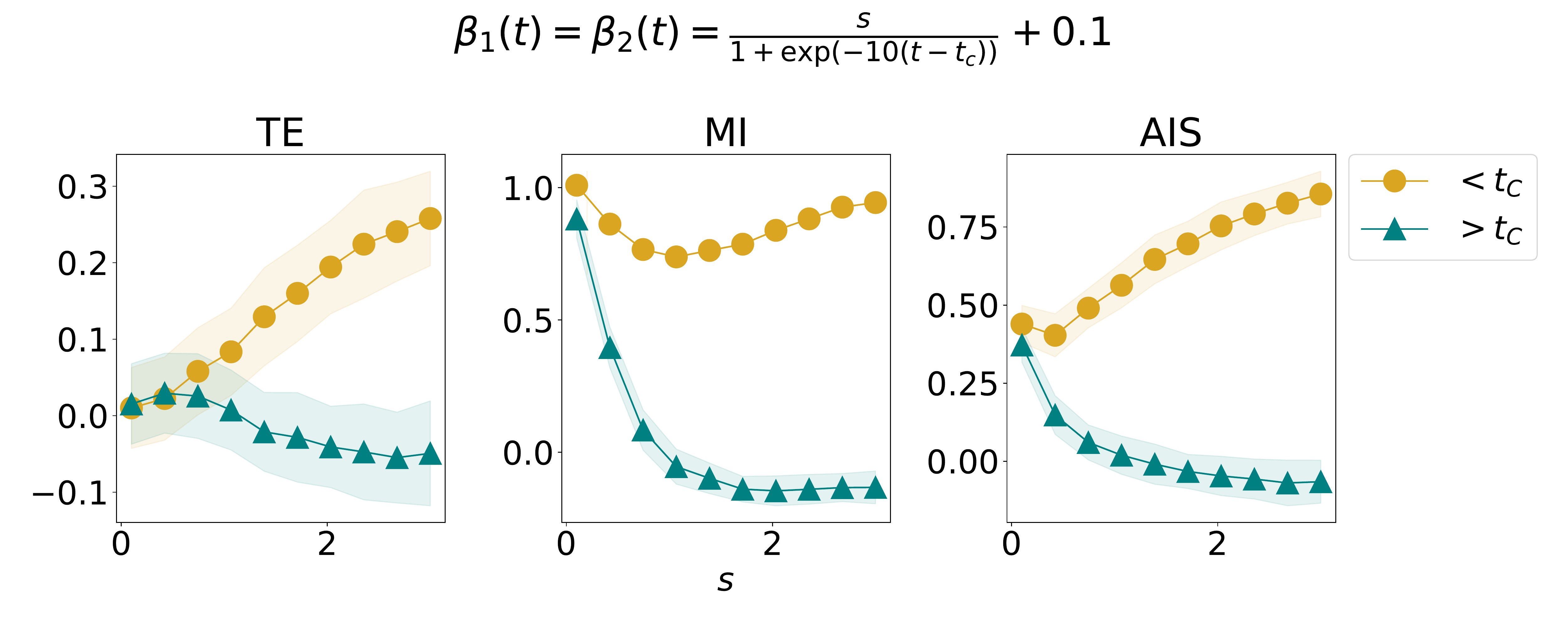}\\
    \includegraphics[width=\linewidth]{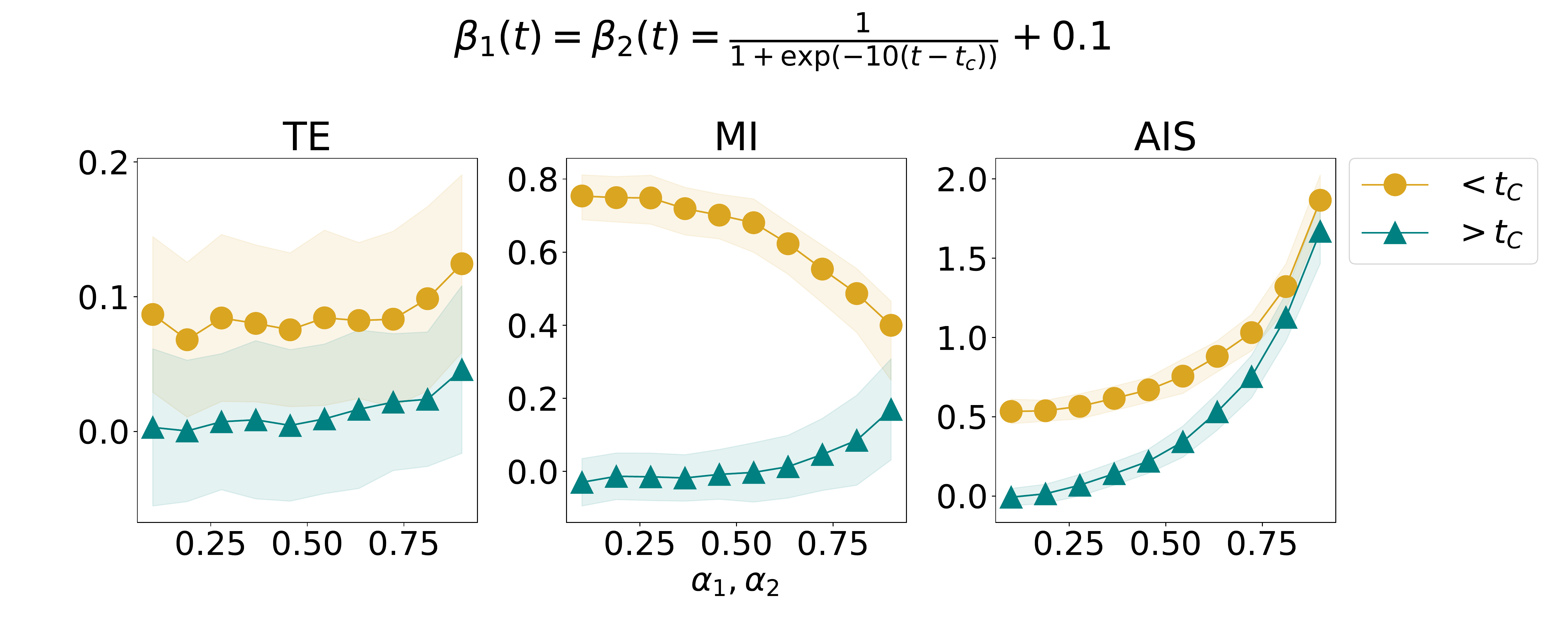}
\caption{Expected values of information dynamical variables when $\beta_1,\beta_2$ are sigmoid functionals of time, parameterised by $s$ and $b$, $c$. Parameter $b$ quantifies the abruptness of the transition, whereas $s$ quantifies the magnitude of change in the two regimes centered at $t_C$. MI, TE, AIS are high in the high-uncertainty regime (high $\beta_1,\beta_2$ regime), and the reverse is true in the low-uncertainty regime. Here we have $C(t)=1=$cost., and $K(t)=1=$cost.}
    \label{AR_driver_sensitivity}
\end{figure}

Lastly, \figref{AR_driver_sensitivity} shows that a change in overall system's uncertainty (the amount of uncorrelated random noise) is detectable with a signature of having low TE, MI and AIS in a high uncertainty regime, and vice versa in a low uncertainty regime. 

\section{Modelling the non-linear relationship between price and spread}\label{app_toy_example}

\begin{figure}[ht]
    \centering
   \hspace{0.5cm} (1)$d=1$\hspace{2cm}  (2)$d=2$ \\\includegraphics[width = 0.4\linewidth]{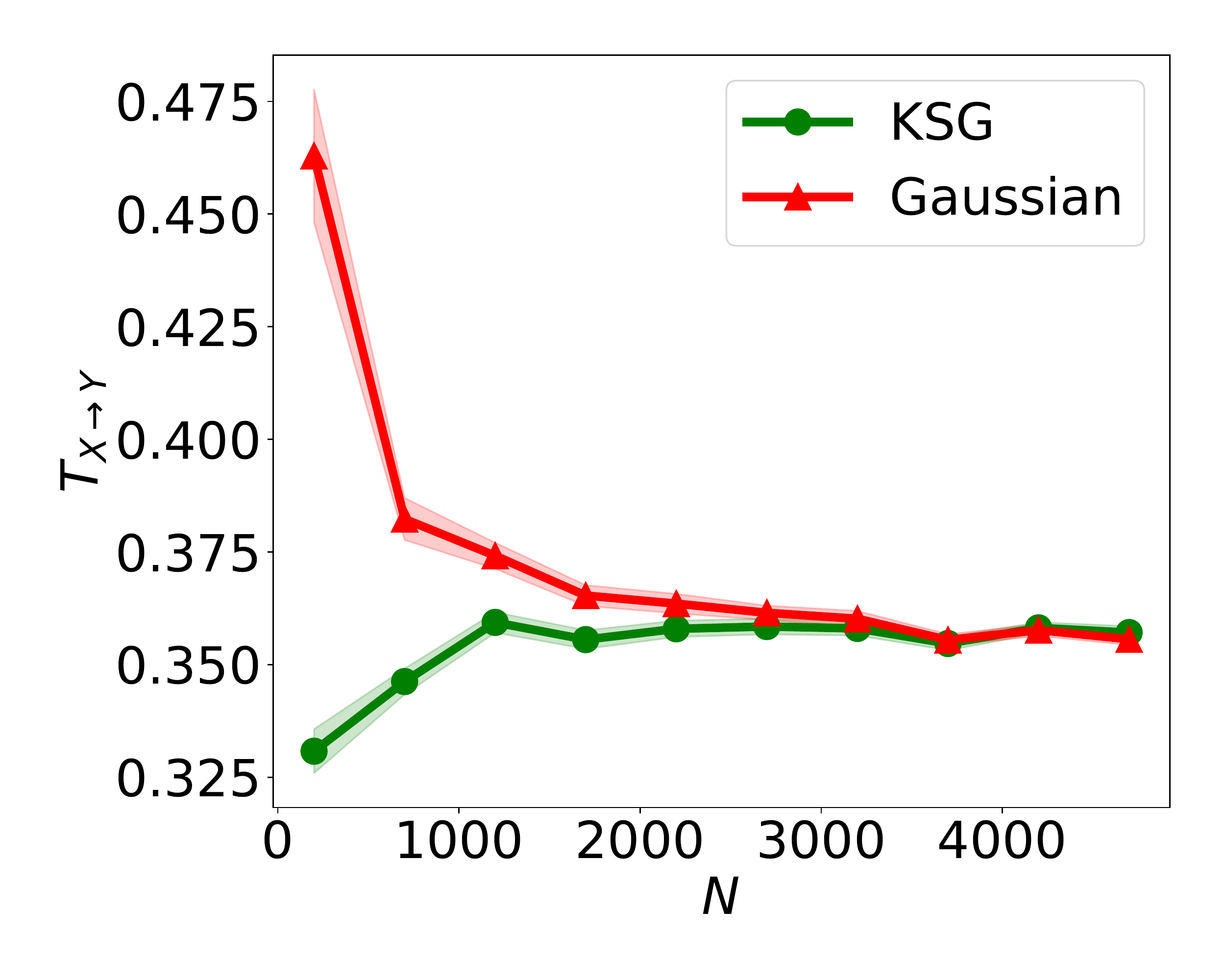}
    \includegraphics[width = 0.4\linewidth]{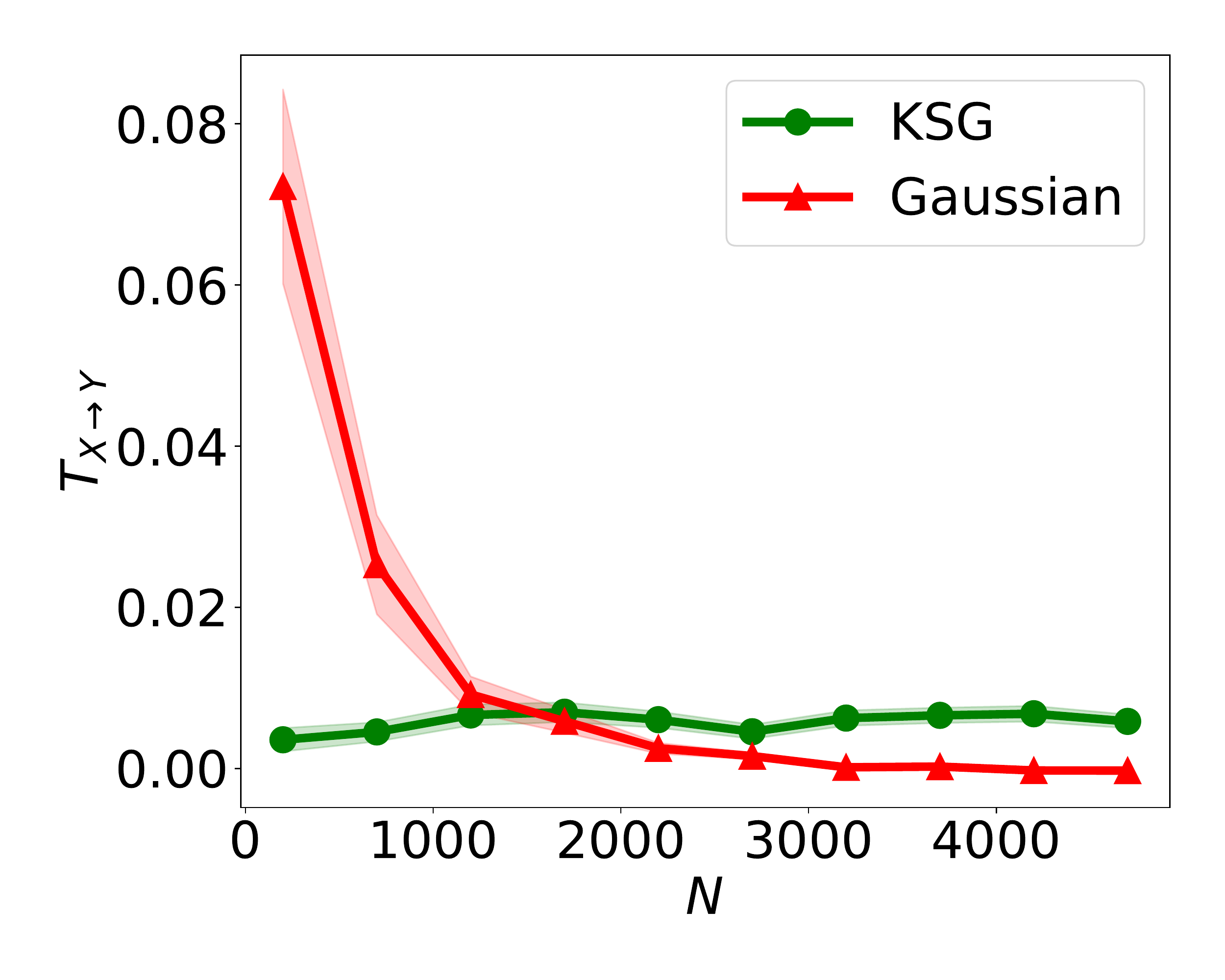}\\
    \caption{(1) $d=1$ i.e., linear relationship between $X,Y$, (2) $d=2$, i.e., non-linear relationship between $X,Y$.}
    \label{fig_ar_gc_te}
\end{figure}

Consider VAR toy example
\begin{eqnarray}\label{eq_ar_def}
X_{t} &=& \alpha_1 X_{t-1} +\beta_1 \varepsilon_{1,t} +K(t)\varepsilon_{t} \\
Y_{t} &=& \alpha_2 Y_{t-1} +\beta_2 \varepsilon_{2,t} +K(t)\varepsilon_{t} +C(t)\left(X_{t-1}\right)^d,\nonumber 
\end{eqnarray}
with $d=1$ and $d=2$. For simplicity, we set $\alpha_1,\alpha_2=0$, $\beta_1,\beta_2=0.1$, $C(t)=1$ and $K(t)=0$. In \figref{fig_ar_gc_te} we show that when the non-linear relation between the variables is present, KSG estimator can detect a large amount of information transfer. In case of linear relation we show a convergence between two measures in the large $N$ limit, while a non-linear relationship is only detectable with a KSG estimated TE.

A more realistic example considers possibly bidirectional coupling between price returns $r$ and spread $s$ when spread is coupled to volatility (variance) of price. Therefore we have coupled time series of the form \eqref{eq_garchspread}.

Note that in this formulation, a link from price to spread is indirect, and in fact price returns affect spread at the two-step delay. Therefore in the following figures we will scan through $\delta\in[1,2]$ and choose the delay for which the observed TE is larger. Depending on parameter values $b,\gamma$ we can have bi- or uni-directional couplings $T_{s\rightarrow r}$ or $T_{r\rightarrow s}$. Each datapoint in the proceeding results is considered at a significance level of $0.01$, and results are obtained from $100$ independent simulations.
 
Parameters $\alpha,\beta,a,b,c ,w,\gamma$ affect stationarity and damping in the system therefore we need to choose such parameter values that system would be stationary throughout. To find a stationary combination of the parameters, we need to consider the moments of the two processes.

First, to compute the unconditional mean values $\sigma^2$ and $s^2$ of volatility and spread, let us consider
\begin{eqnarray*}
\sigma^2&=&Var[r_t]=E[r_t^2]=E[E[r^2_t|{\mathcal F}_{t-1}]]\\
&=&E[w+\alpha r^2_{t-1}+\beta \sigma^2_{t-1}+\gamma s^2_{t-1}]\\
&=&w+(\alpha+\beta) \sigma^2+\gamma s^2
\end{eqnarray*}
and similarly
\begin{eqnarray*}
s^2=E[s_t^2]&=&E[as^2_{t-1}+b\sigma^2_{t-1}+c\epsilon_t^2]\\
&=&as^2+b\sigma^2+c
\end{eqnarray*}

Solving the system we obtain
\begin{eqnarray}\label{eq:meanvalues}
\sigma^2&=&\frac{w(1-a)+\gamma c}{(1-\alpha-\beta)(1-a)-\gamma b}\\
s^2&=&\frac{b\sigma^2+c}{1-a}
\end{eqnarray}
From the first equation we see that unconditional volatility increases with $\gamma$, $c$, and $b$, i.e. there is a contribution of illiquidity to volatility. From the second equation, for large volatility ($\sigma^2\gg b/c$) we obtain the well known proportionality $s\propto \sigma$ between liquidity and volatility. The stationary condition is obtained by imposing that these two quantities are positive, hence, since $a<1$, it must be 
\begin{equation}
(1-\alpha-\beta)(1-a)-\gamma b>0.
\end{equation}

Volatility can be eliminated from the equations \eqref{eq_garchspread} and by recursive substitution we get
\begin{eqnarray*}
\sigma^2_t&=&\frac{w}{1-\beta}+\frac{\alpha}{\beta}\sum_{i=1}^{+\infty}\beta^ir^2_{t-i}+\frac{\gamma}{\beta}\sum_{i=1}^{+\infty}\beta^is^2_{t-i}\\
s_t^2&=&\frac{bw}{1-\beta}+as^2_{t-1}+\frac{b\gamma}{\beta^2}\sum_{i=2}^{+\infty}\beta^is^2_{t-i}+\frac{ba}{\beta^2}\sum_{i=2}^{+\infty}\beta^ir^2_{t-i}+c\epsilon_t^2
\end{eqnarray*}
i.e., the volatility is the sum of an Exponentially Weighted Moving Average (EWMA) of past square returns and of an EWMA of past square spreads and the same holds true for squared spread. Squaring the last two equations in \eqref{eq_garchspread} and taking expectations one gets
\begin{eqnarray*}
E[\sigma^4]=w^2+3\alpha^2E[\sigma^4]+\beta^2E[\sigma^4]+\gamma^2E[s^4]\\
+2w\alpha \sigma^2+2w\beta\sigma^2+2w\gamma s^2\\
+2\alpha\beta E[\sigma^4]+2\alpha\gamma E[\sigma^2s^2]+2\beta\gamma E[\sigma^2s^2]
\end{eqnarray*}
and
\begin{eqnarray*}
E[s^4]=a^2E[s^4]+b^2E[\sigma^4]+3c^2\\
+2abE[\sigma^2s^2]+2acs^2+2bc\sigma^2
\end{eqnarray*}
where we have used the fact that $E[\epsilon_t^4]=E[\varepsilon_t^4]=3$ (because they are Gaussian) and we introduced in the equations the unconditional means $\sigma^2$ and $s^2$ in Eq. (\ref{eq:meanvalues}).

In order to close the system we need an expression for $E[\sigma^2s^2]$. This can be obtained by taking the product of the last two equations in \eqref{eq_garchspread} and taking the expectation, obtaining
\begin{eqnarray*}
E[\sigma^2s^2]=was^2+wb\sigma^2+wc+\alpha a E[\sigma^2s^2]\\
+\alpha b E[\sigma^4]+\alpha c\sigma^2+a\beta E[\sigma^2s^2]+\beta b E[\sigma^4]\\
+\beta c \sigma^2+\gamma a E[s^4]+\gamma b E[\sigma^2s^2]+\gamma c s^2
\end{eqnarray*}
We now have a system of three equations and three unknowns ($E[\sigma^4]$, $E[s^4]$, and $E[\sigma^2s^2$]) whose solution gives the second moments. As usual, these moments will be finite in the parameter region where they are positive. 

The solutions for fourth moments are cumbersome to write in full form. However, we checked that for the parameter combinations that we will study next the expectation of $\sigma^2,s^2,\sigma^4, s^4$ are all well-defined. In particular, we considered the three following cases:
\begin{itemize}
  \item Information transfer from spread to volatility, and, therefore to price returns; no information flow from price returns to spread: $\alpha=0.1,\beta=0.4,a=0.8,b=0.0,c=0.1,\gamma=0.9,\omega=0.1$
  \item Information transfer from volatility, and, therefore, price returns, to spread; no information flow from spread to price returns: $\alpha=0.1,\beta=0.1, a=0.1,b=0.9,c=0.1,\gamma=0.0,\omega=0.1$
  \item Bi-directional coupling from price returns to spread and vice versa: $\alpha=0.1,\beta=0.5, a=0.1,b=0.5,c=0.1,\gamma=0.5,\omega=0.1$.
\end{itemize}

First set of parameter values gives $\sigma^2\approx 0.13,s^2=0.24$,$\sigma^4\approx 0.01+0.04\sigma^2, s^4\approx0.04+0.02s^2+0.24\sigma^2$. All values are positive. For the second set of parameters, we get $\sigma^2\approx 1.1$, $s^2\approx 0.5$, $\sigma^4\approx 0.2+1.4s^2+0.2\sigma^2$, $s^4\approx0.08+0.44s^2$. Lastly,
for the third set of parameters, we find $\sigma^2\approx 1.27$, $s^2\approx0.82$, and $\sigma^4\approx0.2+1.4s^2+0.2\sigma^2$, $s^4\approx 0.08+044s^2$. Therefore all parameter values are suitable.

\begin{figure}[ht]
    \centering Expected link $r\leftrightarrow s$\\
   \includegraphics[width = 0.4\linewidth]{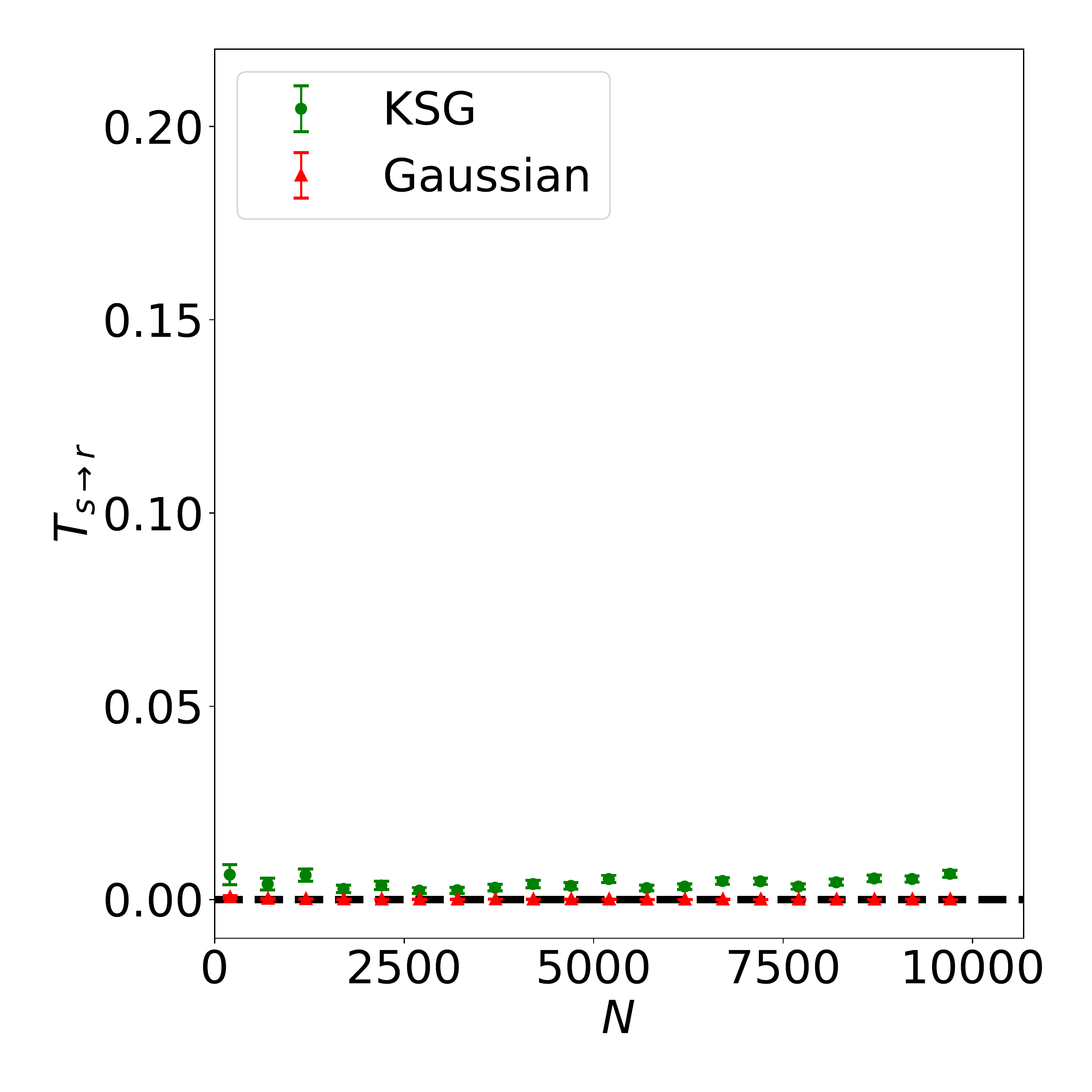}\includegraphics[width = 0.4\linewidth]{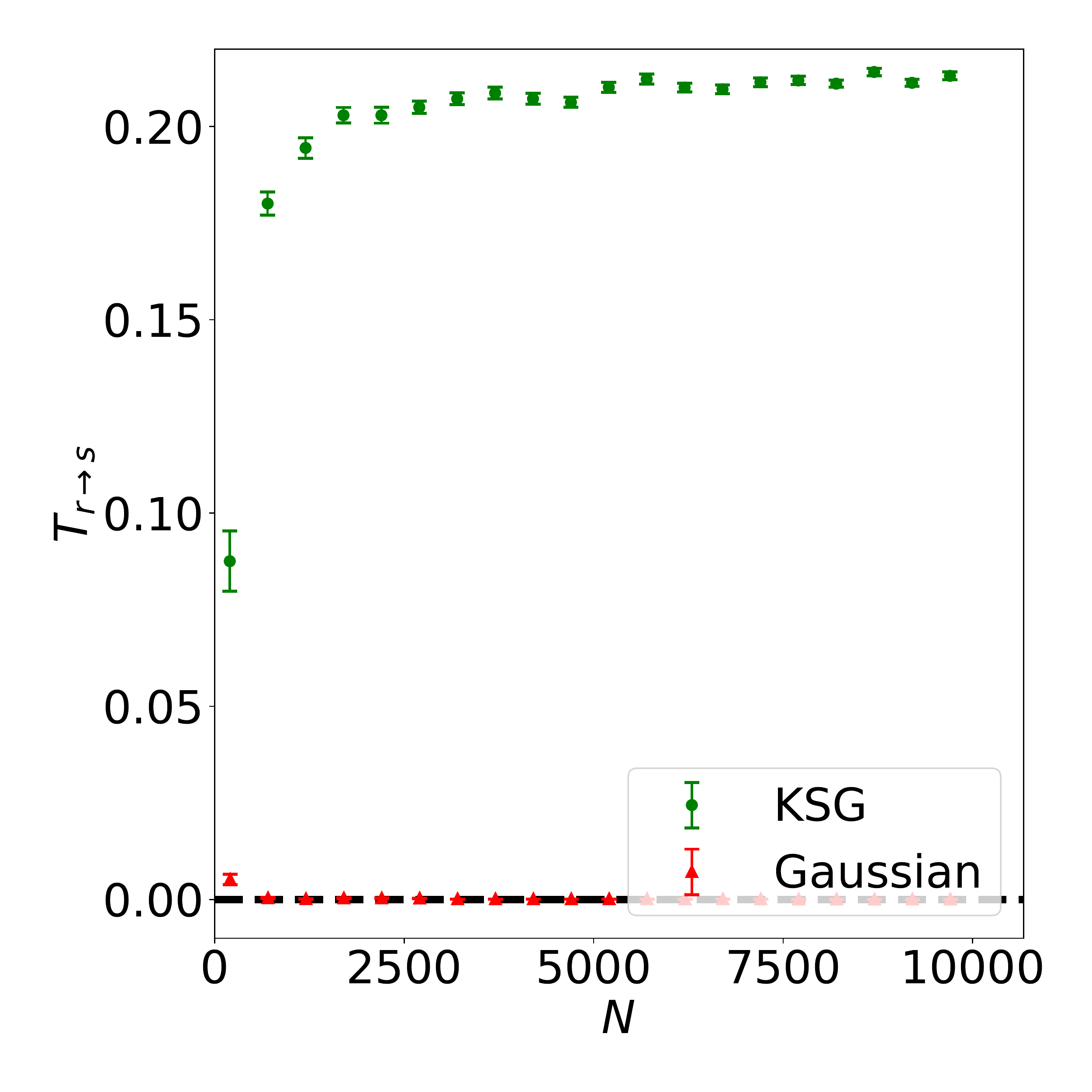}
    \caption{TE from spread to returns (left) and reverse (right), $\alpha=0.1,\beta=0.5, a=0.1,b=0.5,c=0.1,\gamma=0.5$.} 
    \label{fig_returns_spread_bid}
\end{figure}

\subsection{Simulation results} The three cases of coupling between spread and price returns are shown in \figref{fig_spread_to_returns_main}, and \figref{fig_returns_spread_bid}. The first case, shown in \figref{fig_spread_to_returns_main}(top) considers $\gamma=0.9$ and therefore coupling from spread to returns is strong. The figure shows that Gaussian estimate of TE is always close to zero, whereas KSG estimate of TE is clearly larger in the direction from spread to price returns (left figure) as opposed to the reverse (right figure). Note that in some simulations significant non-zero coupling is observed, when Gaussian estimate is used. However, on average, the magnitude of information transfer estimated with Gaussian estimate is smaller in contrast with KSG estimated TE.

The second case, shown in \figref{fig_spread_to_returns_main}(bottom) considers a situation where $\gamma=0.0$, therefore the link from spread to returns should not be observed, whereas parameter $b$, that indicates the strength of a reciprocal coupling is set to a large value $b=0.9$. Here transfer entropy detects a significant flow only from price returns to spread. Similarly to the previous case, Gaussian estimate is also found significant and different from zero in some of the simulations, however the magnitude of TE is smaller than that obtained with KSG estimate.

The last figure, \figref{fig_returns_spread_bid}, considers a case when both $\gamma\neq 0$ and $b\neq 0$. Although the link from spread to returns is much weaker than the reverse, we nevertheless observe significant flow in both directions.

In all cases considered, TE estimates show no apparent size-dependent value drift, and therefore we assume there is no sample size-related bias, as long as $N\geq 1000$.

\subsection{Theoretical values for transfer entropy between price returns and spread}

The mathematical simplicity of our model \eqref{eq_garchspread} allows us to estimate the TE using numerical integration techniques. By comparing the simulation results, obtained using a data-dependent estimator of TE, with the results obtained analytically, we can attest the usability of such estimator for the type of data that could arise from the model considered.

\paragraph{Information transfer from spread to price returns} To proceed, note that transfer entropy from spread to price returns is the difference of two conditional entropy ($H(\cdot|\cdot)$ terms, namely:
\begin{eqnarray*}
T_{s \to r}= H(r_t| r_{t-1}) - H(r_t| r_{t-1}, s_{t-1}).
\end{eqnarray*}
It can also be rewritten in terms of expectation of $ \frac{f(r_t| r_{t-1}, s_{t-1})}{f(r_t| r_{t-1})} $,
where we sample points $\{r_t,r_{t-1},s_{t-1}\}$, $\lbrace r_t,s_t \rbrace$ obtained from \eqref{eq_garchspread}. Using a natural logarithm $\log$ yields a result units of nats, and $f(\cdot)$ denotes probability density function (p.d.f.). 

\begin{figure}
\centering
   a)\includegraphics[width=0.45\linewidth]{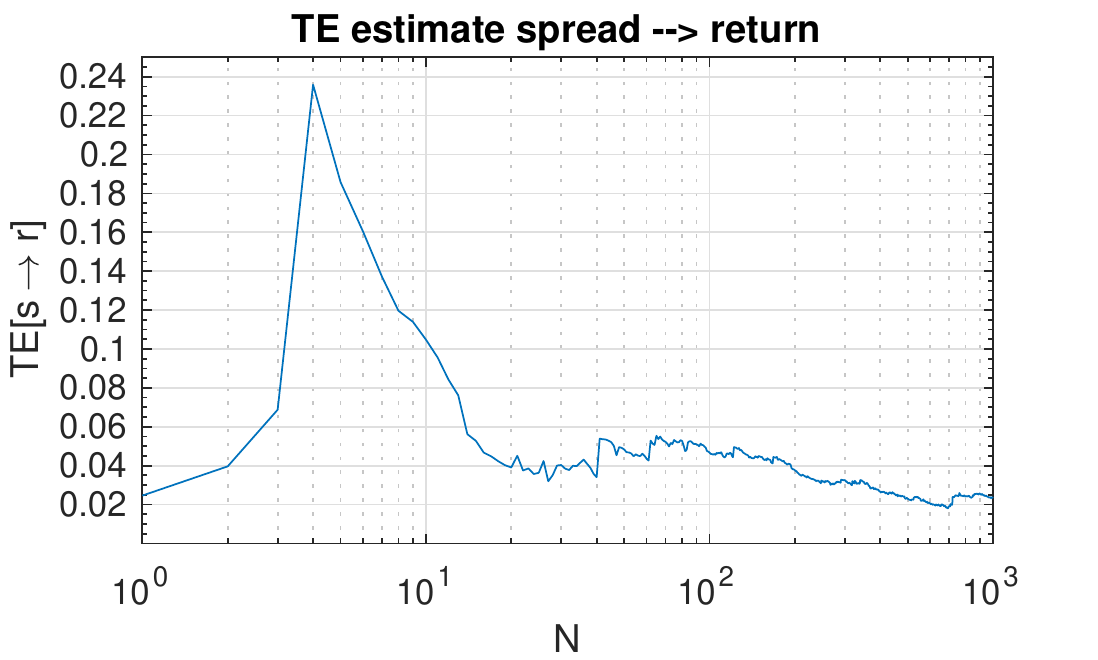}
   b)\includegraphics[width=0.45\linewidth]{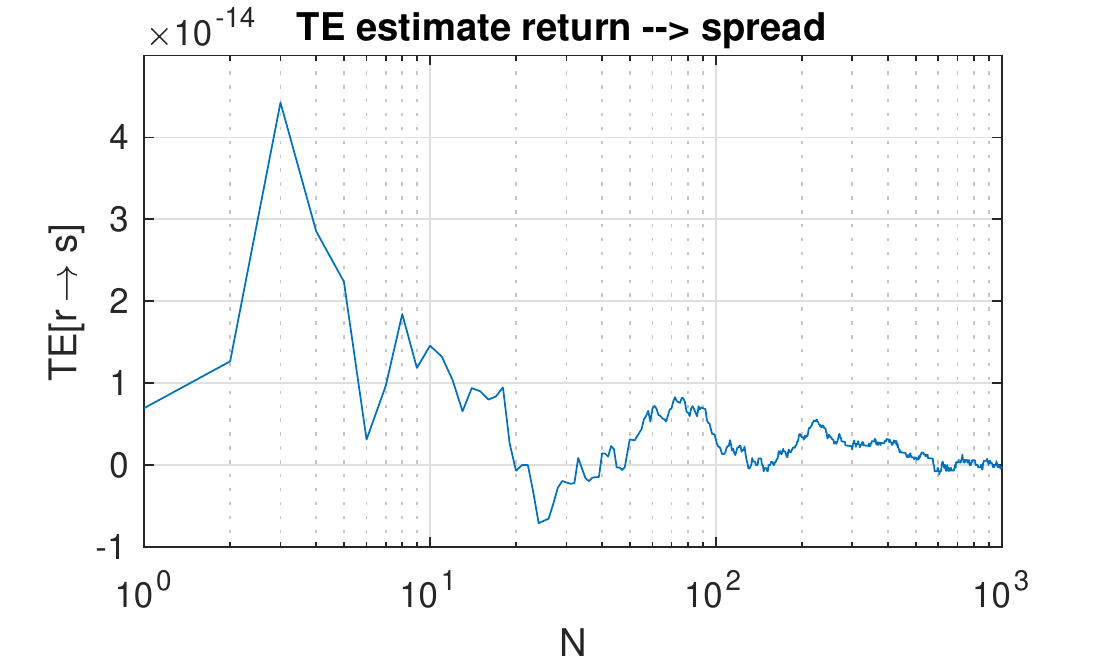}
\caption{Model-based TE from spread to returns (a) and reverse (b), with expected link $s\rightarrow r$. Parameters: $\alpha=0.1,\beta=0.4,a=0.8,b=0.0,c=0.1,\gamma=0.9$. }
\label{model_fig_spread_to_returns}
\end{figure}

First, let us consider the case when $b=0$ (\figref{fig_spread_to_returns_main}) we derive the theoretical distribution densities for model-based transfer entropy from spread to returns. To obtain the p.d.f.\ $f(r_t| r_{t-1}, s_{t-1})$, one can sample values of $f(r_t| r_{t-1}, s_{t-1},\sigma)$ and integrate over the hidden volatility variable:
\begin{eqnarray*}
 f(r_t| r_{t-1}, s_{t-1}) = \int f(\sigma) f(r_t| r_{t-1}, s_{t-1},\sigma) d\sigma. 
\end{eqnarray*}
Here we assume independence of $\sigma$. Further marginalising we obtain the conditional p.d.f.\ of price returns:
\begin{eqnarray*}
f(r_t| r_{t-1}) = \int \int f(\sigma)f(s) f(r_t| r_{t-1}, s,\sigma) d\sigma ds.
\end{eqnarray*}

Finally, in case of our model defined in \eqref{eq_garchspread}, we know the exact probabilistic model:
\begin{eqnarray*}
 f(r_t=x| r_{t-1}, s_{t-1}, \sigma_{t-1}) &=&\\
  \mathcal{N}(x;0, w+\alpha r_{t-1}^2 +\beta \sigma_{t-1}^2+\gamma s_{t-1}^2),
\end{eqnarray*}
where $\mathcal{N}(x;\mu=0,\sigma^2=w+\alpha r_{t-1}^2 +\beta \sigma_{t-1}^2+\gamma s_{t-1}^2))$ denote the Normal probability density function parameterised with $\mu,\sigma$.

Now, compute the quantity $T_{s \to r}$, we perform Monte Carlo sampling:
\begin{eqnarray*}
 f(r_t=x| r_{t-1}, s_{t-1}) = E_{\sigma_i\sim f(\sigma)} [f(r_t=x| r_{t-1}, s_{t-1}, \sigma_{i})]\\
 \approx \frac{1}{N} \sum_{i=1}^N \mathcal{N}(x;0, w+\alpha r_{t-1}^2 +\beta \sigma_{i}^2+\gamma s_{t-1}^2).
\end{eqnarray*}

\begin{eqnarray*}
f(r_t=x| r_{t-1}) = E_{\sigma_i\sim f(\sigma),s_i \sim f(s)} [f(r_t=x| r_{t-1}, s_i, \sigma_{i})]\\
\approx \frac{1}{N} \sum_{i=1}^N \mathcal{N}(x;0, w+\alpha r_{t-1}^2 +\beta \sigma_{i}^2+\gamma s_{i}^2).
\end{eqnarray*}

\begin{figure}
\centering
a)\includegraphics[width=0.45\linewidth]{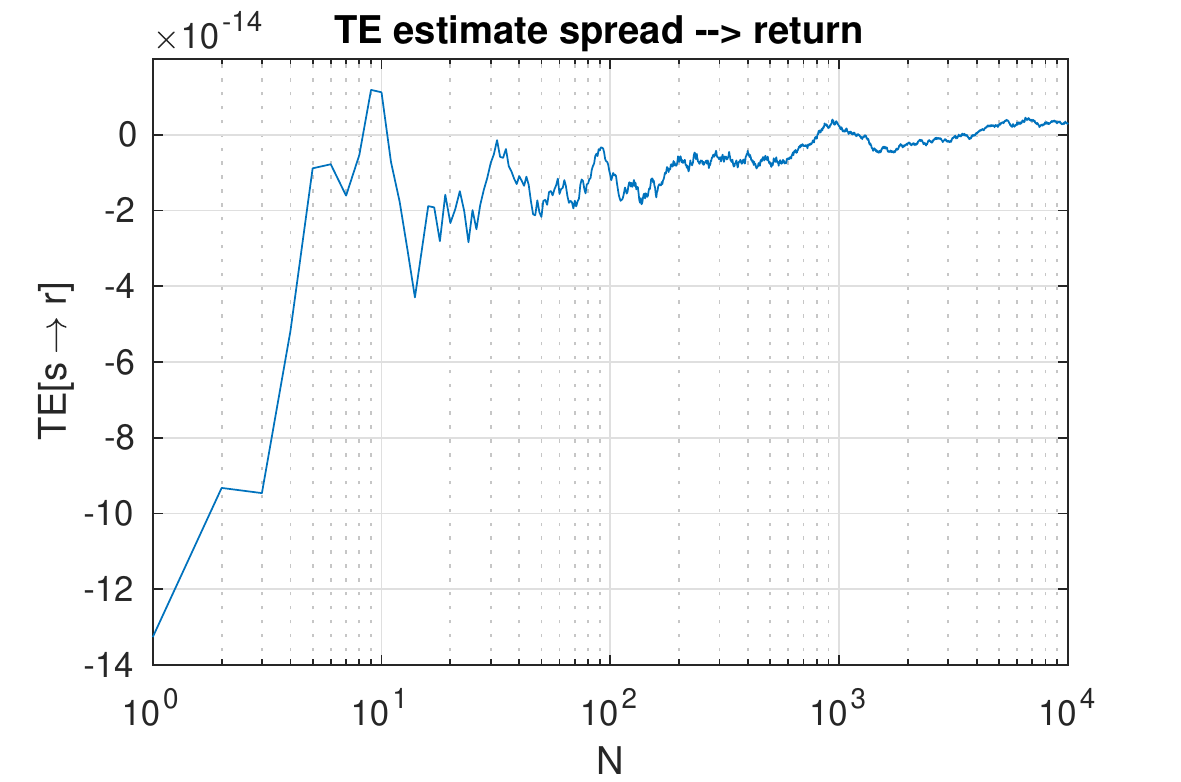}
b)\includegraphics[width=0.45\linewidth]{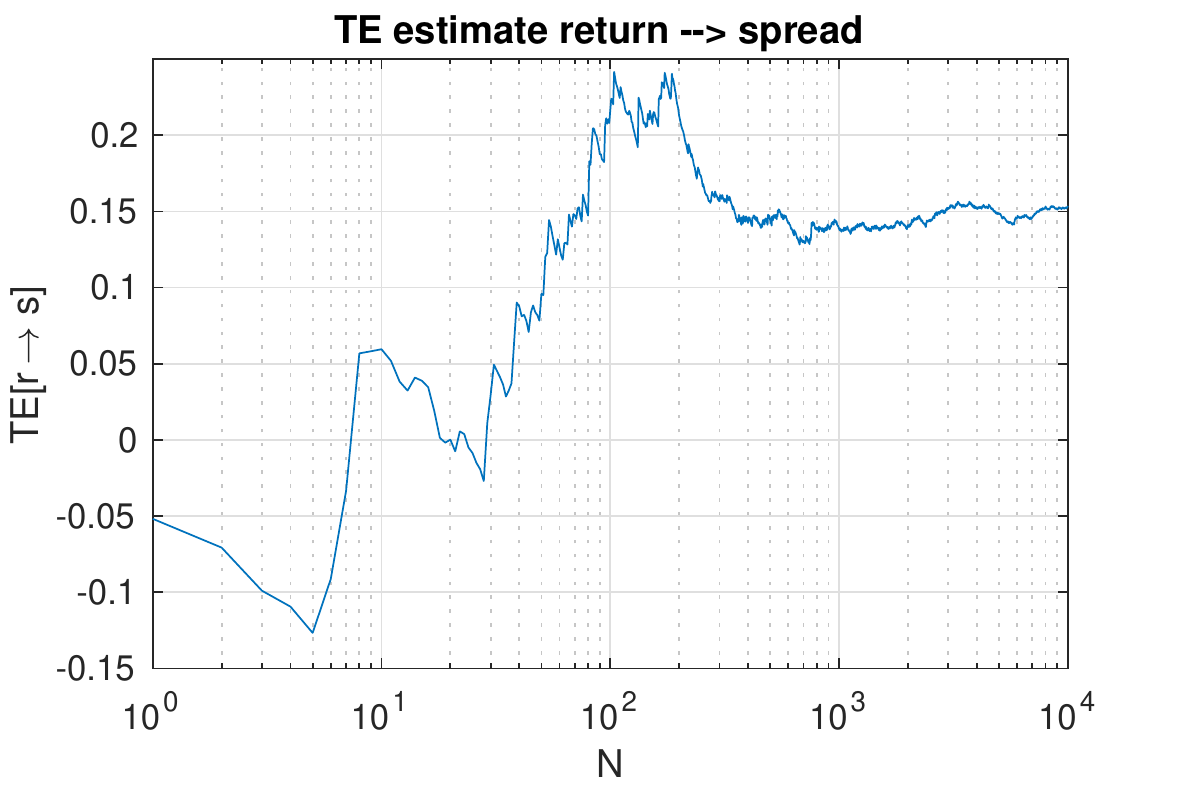}
\caption{Model-based TE from spread to returns (a) and reverse (b) with expected link $r\rightarrow s$. Parameters: $\alpha=0.1,\beta=0.1, a=0.1,b=0.9,c=0.1,\gamma=0.0$. }
\label{model_fig_returns_to_spread}
\end{figure}

On the left panel of \figref{model_fig_spread_to_returns} and right panel of \figref{model_fig_returns_to_spread} we show 
agreement between the transfer entropy obtained from such analytical integration, and the result we found in \figref{fig_spread_to_returns_main} using a K-nearest neighbour data-driven estimator.

\paragraph{Information transfer from price returns to spread} Similarly to the case before, transfer entropy from returns to spread is the difference of two conditional entropies:
\begin{eqnarray*}
T_{r \to s} = H(s_t| s_{t-1}, s_{t-2})] - H(s_t| s_{t-1}, s_{t-2}, r_{t-1}, r_{t-2})],
\end{eqnarray*}
that can also be rewritten in terms of expectation value $ \log \frac{f(s_t| s_{t-1}, s_{t-2}, r_{t-1}, r_{t-2})}{f(s_t| s_{t-1}, s_{t-2})}$.

Here we need to estimate a conditional p.d.f.\ of spread given last two spread values. We do this again via marginalisation of hidden variables:
\begin{eqnarray*}
 f(s_t| s_{t-1}, s_{t-2}) = \nonumber\\ \int\int f(s_t, r_{t-1}, r_{t-2}| s_{t-1}, s_{t-2}) dr_{t-1} dr_{t-2}.
\end{eqnarray*}

The conditional p.d.f. of spread given last two spread and price return values, $f(s_t| s_{t-1}, s_{t-2}, r_{t-1}, r_{t-2})$, is obtained by marginalising over a hidden volatility variable:
\begin{eqnarray}\label{eq_marginalisation}
 f(s_t| s_{t-1}, s_{t-2}, r_{t-1}, r_{t-2}) =\\
 \int f(s_t, \sigma_{t-2}| s_{t-1}, s_{t-2}, r_{t-1}, r_{t-2})) d\sigma_{t-2}.
\end{eqnarray}

For simplicity, let us denote the conditional density of spread with $\phi(x)$
\begin{eqnarray*}
 f(s_t = y | s_{t-1}, s_{t-2}, r_{t-2}, \sigma_{t-2}) = \phi(y).
\end{eqnarray*}
Then the value of spread at time $t$ is connected to the conditional variables as
\begin{eqnarray*}
s_t^2 = \underbrace{a s_{t-1}^2 + b(w+\alpha r_{t-2}^2 +\beta \sigma_{t-2}^2+\gamma s_{t-2}^2)}_{c^{*}_t} + c \epsilon^2,
\end{eqnarray*}
where $\epsilon^2$ is chi-square distribution. Note that spread can be rewritten as
\begin{eqnarray*}
s_t = \sqrt{c^{*}_t + c \epsilon^2},
\end{eqnarray*}
where $\sqrt{c \epsilon^2}$ is Nakagami distribution. Then we have
\begin{eqnarray*}
\phi(x) = \frac{2\sqrt{0.5}}{\Gamma(0.5) \sqrt{c}} \exp \left( \frac{-x^2}{2c}  \right),
\end{eqnarray*}
and $c^{*}_t + c \epsilon^2$ is shifted Gamma distribution. Under the monotone transformation of random variable $y=g(x)=\sqrt{c^{*}_t+x}$, we can compute the p.d.f. of $s_t$ as:
\begin{eqnarray*}
\phi(y) = f_x(g^{-1}(y))\left|\frac{d}{dy} (g^{-1}(y))\right|,
\end{eqnarray*}
where,
\begin{eqnarray*}
f_x(x) = \frac{1}{\Gamma(1/2)\sqrt{2c}}x^{-1/2}\exp\left( -\frac{x}{2c} \right),
\end{eqnarray*}
and
\begin{eqnarray*}
g^{-1}(y) = y^2 - c^{*}_t
\end{eqnarray*}
Finally we get conditional probability density for spread $ f(s_t = y | s_{t-1}, s_{t-2}, r_{t-2}, \sigma_{t-2})=\phi(y)$
\begin{eqnarray*}
\phi(y) =  \frac{2y}{\Gamma(1/2)\sqrt{2c(y^2 - c^{*}_t)}}\exp{\left( -\frac{y^2 - c^{*}_t}{2c} \right)} \bold{1}_{ [{\sqrt{c^{*}_t},\infty }\big) } (y),
\end{eqnarray*}
where $c^{*}_t= a s_{t-1}^2 + b(w+\alpha r_{t-2}^2 +\beta \sigma_{t-2}^2+\gamma s_{t-2}^2)$ and $\bold{1}_{\mathcal{A}}$ stands for an indicator function defined over a subset of real numbers, $\mathcal{A}$. Now we can get a numerical Monte Carlo estimate of the model-based transfer entropy as
\begin{eqnarray*}
 f(s_t=y| s_{t-1}, s_{t-2}, r_{t-1}, r_{t-2}) =\\
E_{\sigma_i \sim f(\sigma_{t-2})}[f(s_t| s_{t-1}, s_{t-2}, r_{t-1}, r_{t-2}, \sigma_i))=\\
\frac{1}{N}\sum_{i=1}^N \phi(y;c,c^{*}_t= a s_{t-1}^2 + b(w+\alpha r_{t-2}^2 +\beta \sigma_{i}^2+\gamma s_{t-2}^2)).
\end{eqnarray*}
and $f(s_t=y| s_{t-1}, s_{t-2})$ is an expectation
\begin{eqnarray*}
 f(s_t=y| s_{t-1}, s_{t-2})=\\
E_{r_i,r_j \sim f(r_{t-1},r_{t-2})}[f(s_t=y| s_{t-1}, s_{t-2}, r_{t-1}, r_{t-2})).
\end{eqnarray*}
 In \figref{model_fig_spread_to_returns} and \figref{model_fig_returns_to_spread} we show
agreements of transfer entropy obtained from such analytical integration, in comparison to transfer entropy values from returns to spread with KSG estimates shown in corresponding panels of \figref{fig_spread_to_returns_main}. For the case shown in \figref{fig_returns_spread_bid}, where we have bidirectional flow, namely, $b>0$ and $\gamma>0$, deriving theoretical density distributions requires special attention since we need to account for dependence between spread and volatility in marginalisation step, \eqref{eq_marginalisation}, that falls out of the scope of current work.

\end{document}